\documentclass[3p,times,authoryear]{elsarticle} %twocolumn

\usepackage{ecrc}

\volume{00}

\firstpage{1}

\journalname{Life Sciences in Space Research}

\runauth{Liu \& Dobynde et al.}

\jid{LSSR}

\jnltitlelogo{{\large Life Sciences in Space Research}}

\CopyrightLine{2011}{Published by Elsevier Ltd.}

\usepackage{amssymb}
\usepackage{multirow, url, color}
\usepackage{bbding, amssymb} % symbols
\usepackage{ulem} % underline for emphasis
\usepackage{verbatim} % display LaTeX commands
\usepackage{enumerate}
\usepackage{amsmath}
\usepackage{booktabs} % Table Top/Mid/Bottom Rule
\usepackage{graphicx}
\usepackage[unicode,colorlinks]{hyperref}
\usepackage{orcidlink}
\usepackage{ifthen,afterpage}

\usepackage[figuresright]{rotating}

\begin{document}

\begin{frontmatter}
%% Title, authors and addresses
%% use the tnoteref command within \title for footnotes;
%% use the tnotetext command for the associated footnote;
%% use the fnref command within \author or \address for footnotes;
%% use the fntext command for the associated footnote;
%% use the corref command within \author for corresponding author footnotes;
%% use the cortext command for the associated footnote;
%% use the ead command for the email address,
%% and the form \ead[url] for the home page:
%%
%% \title{Title\tnoteref{label1}}
%% \tnotetext[label1]{}
%% \author{Name\corref{cor1}\fnref{label2}}
%% \ead{email address}
%% \ead[url]{home page}
%% \fntext[label2]{}
%% \cortext[cor1]{}
%% \address{Address\fnref{label3}}
%% \fntext[label3]{}

\dochead{}
%% Use \dochead if there is an article header, e.g. \dochead{Short communication}
%% \dochead can also be used to include a conference title, if directed by the editors
%% e.g. \dochead{17th International Conference on Dynamical Processes in Excited States of Solids}

\title{Time-Dependent Radiation Quality Factor $\langle Q \rangle$ of Galactic Cosmic Rays in Deep Space and Shielded Environments: Modeling and Measurements} % Time-Dependent

%% use optional labels to link authors explicitly to addresses:
%% \author[label1,label2]{<author name>}
%% \address[label1]{<address>}
\author[2]{Weihao Liu\orcidlink{0000-0002-2873-5688}\corref{cofirst}}
\author[2]{Mikhail Dobynde\orcidlink{0000-0001-9523-5002}\corref{cofirst}}
\cortext[cofirst]{These authors contribute equally to this work.}
\author[2]{Jingnan Guo\orcidlink{0000-0002-8707-076X}\corref{corr}}
\cortext[corr]{Corresponding author}
\ead{jnguo@ustc.edu.cn}
\author[3]{Jordanka Semkova\orcidlink{0000-0002-2366-2176}}
\author[3]{Krasimir Krastev}
% \author[4]{Cary Zeitlin\orcidlink{0000-0002-1737-141X}}

% \address[1]{Department of Climate and Space Sciences and Engineering, University of Michigan, Ann Arbor, MI 48109, USA}
\address[2]{National Key Laboratory of Deep Space Exploration/School of Earth and Space Sciences, University of Science and Technology of China, Hefei 230026, People's Republic of China}
\address[3]{Space Research and Technology Institute, Bulgarian Academy of Science, Sofia 1113, Bulgaria}
% \address[4]{Leidos Inc., Houston, TX 77042, USA}

\begin{abstract}
%% Text of abstract
Understanding the long-term variation of the galactic cosmic ray (GCR) radiation environment is critical for assessing radiation risks in space exploration missions. 
In this study, we systematically model the linear energy transfer (LET) spectra of GCRs and the corresponding radiation quality factor, $\langle Q \rangle$, in deep space and shielding environments. The Badhwar–O'Neill 2020 (BON20) model is used to represent GCR fluxes under different solar modulation potentials ($\phi$), which characterize the level of solar activity. GCR interactions with spherical shielding of different thicknesses are simulated to obtain the LET spectra, absorbed dose, dose equivalent, and $\langle Q \rangle$. 
We present a comprehensive dataset of these quantities for a range of $\phi$ values and shielding thicknesses. The results show that $\langle Q \rangle$ depends strongly on the shielding thickness but only weakly on solar activity. Furthermore, model predictions are validated against long-term measurements from the Cosmic Ray Telescope for the Effects of Radiation (CRaTER) orbiting the Moon, as well as the Liulin-MO detector on board the ExoMars Trace Gas Orbiter (TGO) orbiting Mars. 
In this comparison, we account for factors for anomalous cosmic ray (ACR) contributions and the radial gradients of both GCRs and ACRs, applying scaling factors of 6.3\% at 1 AU and 11.0\% at 1.5 AU to the calculated absorbed dose rate. With these corrections included, the modeled absorbed dose rate and $\langle Q \rangle$ exhibit consistent temporal variations with the observations under both thin and thick shielding conditions. 
Moreover, we investigate the distinct temporal evolution of $\langle Q \rangle$ for light and heavy GCR nuclei, revealing how solar modulation influences the elemental radiation quality factor across GCR species. These results offer new insights into the temporal and environmental dependence of the space radiation quality factor, with implications for radiation dose estimate and crewed mission design. 
\end{abstract}

\begin{keyword}
    Galactic cosmic rays \sep 
    Linear energy transfer \sep 
    Radiation quality factor \sep 
    Radiation transport modeling 
%% keywords here, in the form: keyword \sep keyword
%% PACS codes here, in the form: \PACS code \sep code
%% MSC codes here, in the form: \MSC code \sep code
%% or \MSC[2008] code \sep code (2000 is the default)
\end{keyword}

\end{frontmatter}
    
%% 3p, twocolumn Adjustment
% \ifthenelse{\boolean{@twocolumn} 
%     \AND \equal{\jtype}{3}}{
%     \vspace*{0em}
%     \afterpage{\vspace*{0em}}}{}
% \clearpage
%% Start line numbering here if you want
% \linenumbers

%% main text
\section{Introduction} \label{sec:intro_01}

Deep space exploration has been a central focus of space agencies worldwide for decades. Space radiation, however, can present severe risks not only for the operational lifespan of satellites and the performance of onboard instruments but also for the health of astronauts \citep{zheng2019space, jun2024review}. 
In the heliosphere, the radiation environment is mainly composed of the background galactic cosmic rays (GCRs), anomalous cosmic rays (ACRs) and occasional solar energetic particle (SEP) events \citep[e.g.,][]{kudela2009energetic, guo2024particle}. GCRs are the slowly varying background energetic particles that originate outside the solar system. %They have a wide energy distribution that ranges from below $10^{6}\;\mathrm{eV\; nuc^{-1}}$ to over $10^{18}\;\mathrm{eV\; nuc^{-1}}$ with a peak around $10^{9}\;\mathrm{eV\; nuc^{-1}}$, after which their spectra follow power-law distributions \citep{biermann2001introduction}. 
In the inner heliosphere, GCR particles contain $\sim$98\% nuclei and $\sim$2\% electrons and positrons, in which the nuclei consist of 87\% protons, 12\% helium ions and $\sim$1\% heavier ions \citep{simpson1983}. % They undergo persistent solar modulation with a period of 11 years, during which the GCR fluxes are influenced by the varying solar activity, heliospheric magnetic field strength, and solar wind, resulting in reduced fluxes at solar maximum and enhanced fluxes at solar minimum \citep[e.g.,][]{cliver20131859, potgieter2013cosmic, chowdhury2016study}. 
On the other hand, ACRs, typically in the energy range of a few to $\sim$100 MeV/nuc, are generally believed to be produced from interstellar neutrals that enter the heliosphere and become ionized, transported outward as pickup ions, and accelerated at the termination shock \citep[e.g.,][]{garcia1973new, mcdonald1974anomalous}. 
Alternatively, SEPs originate from solar eruptive events and are usually observed over a wide range of energies, from suprathermal (a few keVs) to relativistic (a few GeVs) \citep{Reames-2013, desai2016large}. While SEPs contribute to sporadic radiation hazards and ACRs also have slight influences, radiation induced by GCRs, especially those with high-energy and high-charge ions, is the dominant source of background ionizing radiation in interplanetary space and represents a major concern for long-duration human and robotic missions beyond Earth's magnetosphere \citep[e.g.,][]{cucinotta2013safe, norbury2019advances, guo2021review, cucinotta2024non, dobynde2024guidelines}. 

% Para 2: Radiobiology Perspective: LET and <Q> Factor
To evaluate radiation levels, specific physical and biological quantities are employed. The absorbed dose is defined as the ratio of the deposited energy in a material to its mass and is expressed in Grays ($1\; \mathrm{Gy} = 1\; \mathrm{J\; kg^{-1}}$). Division by the accumulation time yields the absorbed dose rate ($D$). Advanced radiation detectors, often using coincidence techniques, measure the energy deposition distribution within a detector to produce the linear energy transfer (LET, denoted as $L \equiv \mathrm{d}E/\mathrm{d}x$, with the energy $\mathrm{d}E$ deposited along the path of $\mathrm{d}x$) spectrum, which demonstrates the distribution of energy loss per unit of path length \citep[e.g.,][]{zeitlin2013measurements, zeitlin2019quality}. According to the commonly used recommendations of the International Commission on Radiological Protection (ICRP) \cite{icrp60}, the radiation quality factor, $Q=Q(L)$, is a function of LET in water ($L$) and serves as a biological weight to relate absorbed dose rates to dose equivalent rates ($H$), the latter being expressed in Sieverts ($1\; \mathrm{Sv} = 1\; \mathrm{J\; kg^{-1}}$). The averaged radiation quality factor, $\langle Q \rangle$, can then be evaluated given the energy-integrated absorbed dose and dose equivalent: $\langle Q \rangle = H/D$. It represents the overall biological effectiveness of the radiation environment, with higher values indicating a greater potential of biological damage per unit of absorbed dose. 

% Para 3: Highlight the Meaning of <Q> and Our Study of Its Dependence on the Solar Activity / Shielding
In the heliosphere, GCR particles undergo persistent solar modulation with an 11-year cycle, during which GCR fluxes are influenced by varying solar activity, heliospheric magnetic field strength, and solar wind, resulting in reduced fluxes at solar maximum and enhanced fluxes at solar minimum \citep[e.g.,][]{cliver2013solar, potgieter2013cosmic, chowdhury2016study}. Consequently, the LET spectrum, which depends on the composition and energy distribution of GCRs, also varies gradually throughout the solar cycle \citep{looper2013radiation, zeitlin2016CRaTER}. 
Meanwhile, spacecraft with different structural and material configurations experience distinct shielding environments in different periods, reporting different radiation doses in their measurements. 
Since the overall radiation quality factor, $\langle Q \rangle$, is closely linked to the LET spectrum that varies with solar modulation and shielding thicknesses, it is essential to examine the long-term radiation environment and understand the dependence of $\langle Q \rangle$ on these factors to better assess radiation risks and design protective measures. 
    
% Para 4: Missions / Measurements
Over the past 20 years, multiple space missions have been launched to study the radiation environment in the inner heliosphere. These missions have collectively improved our understanding of space radiation and provided a substantial benchmark for radiation transport models \cite[e.g.,][]{looper2013radiation, matthia2016martian, matthiae2017radiation, guo2019atris, simonsen2020nasa, ZhangJ2022JGR, liu2023modeling, lyu2024long}. Among these, several radiation detectors have provided valuable measurements:
\begin{itemize}
    \item[1.] The Cosmic Ray Telescope for the Effects of Radiation \citep[CRaTER,][]{spence2010, mazur2011new} on board the Lunar Reconnaissance Orbiter \citep[LRO,][]{chin2007lunar, vondrak2010lunar}, launched in 2009 and orbiting the Moon, has offered invaluable data on the lunar radiation environment, including the LET spectrum and absorbed dose rate, for over 15 years \citep[e.g.,][]{schwadron_lunar_2012, zeitlin2013measurements, zeitlin2019quality, looper2020long}. 
    \item[2.] The Lunar Lander Neutron and Dosimetry \citep[LND,][]{wimmer2020lunar} experiment on board the Chang'E-4 mission, launched in 2019, has contributed significantly to our understanding of the lunar surface radiation dose and provided insight into the radiation environment on the far side of the Moon \citep[e.g.,][]{zhang2020first, Xu_2022}. 
    \item[3.] As a secondary payload as part of the Artemis I mission \citep{smith2020artemis, rahmanian2023galactic}, launched in 2022, the BioSentinel CubeSat \citep{ricco2020biosentinel} has entered a permanent heliocentric orbit and transmitted physical measurements of the lunar and deep-space radiation environment in the inner heliosphere \citep[e.g.,][]{george2024space, rahmanian2024galactic}. 
    \item[4.] Farther away, the Radiation Assessment Detector \citep[RAD,][]{hassler2012} on the Mars Science Laboratory \citep[MSL,][]{grotzinger2012mars} mission was launched in 2012. It has been crucial in quantifying radiation exposure on the Mars's surface and delivering critical data for future manned missions to Mars \citep[e.g.,][]{zeitlin2013, hassler2014, ehresmann2014, guo2021review}. 
    \item[5.] The Liulin-MO detector \citep{Semkova2018} on board the ExoMars Trace Gas Orbiter \cite[TGO,][]{vago2015esa} was launched in 2016. It has been orbiting Mars and measuring particle fluxes, LET spectra, and absorbed dose rates since 2018 \citep[e.g.,][]{semkova2021results, semkova2023observation}.  
\end{itemize}

% Para 5: Models/Simulations
In addition to the \textit{in-situ} observations, numerous modeling studies have analyzed the deep-space radiation environment under various shielding conditions and its biological effects on human tissues. These studies are generally based on GCR models and particle transport codes \cite[e.g.,][]{deangelis2006modeling, cucinotta2010space, kim2014HZETRN, slaba2017optimal, banjac2018, Khaksarighiri2020, dobynde2021beating, zaman2022modeling, li2023impact, charpentier2024aramis}. 
A recent study by \cite{liu2024comprehensive} provided a comprehensive analysis of several commonly-used GCR models, highlighting their strengths and limitations. Based on this study, we adopt the Badhwar‐O'Neill 2020 (BON20) model \citep{slaba2020badhwar} as the GCR model in this study, since it provides reasonably accurate proton and helium GCR fluxes in deep space. 
In addition, various particle transport codes have been utilized to investigate the behavior of energetic particles in deep space as well as their radiation transport and effects. Notable examples include the GEometry ANd Tracking \citep[GEANT4,][]{agostinelli2003} program, the FLUctuating KAscad \citep[FLUKA,][]{andersen2004fluka, battistoni2015overview} code, the High charge (Z) and Energy TRaNspor \cite[HZETRN,][]{wilson1991, wilson2014advances, slaba2010faster} code, the Particle and Heavy Ion Transport code System \cite[PHITS,][]{niita2006phits, sato2018features}, and the Monte Carlo N–Particle transport code \cite[MCNP6,][]{goorley2012initial, kulesza2022mcnp}. Many of these tools are Monte Carlo-based codes and enable detailed analyses of GCR particle interactions with different phantoms under various conditions. They provide GCR particle spectra for different shielding thicknesses and the corresponding LET spectra, allowing the derivation of $\langle Q \rangle$ values by combining the GCR models with particle transport codes. 

% Para 6: Focus of this Work & Paper Structure
% In this work, we aim to explore the variation of the radiation quality factor, $\langle Q \rangle$, under different solar activities and shielding conditions. 
In this work, we model the long-term LET spectrum and the corresponding $\langle Q \rangle$ values of GCRs in deep space and under shielding conditions, with the aim of quantifying variations of $\langle Q \rangle$ with solar activity and shielding thicknesses. The GCR fluxes are provided by the BON20 model with modifications considering additional contributions by ACRs when calculating the radiation dose, and GCR particle interactions with a water phantom under various shielding conditions are simulated using GEANT4. For validation, we compare our modeled results with measurements from CRaTER and TGO/Liulin-MO, as these detectors can provide long-term deep-space and/or shielded dose rates together with LET spectra. 
In Section \ref{sec:method_02}, together with \ref{sec:appendCal}, we describe our modeling setups and correction factors used in dose calculations, as well as instrumental details about CRaTER and TGO/Liulin-MO. Section \ref{sec:result_03} presents the modeled $\langle Q \rangle$ values for both deep space and shielding environments, followed by model--data comparisons and physical interpretations. Further discussions on elemental contributions to radiation dose, as well as on the differing $\langle Q \rangle$ variations between lighter and heavier elements over the solar cycle are provided in Section \ref{sec:discuss_04}. Finally, our main conclusions are summarized in Section \ref{sec:sumcon_05}.

\section{Methodology and Data} \label{sec:method_02}

\subsection{The BON20 Model for Deep-Space GCRs} \label{sec:method_02_GCR01}
    
    As inputs to our radiation calculations, we adopt the GCR particle spectra provided by the BON20 model \citep{slaba2020badhwar}. BON20 numerically solves the one-dimensional Parker transport equation \citep{parker1965, gleeson1967cosmic, gleeson1968solar, F1971JGR....76..221} using a finite-difference method under the assumption of a quasi-steady, spherically symmetric heliosphere. The model provides energy spectra for different GCR particle species modulated to 1 AU under varying solar activity conditions. In this work, we consider GCR particles ranging from hydrogen (H) to iron (Fe) nuclei, i.e., atomic number $Z=1$–26. 
    
    In BON20, the level of solar activity is characterized by the solar modulation potential $\phi$, which reflects the averaged energy loss of GCRs as they propagate from the interstellar medium into the inner heliosphere \citep{gleeson1968solar}. The value of $\phi$ can be derived from the Sunspot Number (SSN) or oxygen fluxes measured by the Cosmic Ray Isotope Spectrometer \cite[CRIS\footnote{\url{https://izw1.caltech.edu/ACE/ASC/level2/lvl2DATA_CRIS.html} \label{fnt:acecris}},][]{Stone1998} on board the Advanced Composition Explorer \cite[ACE,][]{stone1998advanced} spacecraft, both serving as proxies for solar activity. While BON20 performs slightly better when driven by the ACE/CRIS data, we adopt the SSN as input owing to its long-term availability and better prediction capability. The monthly SSN data are obtained from the World Data Center SILSO\footnote{\url{https://www.sidc.be/SILSO/datafiles} \label{fnt:ssn}}, Royal Observatory of Belgium, Brussels \citep{sidc}. 
    
    The modulation potential $\phi(t)$ is incorporated into the transport equation by varying the diffusion coefficient, thereby linking solar activity directly to the particle transport process. This approach enables BON20 to reproduce the observed GCR spectra at the heliocentric distance ($r$) of 1 AU with improved consistency across particle species and over different phases of the solar cycle. In subsequent calculations, the BON20-derived fluxes are treated as deep-space GCR particle fluxes, denoted as $J_{Z}(E)$ for different species, where $E$ is the kinetic energy per nucleon. 
    
    Note that BON20 includes GCRs only, while ACRs in the background also contribute to the radiation environment. Currently, only a few models, such as the Cosmic-Ray Effects on MicroElectronics \citep[CR\`EME\footnote{\url{https://creme.isde.vanderbilt.edu}},][]{tylka1997creme96, adams2012creme}, incorporate both GCRs and ACRs and are publicly available. However, CR\`EME provides fluxes with a yearly cadence, making it less suitable for our temporal (monthly) resolution analysis. 
    
    Instead of directly using the CR\`EME fluxes in the following analysis, we use the CR\`EME and BON20 comparison results to account for ACR contributions. We first refer to the long-term model comparison by \cite{liu2024comprehensive}, which shows that CR\`EME generally overestimates the absorbed dose rate by 2.3\%, while BON20 underestimates it by 5.9\% with respect to CRaTER measurements at $r=1$ AU. 
    Accordingly, we apply a scaling factor of $1/(1-0.059) \approx 1.063$ to the calculated radiation dose to account for ACR contributions at $r=1$ AU hereafter, effectively removing the bias relative to long-term CRaTER measurements. 
    Furthermore, since BON20 fluxes are modulated to 1 AU, radial gradients of both GCRs and ACRs must be considered for better radiation dose estimates elsewhere in the inner heliosphere, such as near Mars at $r = 1.5$ AU. The analysis and derivation of this correction factor are detailed in \ref{sec:appendCal}, demonstrating that factors of 11.0\% and 9.7\% are respectively applied to the absorbed dose rate and the dose equivalent rate calculated at 1.5 AU relative to the original BON20 results. 

    % Accordingly, we scale the absorbed dose and dose equivalent in our calculations to account for the ACR contributions hereafter. Furthermore, BON20 fluxes are modulated to 1 AU. Observations suggest a radial gradient of 2–5\%/AU for GCR fluxes in the inner heliosphere \citep[e.g.,][]{gieseler2016spatial, vos2016global, honig2019multi, Roussos2020}. Hence, a representative 3.0\%/AU radial gradient is adopted to correct the calculated absorbed dose and dose equivalent when applied to Mars. 

\subsection{Modeling the LET Spectrum and \texorpdfstring{$\langle Q\rangle$}{\it Q} Factor} \label{sec:method_02_G4simu02}

    % Calculating the average quality factor $\langle Q\rangle$ from modeled results requires us to calculate total absorbed dose D and total dose equivalent H in the detector. Dose equivalent is calculated for each segment of particle track by multiplying absorbed dose along this track segment by the LET-dependent quality factor on this track segment. For analyzing measured radiation fields in space missions, the most commonly used Q(LET) dependencies is defined by ICRP60 \cite{icrp60}. Thus radiation exposure could be described with LET dose spectra, which being convoluted with Q(LET) gives dose equivalent value. In the same radiation environment, LET spectra is different in different materials.  LET in space is often measured in thin silicon slabs and converted to that in a water slab, so we use a thin water layer as the detector in our modeling.
    Calculating the average quality factor, $\langle Q\rangle$, from the modeled results requires computing both the total absorbed dose $D$ and the total dose equivalent $H$ in the detector. The dose equivalent is obtained for each segment of a particle track by multiplying the absorbed dose along that segment by the LET-dependent quality factor, $Q(L)$, which defined in \cite{icrp60} is adopted in this work. 
    % For analyzing measured radiation fields in space missions, the most commonly used relationship between $Q$ and LET is that defined by ICRP Publication 60 \citep{icrp60}. In this formalism, the radiation exposure can be represented by the LET dose spectrum, which, when convolved with $Q(\mathrm{LET})$, yields the total dose equivalent. 
    The LET spectrum depends on the detector material, and LET in space is commonly measured in thin silicon slabs and converted to its equivalent in water. In this study, we employ a thin water spherical layer as the detector. 

    The transport of particles through the shielding material is simulated using the Monte Carlo radiation transport code GEANT4\footnote{\url{https://geant4.web.cern.ch/} \label{ftn:geant4}} \citep{agostinelli2003}. The particle–matter interactions are described by the QGSP\_INCLXX\_HP physics model, which combines the Quark–Gluon String Precompound (QGSP) model for particles with energies greater than 10 GeV, the Li\`ege Intranuclear Cascade (INCLXX) model for energies below 10 GeV and the High Precision Neutron (HP) model for neutrons with energies below 20 MeV. 
    
    % To calculate LET-spectra in a water detector under different shielding conditions, we used a two-step approach. First, we calculate the radiation environment inside the aluminum shielding, and then we calculate the resulting LET spectra in a thin water detector. Both steps are organized in the response function concept: the energy range is divided into bins and for one thread of modeling with a specific particle type and energy is selected randomly from this energy bin, i.e. energy distribution inside the bin is flat. The energy grid is in log scale, with bin edges at 10$^{0.9:.1:5.1}$ for incoming GCRs and 10$^{-9.1:.1:6.1}$ for incoming secondary radiation. To calculate resulting flux of the LET-spectrum we convolute response functions with bin-integrated GCR particle flux. 
    To compute the LET spectra in a water detector under different shielding conditions, we adopt a two-step approach. First, we calculate the radiation environment inside the aluminum shielding; then we calculate the resulting LET spectra in a thin water detector. Both steps are formulated using the response-function method \cite[e.g., detailed in][]{guo2019atris}, in which the energy range is divided into logarithmically spaced bins. Within the corresponding bin, one particle type and energy are selected randomly for each Monte Carlo run, assuming a uniform energy distribution inside the bin. The energy bins for the primary GCRs range from $10^{0.9}$ to $10^{5.1}$ MeV with a logarithmic step of 0.1 orders of magnitude, while for secondary particles, the bins range from $10^{-9.1}$ to $10^{6.1}$ MeV with the same step size. Resulting fluxes of the LET spectrum are obtained by convolving the response functions with the bin-integrated GCR particle fluxes. 
    
    \begin{figure*}[ht!]
    \centering {\hspace{0.025\hsize}\includegraphics[width=0.925\hsize]{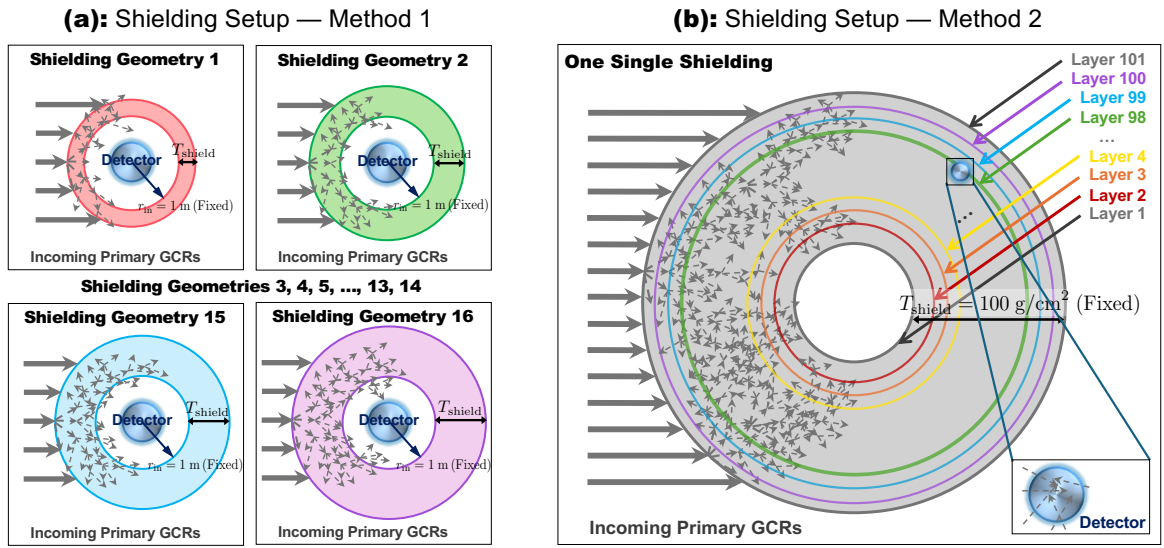}}
    \caption{Schematic diagram of the shielding setup adopted in this study. The layers shown here correspond to two-dimensional plane cuts. 
        (a) In the first method, we employ 16 distinct spherical shells, each with a specific shielding thickness ($T_\mathrm{shield}$). The inner radius of the shielding layer ($r_\mathrm{in}$, shown in the plot) is fixed at 1 m. 
        (b) In the second method, we use one single spherical shell configuration with a total thickness of 100 g/cm$^2$ and split it into 101 concentric layers. 
        In both approaches, isotropic primary GCRs are injected into the system. Particle fluxes inside the shielding (Method 1) or at the inner boundary of each layer (Method 2, also see a zoomed-in view at its lower-right corner) are used for dose calculations in a 1 mm thick spherical water ``detector" with an inner radius of 175 mm, which is located at the center of the shielding sphere.} \label{fig00:Model}
    \end{figure*}
    
    To model spacecraft shielding, we use a spherical geometry approximation which allows isotropic irradiation to be modeled using a unidirectional particle beam \cite{dobynde2021radiation}. %The spherical symmetry of the setup allows us to use unidirectional beam irradiation to reproduce isotropic particle incidence on the outer layer of shielding. 
    Two configurations of the shielding setup are employed, as demonstrated in Figure~\ref{fig00:Model}. %We made two sets of calculations. 
    In the first one (Method 1), we consider 16 different aluminum shells with a thickness of 0.1, 1, 5, 10, 15, 20, 25, 30, 35, 40, 50, 60, 70, 80, 90 and 100 g/cm$^{2}$ (see Figure~\ref{fig00:Model}(a)) and calculate corresponding flux-response functions inside the shielded region for each shell. The inner radius of the shell, $r_\mathrm{in}$, is fixed at 1 m. 
    Subsequently, the energy deposition and LET spectra are calculated in the ``detector" which is approximated by a spherical water layer with an inner radius of 175 mm and a thickness of 1 mm. The LET spectrum is computed for all relevant particle species, including primary GCRs, fragmented nuclei, secondary protons, electrons, neutrons and gamma rays at the grid with energy bin edges from $10^{-9.1}$ to $10^{6.1}$ MeV with a step size of 0.1 orders of magnitude. 
 
    The second configuration (Method 2) considers only a single spherical aluminum layer with a total surface density of 100 g/cm$^{2}$. Flux-response functions are calculated inside the aluminum layer for 101 radial distances with a step of 1 g/cm$^{2}$, as illustrated in Figure~\ref{fig00:Model}(b). 
    At each layer, inward-directed particles are considered for subsequent LET calculations (in the same manner as described above).      
    
    % The transportation of particles through the shielding is modeled using the Monte Carlo GEANT4 code \cite{agostinelli2003}. The particle-matter interaction is described with QGSP\_INCLXX\_HP physics model, which includes Quark-Gluon String Precompound model for particles with energies higher than 10 GeV, Leige INC (INCLXX) model for energies lower than 10 GeV and the High Precision Neutron model for energies lower than 20 MeV.

    % Next we calculate energy deposition and LET spectra in a spherical water layer with a 175 mm inner radius and a 1 mm thickness. The LET spectrum is calculated for primary GCRs, protons, gamma-rays, and electrons at the grid with bin edges at 10$^{-9.1:0.1:6.1}$. 
    % Subsequently, the energy deposition and LET spectra are calculated in a spherical water layer with an inner radius of 175 mm and a thickness of 1 mm. The LET spectrum is computed for all relevant particle species, including primary GCRs, fragmented nuclei, secondary protons, electrons, neutrons and gamma rays at the grid with energy bin edges from $10^{-9.1}$ to $10^{6.1}$ MeV with a step size of 0.1 orders of magnitude. 

\subsection{Measuring Absorbed Dose Rates and LET Spectra} \label{sec:method_02_obs03}
    
    In this study, we use data from two detectors, which operate under distinct shielding conditions, to characterize the long-term GCR radiation environment in the inner heliosphere. 
    
    Launched in June 2009 into a 50 km lunar orbit, CRaTER \citep{spence2010, mazur2011new} on board the LRO \citep{chin2007lunar, vondrak2010lunar} carries three pairs of silicon detectors, separated by tissue-equivalent plastic (TEP) blocks. Each pair consists of a thin detector (0.15 mm for D1, D3 and D5) and a thicker detector (1.0 mm for D2, D4 and D6) with low and high amplifier gains, respectively, and providing absorbed dose rate measurements at lunar orbit. 
    When instruments on board LRO are pointed toward the lunar nadir, the D1 and D2 pair (hereafter ``D1 and D2'' denotes coincidence logic) faces deep space with minimal shielding of 0.22 g cm$^{-2}$, while D5 and D6 face the Moon, and D3 and D4 are positioned in the middle between two TEP blocks. Among them, the D1 and D2 pair features a wide field of view ($169^\circ$ out of the full angle), a large geometric factor (24.15 cm$^2$ sr), and a low threshold energy for incoming particles (e.g., 12.7 MeV for protons) and measures the radiation environment closest to deep-space scenario without shielding \citep{spence2010}. Therefore, data from the D1 and D2 pair are used in our study and are publicly available in the CRaTER data center\footnote{\url{https://crater-web.sr.unh.edu/} \label{fnt:crater}}. 
    In this work, we remove transient spikes corresponding to SEP events, and the remaining absorbed dose rate primarily represents the GCR contribution. Note that the data released online are reported as absorbed dose rates in water on the lunar surface. We convert the acquired data to those in water at the Moon's orbit, accounting for contributions from orbital altitude and secondary particles, the latter being 8.6–19.9\% as reported by \cite{schwadron_lunar_2012} and \cite{zaman2020absorbed}. A detailed description of the CRaTER data conversion is presented in Section 2.3 of \cite{liu2024comprehensive}. 
    
\begin{figure}[ht!]
    \centering {\includegraphics[width=0.61\hsize]{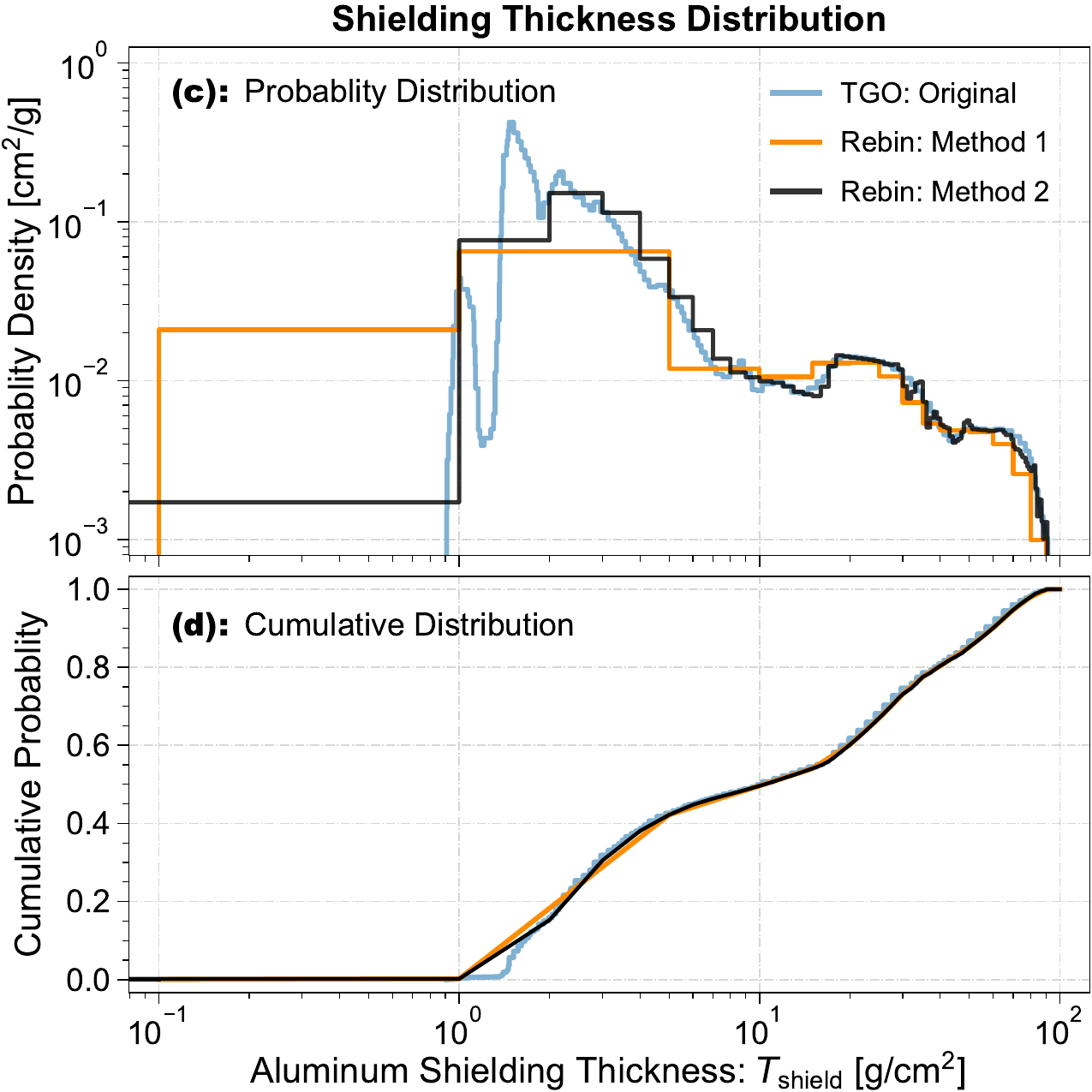}}
    \caption{Shielding thickness distribution of TGO/Liulin-MO. 
        (a) Probability distribution function. 
        (b) Cumulative distribution function. 
        In both panels, the original TGO/Liulin-MO shielding thickness distribution taken from \cite{semkova2021results} is shown in blue. The rebinned distribution functions corresponding to the first and second methods are shown in orange and black, respectively.} \label{fig01:TGOthick}
    \end{figure}

    In addition, we use measurements from the Liulin-MO\footnote{\url{https://esa-pro.space.bas.bg/datasources} \label{fnt:tgo}} dosimeter on board TGO \citep{semkova2015radiation, Semkova2018}. Launched in March 2016 and inserted into a 400 km Mars's orbit in April 2018 \citep{vago2015esa}, TGO carries the Liulin-MO dosimetry detector, which consists of two perpendicular pairs of silicon photodiodes and corresponding amplifiers, each converting energy deposited by charged particles or photons into voltage pulses. 
    Accounting for the shielding from the spacecraft structure and residual fuel, Liulin-MO has an aluminum-equivalent shielding ranging from $\sim$1 to 100 g cm$^{-2}$, with an average value of $\sim$20 g cm$^{-2}$ \citep{semkova2021results}. The corresponding probability density function of the shielding distribution is shown in Figure \ref{fig01:TGOthick}(a). 
    In our analysis, we resample the shielding thickness distribution for the 16 or 100 shielding layers described in Section \ref{sec:method_02_G4simu02}. We first derive the cumulative distribution profile from the original probability density function and perform resampling along that profile, as shown in Figure \ref{fig01:TGOthick}(b). This procedure yields a rebinned shielding thickness distribution for both methods, as shown in Figure \ref{fig01:TGOthick}(a). The resulting distribution is used as weighting factors for different shielding layers when constructing the modeled spectra under TGO's shielding conditions. 

\section{Results} \label{sec:result_03}

\subsection{LET Spectrum and \texorpdfstring{$\langle Q\rangle$}{\it Q} Calculations for GCRs in Deep Space} \label{sec:result_03_Qlong_01}
    
    Using the GCR spectra derived from BON20 and the response functions from GEANT4 simulations, we first present the modeled GCR spectra and corresponding LET functions in deep space for October 2019. This period, occurring near the solar minimum between Solar Cycles 24 and 25, represents a condition of low solar modulation ($\phi=398$ MV) and thus elevated GCR fluxes. The analysis presented in this section serves as a reference case for subsequent calculations. 

    \begin{figure*}[ht!]
    \centering {\includegraphics[width=0.99\hsize]{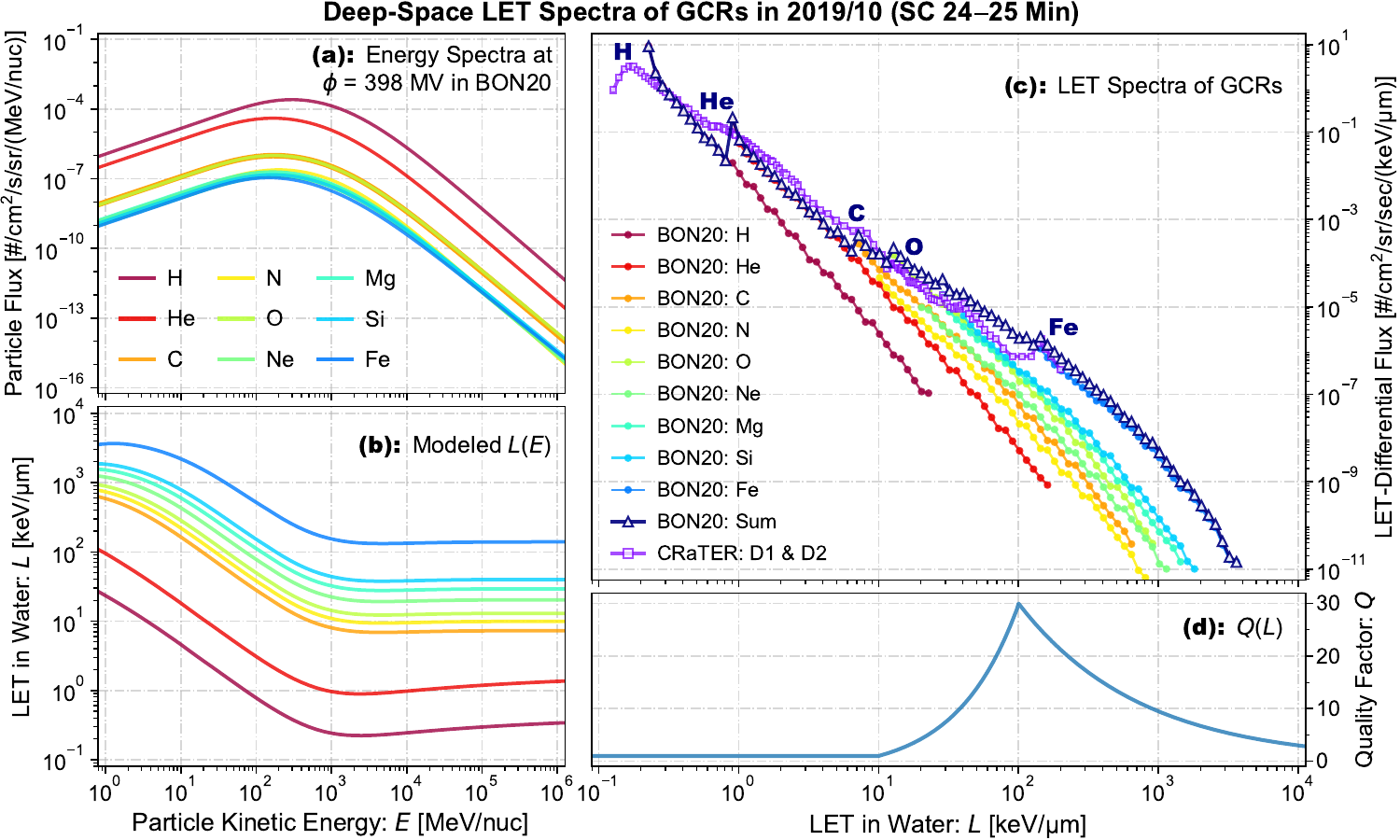}}
    \caption{Deep-space GCR spectra in October 2019 (solar minimum between SCs 24 and 25) and the corresponding LET spectra. 
        (a) GCR energy spectra of different species in BON20, with the modulation potential $\phi=398$ MV used as the model input for this period. 
        (b) LET distributions ($L \equiv \mathrm{d}E/\mathrm{d}x$) in water as functions of particle kinetic energy $E$ for different species of GCRs. Panels (a) and (b) share the same legend shown at the bottom-left corner of panel (a). 
        (c) Converted LET spectra of GCRs in deep space, derived from the results shown in panels (a) and (b). Curves with dotted points denote the modeled LET spectra for individual particle species; the dark-blue curve with triangles represents the LET spectrum summed over 26 elements (from H to Fe) in BON20, and the purple curve with squares shows measurements from the CRaTER D1 and D2 pair. Characteristic peaks in the LET spectrum for specific elements (H, He, C, O, and Fe) are marked. Detailed notations are listed at the bottom-left corner of this panel. 
        (d) $\langle Q \rangle$ as a function of $L$ (for LET), taken from \cite{icrp60}.} \label{fig02:allspectra201910}
    \end{figure*}
    
    Based on the monthly SSN in October 2019, we adopt a modulation potential of $\phi = 398$ MV for BON20 and obtain the differential energy spectra of individual GCR species, as shown in Figure \ref{fig02:allspectra201910}(a). 
    % As expected, lighter ions such as protons and helium dominate, while heavier nuclei contribute increasingly at higher linear energy transfer values due to their larger charge and mass. 
    Then, Figure \ref{fig02:allspectra201910}(b) presents the corresponding energy dependence of the LET in water, representing the energy loss per unit path length as GCR particles interact with the water sphere detector as illustrated in Figure \ref{fig01:TGOthick}. 
    For each ion species, $L(E)$ rises sharply toward low energies, reflecting enhanced ionization losses near the end of the particle range \citep[also known as the Bragg peak,][]{bragg1905xxxix}. In fact, the magnitude of $L$ scales approximately with $Z^2/A$, where $A$ is the atomic mass, resulting in substantially higher LET values for heavier ions such as iron. 

    Combining the BON20 energy spectra and the $L(E)$ relationships, we derive the modeled LET spectra in deep space, as shown in Figure \ref{fig02:allspectra201910}(c). The dotted curves correspond to individual species, while the dark-blue solid curve with triangles represents the total LET spectrum, summed over elements from H to Fe. For comparison, we also plot the GCR LET spectrum measured by the CRaTER D1–D2 detector pair (purple curve). Although data of the D1 and D2 pair correspond to a shielding thickness of 0.22 g/cm$^2$, we include them here as a reference for a near–deep-space environment. 
    Overall, the modeled total LET spectrum agrees well in both shape and magnitude with the CRaTER measurements over the range 0.3–200 keV/\textmu m. The modeled spectrum exhibits an extended high-LET tail due to contributions from heavy nuclei such as irons, which are largely absent in measurements because of the limited coincidence statistics. At the low-LET end, the modeled spectrum shows a gradual rise followed by obvious discontinuities at the characteristic peaks, which likely reflects distinct transitions between species and differences in the energy-loss behavior of light and heavy ions and is also found in a study by \cite{naito2022considerations}. 
    
    Finally, Figure \ref{fig02:allspectra201910}(d) shows the relationship between the radiation quality factor $Q$ and LET as recommended by the ICRP 60 report \citep{icrp60}. This function is subsequently applied to calculate the dose equivalent from the modeled LET spectra. For instance, here, by integrating the modeled LET spectrum, we obtain an absorbed dose rate of 192.7 mGy/year in water; combining with the $Q(L)$ function gives a dose equivalent rate of 1077.9 mSv/year, yielding an averaged $\langle Q \rangle$ of $\sim$5.6.

\subsection{Simulations of Shielding Environments} \label{sec:result_03_shield_02}
    
    \begin{figure*}[ht!]
    \centering {\includegraphics[width=1.0\hsize]{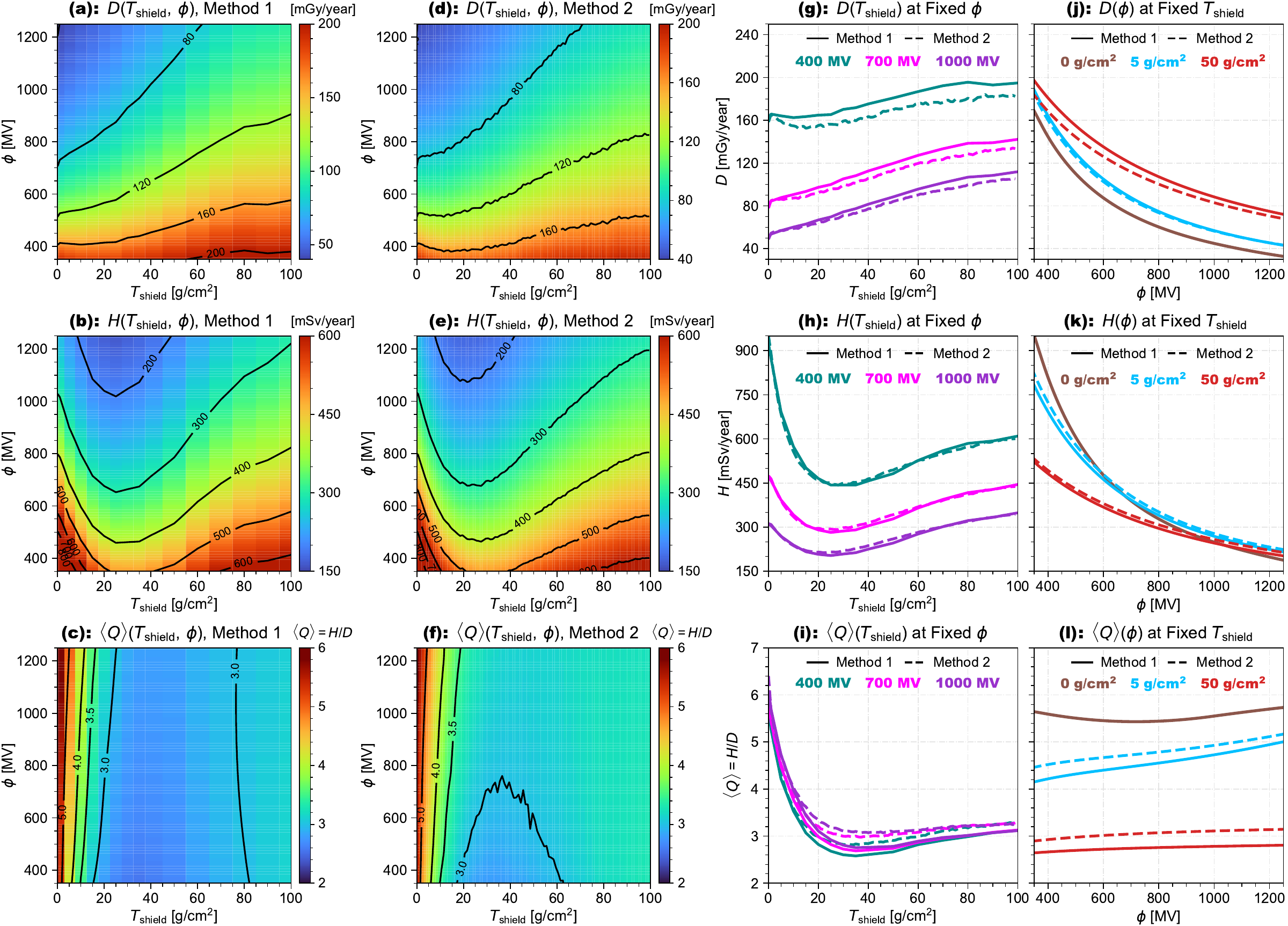}}
    \caption{Absorbed dose rate ($D$), dose equivalent rate ($H$), and the radiation quality factor $\langle Q \rangle$ under shielding conditions. 
        (a)–(c) $D$, $H$, and $\langle Q \rangle$ obtained via the first method, shown as functions of the aluminum shielding thickness ($x$-axis) and solar modulation potential ($y$-axis). 
        (d)–(f) Same parameters plotted in a same style as in panels (a)–(c), but obtained via the second method. 
        (g)–(i) $D$, $H$, and $\langle Q \rangle$ as functions of the aluminum shielding thickness ($x$-axis) for fixed modulation potentials: $\phi=400$ MV (green), 700 MV (magenta), and 1000 MV (purple). %In each panel, results from the first and second methods are plotted as solid lines and dashed lines, respectively. 
        (j)–(l) $D$, $H$, and $\langle Q \rangle$ as functions of the solar modulation potential ($x$-axis) for fixed aluminum shielding thickness values: $T_\mathrm{shield}=0$ g/cm$^2$ (i.e., deep space, in brown), 5 g/cm$^2$ (blue), 50 g/cm$^2$ (red). In each panel from (g) to (l), results from the first and second methods are shown as solid lines and dashed lines, respectively.} \label{fig03:DHQmap}
    \end{figure*}
    
    Taking the deep-space GCR fluxes derived from BON20 as incident particle spectra, we employ two configurations as described in Section \ref{sec:method_02_G4simu02} in GEANT4 to model the absorbed dose, dose equivalent and radiation quality factor under different aluminum shielding thicknesses. The simulations are performed for a full range of solar modulation potentials (0–2000 MV) and results for 350–1250 MV are shown to represent different phases of the solar cycle. We show the results for both simulation approaches in Figure \ref{fig03:DHQmap} and outline the main findings below. 
    
    \begin{itemize}
        \item[1.] Panels (a)–(c) and (d)–(f) show values of $D$, $H$, and $\langle Q\rangle$ as functions of the aluminum shielding thickness and solar modulation potential, derived from both simulation methods, respectively. For direct comparison, the same color scale and contour style are used for the corresponding quantities in both methods. Although the model configuration setup and the shielding thickness resolution are different, both methods exhibit consistent trends in the variation of $D$, $H$, and $\langle Q\rangle$. Notably, panels (c) and (f) reveal that $\langle Q\rangle$ depends strongly on the shielding thickness but only weakly on the modulation potential, i.e., the solar activity. 
        
        \item[2.] To highlight these variations, we plot $D$, $H$, and $\langle Q\rangle$ as functions of aluminum shielding thickness for fixed modulation potentials in panels (g)–(i), respectively. In both methods, for a given $\phi$, the absorbed dose rate rises gradually with increasing shielding thicknesses, reflecting contributions from secondary particles. 
        In contrast, the dose equivalent decreases sharply up to 10–30 g/cm$^2$ (depends on $\phi$ and the method), followed by a gradual increase at larger thicknesses. In thin shielding ($<$10 g/cm$^2$), $H$ drops due to the progressive attenuation of primary GCRs, and the characteristic minima indeed arise from the competition between primary GCR attenuation and secondary particle production. In thick shielding ($>$30 g/cm$^2$), $H$ rises because of the buildup of secondary neutrons, gamma rays and light ions. 
        Accordingly, $\langle Q\rangle$ exhibits a similar pattern, i.e., falling drastically to a minimum at $\sim$10–30 g/cm$^2$, then rising gradually toward $\sim$3 at higher thicknesses. 
        Although both methods capture these variations consistently, slight differences ($\sim$10\%) can be found in their results, as also illustrated in panels (j)–(l). It is likely because Method 2 includes a small contribution of secondary particles generated in the inner layers that are first directed outward and inward again. Note that only inward fluxes are taken at each layer of the subsequent LET calculations. 
        
        \item[3.] Similarly, we show $D$, $H$, and $\langle Q\rangle$ as functions of the modulation potential in deep space (0 g/cm$^2$) and for cases with thin (5 g/cm$^2$) and thick (50 g/cm$^2$) shielding. As the modulation potential increases (i.e., from solar minimum to maximum), both $D$ and $H$ decrease because of the reduced GCR fluxes under stronger solar activity. The value of $\langle Q\rangle$ varies modestly with $\phi$, showing a fluctuation of $\sim$1 and again indicating that shielding plays a more dominant role than solar modulation in shaping the radiation quality factor. 
    \end{itemize}
    
    To summarize, both simulation methods show consistent trends in $D$, $H$ and $\langle Q\rangle$ as functions of aluminum shielding and solar modulation potential. The tabulated results presented above can serve as quick references for estimating the absorbed dose rate, dose equivalent rate and $\langle Q\rangle$ given specific values of shielding thickness and modulation potential. Quantitative differences in thin shielding probably originate from albedo interactions included in Method 2. 
    As reported by \cite{schwadron_does_2014}, $\langle Q\rangle$ reaches $\sim$6 for $\phi=400$ MV at a thin shielding of 1 g/cm$^2$, consistent with the results obtained from both methods shown in panel (i). For the subsequent analysis, we adopt the results based on Method 1, as it provides a more realistic treatment for each given shielding thickness, while we note Method 2 may serve as a useful approximation for rapid estimates across various shielding thicknesses. 
    % Therefore, we adopt the results of Method 1 for subsequent comparison with measurements and further analysis. This choice does not imply that Method 2 is inaccurate; it captures additional contributions of albedo-produced secondary particles, which are important in realistic particle interactions. 

\subsection{Validations against CRaTER and TGO Measurements} \label{sec:result_03_shield_03}
    
    We next validate our modeled results by comparing them with CRaTER and TGO measurements in different shielding environments. 

    \begin{figure}[ht!]
    \centering {\includegraphics[width=0.61\hsize]{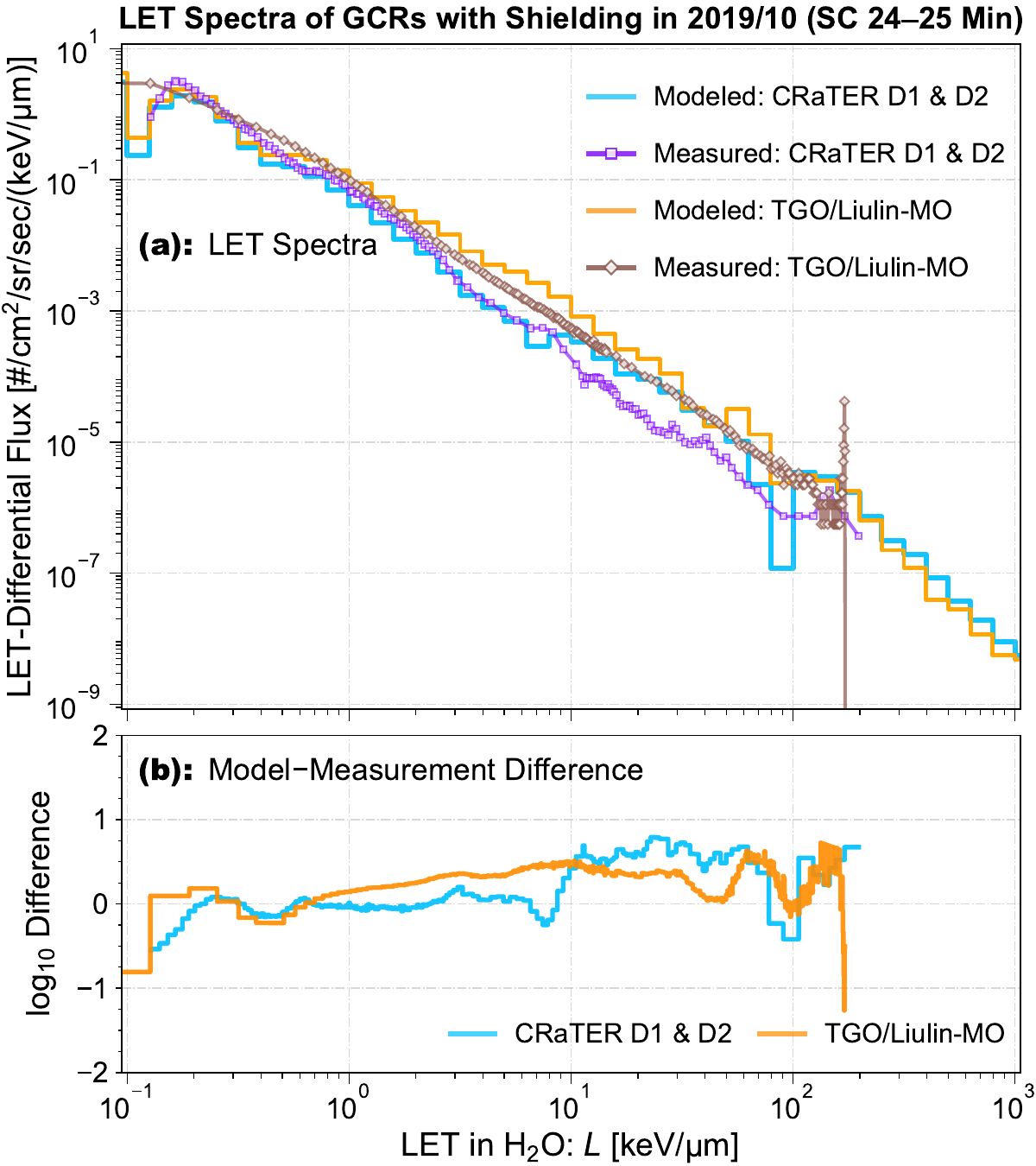}}
    \caption{LET spectra of GCRs from both modeling and measurements in October 2019 (solar minimum between SCs 24 and 25). 
        (a) Modeled and measured LET spectra. The modeled LET spectrum for the CRaTER D1 and D2 pair with a thin aluminum shielding thickness of 0.22 g/cm$^2$ is shown in blue, with the corresponding measured spectrum plotted as a purple curve with squares. The modeled LET spectrum for TGO/Liulin-MO, based on the shielding thickness distribution in Figure \ref{fig01:TGOthick}, is shown in orange, with the corresponding measurement plotted as a brown curve with diamonds. 
        (b) Logarithmic differences between the modeled and measured LET spectra for CRaTER (blue) and TGO/Liulin-MO (orange).} \label{fig04:shield}
    \end{figure}
    
    To start with, we examine the LET spectra. As an example, in Figure \ref{fig04:shield}(a), we compare the modeled and measured LET spectra for CRaTER and TGO/Liulin-MO in October 2019, corresponding to the solar minimum between SCs 24 and 25. The modeled spectrum for the CRaTER D1 and D2 pair with a thin aluminum shielding of 0.22 g/cm$^2$ (blue) closely reproduces the measured spectrum (purple, with squares), remaining within one order of magnitude. Similarly, the modeled LET spectrum for TGO/Liulin-MO (orange), computed using the rebinned shielding thickness distribution shown in Figure \ref{fig01:TGOthick}, overall agrees well with the measured data (brown, with diamonds). 
    At high LET values, the measured Liulin-MO data show a noticeable peak in the last several LET channels. This occurs because these channels include particles with LETs exceeding their nominal upper limits, leading to partial signal saturation in the instrument electronics. As a result, particles with LET $>$170 keV/\textmu m are redistributed among channels spanning roughly 165–177 keV/\textmu m, with the last Liulin-MO channel corresponding to 177 keV/\textmu m \citep[see also][]{semkova2021results}. 
    
    To show the quantitative difference, in Figure \ref{fig04:shield}(b), we plot the logarithmic differences between the modeled and measured LET spectra for both CRaTER (blue) and TGO/Liulin-MO (orange). The modeled and measured spectra agree well at a low LET ($<$10 keV/\textmu m), while slightly larger deviations occur at a high LET ($>$100 keV/\textmu m), likely due to coincidence statistics and instrumental response limitations. 
    
    \begin{figure}[htp!]
    \centering {\includegraphics[width=0.6\hsize]{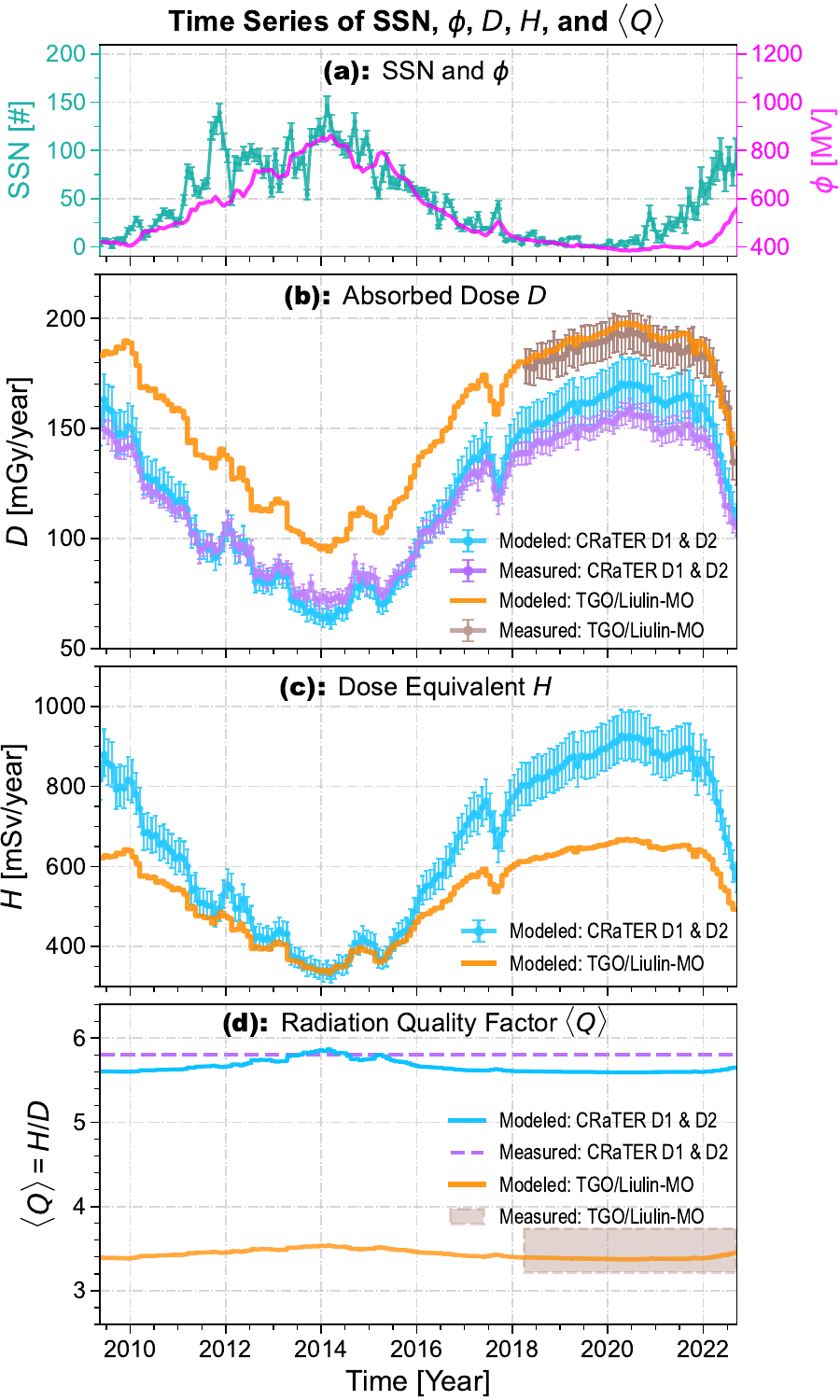}}
    \caption{Time evolution of the sunspot number (SSN), solar modulation potential ($\phi$) used in BON20, absorbed dose rate, dose equivalent rate, and radiation quality factor. 
        (a) SSN (green) and $\phi$ (magenta), corresponding to the left and right $y$-axes, respectively. 
        (b) Modeled and measured absorbed dose rates in water. The blue curve represents the modeled in-orbit absorbed dose rate for the CRaTER D1 and D2 pair, with vertical bars indicating uncertainties due to lunar albedo particle contributions. The purple curve shows measurements from the CRaTER D1 and D2 pair at the orbit of the Moon, plotted in a monthly cadence, with vertical bars for the standard variation of daily measurements within each month. The orange and brown curves indicate the modeled and measured TGO/Liulin-MO absorbed dose rate at Mars's orbit, with brown vertical bars denoting the standard variations of measurements. 
        (c) Modeled in-orbit dose equivalent rates in water for the CRaTER D1 and D2 pair (blue) and TGO/Liulin-MO (orange). The vertical bars for the CRaTER D1 and D2 pair accounts for uncertainties due to varying lunar albedo particle contributions. The Martian albedo contribution is also included in the model, but the uncertainty is not shown (see more explanations in the text). 
        (d) Radiation quality factor $\langle Q\rangle$. %The blue and dashed purple curves represent $\langle Q\rangle$ from our model and the CRaTER D1 and D2 pair \cite[5.8 from][]{zeitlin2013measurements}, respectively. The orange curve and brown shaded region represent $\langle Q\rangle$ from our model and TGO/Liulin-MO measurements \cite[$3.49\pm0.26$ from][]{semkova2021results}, respectively.
        The blue curve represents our modeled $\langle Q\rangle$ corresponding to the environment where the CRaTER D1 and D2 pair is exposed to, while the dashed purple curve shows the value derived from the CRaTER D1 and D2 pair measurements \cite[5.8, from][]{zeitlin2013measurements}. The orange curve represents our modeled $\langle Q\rangle$ for the TGO/Liulin-MO environment, and the brown shaded region shows the corresponding measurements \cite[$3.49\pm0.26$, from][]{semkova2021results}.} \label{fig05:DHQtime}
    \end{figure}
    
    We then examine the temporal variations of the absorbed dose rate, dose equivalent rate, and averaged radiation quality factor, and present their evolutions alongside the SSN and $\phi$ values in Figure \ref{fig05:DHQtime}. 
    Panel (a) shows the monthly SSN and the corresponding solar modulation potential adopted in the BON20 model, together illustrating the evolution of solar activity and associated modulation of GCRs. 
    Panel (b) compares the modeled and measured absorbed dose rates $D$ for CRaTER and TGO/Liulin-MO. For the CRaTER D1 and D2 pair, the modeled absorbed dose rate (blue) generally aligns with the measured trend (purple), and the vertical bars in the modeled curve represent uncertainties arising from lunar albedo contributions \citep[8.6–19.9\%;][]{schwadron_lunar_2012, zaman2020absorbed}, with the central value corresponding to the median albedo contribution of 14.25\%. 
    Both the modeled and measured $D$ values decrease during solar maximum and increase toward solar minimum, consistent with the expected GCR modulation. A generally closer match is obtained for CRaTER before 2019, while noticeable discrepancies appear afterward, likely because the BON20 calibration precedes this period and 2019–2021 appears to be at an unusually quiet solar minimum. 
    For TGO/Liulin-MO, the modeled (orange) and measured (brown) absorbed dose rates at Mars's orbit show overall aligned temporal variations. In the modeled results, in addition to the correction factors for the presence of ACRs and the radial gradients of GCRs and ACRs in \ref{sec:appendCal}, albedo particle contributions are accounted for using the fractions presented in Section 3.3 of \cite{liu2023modeling}, which vary in the range of 4–8\% depending on the solar modulation at an altitude of 400 km around Mars. As the modeled albedo particle contributions in \cite{liu2023modeling} provide specific values, the associated uncertainties are not shown here. However, additional uncertainties may arise from the variability of the model inputs (e.g., SSN) and from the regolith composition and physical properties of the Martian surface \citep[e.g.,][]{rostel2020}, which influence the generation and transport of albedo particles and is beyond the scope of this study. The measured data are converted to dose rates in water with a factor of 1.333, as derived in Appendix B of \cite{schwadron_lunar_2012}. Both modeled and measured values for TGO/Liulin-MO are systematically higher than those for CRaTER, due to (1) enhanced contributions from secondary particles under the thicker shielding environment \citep[see also, e.g.,][]{dobynde2019ray, Xu_2022, ZhangJ2022JGR, liu2023modeling}, and (2) radial gradient of the GCR and ACR fluxes between Earth and Mars, as detailed in \ref{sec:appendCal}. 
    
    We plot the modeled dose equivalent rates $H$ in panel (c), which exhibit similar temporal variations as the absorbed dose rate. %but with larger absolute amplitudes owing to the LET-dependent weighting of radiation quality. 
    While the absorbed dose rate for TGO/Liulin-MO is systematically higher than that for CRaTER, the corresponding dose equivalent rate is lower, particularly near solar minimum, with a reduction of about 30\%. This consistent decrease reflects the shielding effects that preferentially attenuate high-LET components contributing more strongly to the dose equivalent. 
    
    In panel (d), we compare the modeled and measured $\langle Q\rangle$ values for GCRs under CRaTER and TGO shielding conditions. 
    Note that the acquired measured LET spectra (Figures \ref{fig02:allspectra201910}(c) and \ref{fig04:shield}(a)) cover a limited LET range due to coincidence statistics. Hence, they are used only for comparison with the modeled LET spectrum, while literature values of $\langle Q\rangle$ derived by the instrument teams are adopted for the following quantitative comparisons. 
    As shown, our modeled $\langle Q\rangle$ for the CRaTER D1 and D2 pair (blue) closely matches the measurement-derived value of 5.8 \citep{zeitlin2013measurements}. Besides, our modeled $\langle Q\rangle$ for TGO/Liulin-MO (orange), which includes the contribution by neutrons and gamma rays based on \cite{mitrofanov2018fine} and \cite{semkova2023comparison}, agrees well with the measured $3.49 \pm 0.26$ \citep{semkova2021results}. We also find a modest variation of $\langle Q\rangle$ during SC 24 maximum in our modeled results, which is up to 0.4 for the CRaTER D1 and D2 pair and up to 0.2 for TGO/Liulin-MO, indicating that $\langle Q\rangle$ is relatively insensitive to solar modulation. This trend in both CRaTER and TGO further confirms that shielding properties play a more dominant role than heliospheric modulation in determining the radiation quality factor for GCRs, consistent with a recent study by \cite{naito2025charge}. 

    % \textbf{In the end, we can provide a table listing the absorbed dose, dose equivalent, and $\langle \mathbf{Q} \rangle$ values in different cases: (a) modeled in deep space; (b) modeled with 10 g/cm2 Al shielding; (c) modeled with the TGO distribution function; (d) measured by CRaTER; (e) measured by TGO in 2019/10 (and, e.g., 2024/10 or another month in 2024, for the solar maximum).}

\section{Discussion} \label{sec:discuss_04}

    While the GCR fluxes vary with the solar modulation, the resulting $\langle Q \rangle$ exhibits only a weak dependence on it. To understand why $\langle Q \rangle$ is relatively insensitive to solar activity, we next investigate and discuss the contributions from individual GCR elements from a modeling perspective. 
    % In the following, we investigate the reason for $\langle Q \rangle$ insensitive to the solar activity from the model perspective. 
    
    \subsection{Radiation Dose Contribution from Individual GCR Elements} \label{sec:discuss_04_rad_01}
    
    To quantify the elemental contributions to the overall radiation environment, we decompose the absorbed dose rate, dose equivalent rate, and radiation quality factor into contributions from individual GCR elements from hydrogen to iron, denoted as $D_Z$, $H_Z$ and $\langle Q \rangle_Z$, respectively. 
    
    \begin{figure*}[ht!]
    \centering {\includegraphics[width=0.99\hsize]{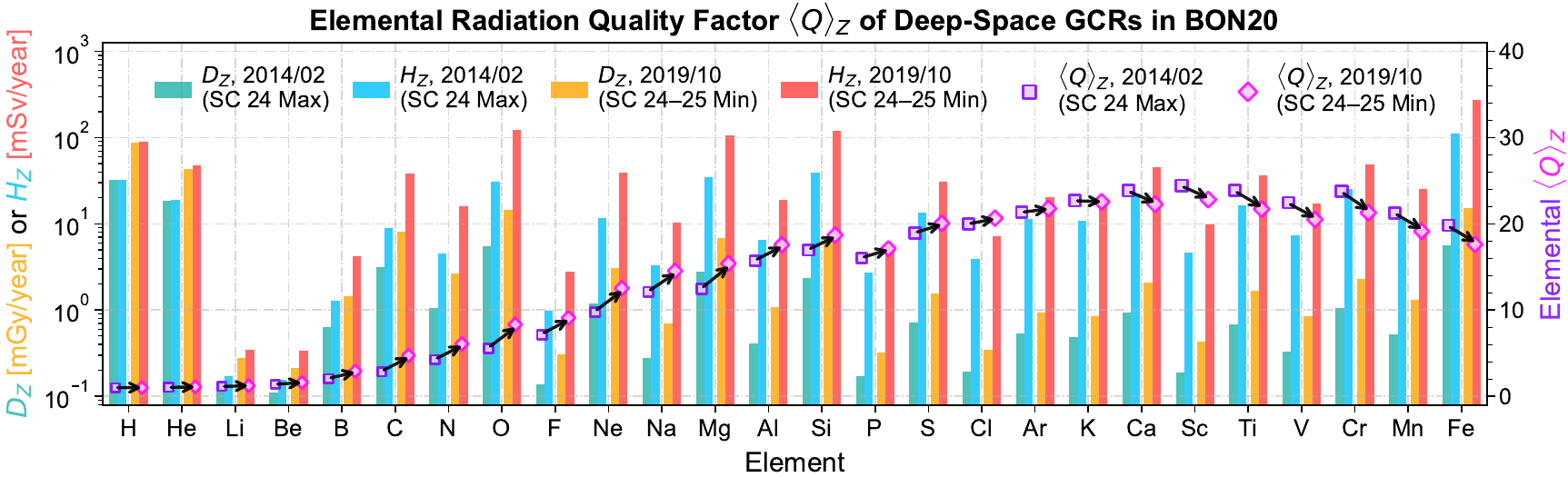}}
    \caption{Elemental absorbed dose rate ($D_Z$), dose equivalent rate ($H_Z$), and radiation quality factor ($\langle Q \rangle_Z$) of deep-space GCRs derived from the BON20 model. 
        $D_Z$ and $H_Z$ correspond to the left $y$-axis (logarithmic scale), while $\langle Q \rangle_Z$ corresponds to the right $y$-axis. 
        For each element from H to Fe, absorbed dose and dose equivalent rates are shown as green and blue vertical bars for February 2014 (SC 24 maximum), and as orange and red vertical bars for October 2019 (SC 24–25 minimum), respectively. The corresponding $\langle Q \rangle_Z$ in these two periods are plotted as purple squares and pink diamonds, respectively, with marker sizes scaled according to their values. Black arrows connect the calculated $\langle Q \rangle_Z$ scatter points during these two periods for each element to better illustrate the variation of $\langle Q \rangle_Z$ with solar activity.} \label{fig06:Qelement}
    \end{figure*}
    
    Figure \ref{fig06:Qelement} shows these elemental quantities of deep-space GCRs derived from BON20 for two representative periods: February 2014 (SC 24 maximum) and October 2019 (SC 24–25 minimum). 
    In the figure, $D_Z$ and $H_Z$ are plotted as vertical bars corresponding to the left $y$-axis on a logarithmic scale, while $\langle Q\rangle_Z$ values are shown as scatter points overlaid on the respective bars and aligned to the right $y$-axis. Also, we use black arrows to connect the $\langle Q\rangle_Z$ values between the two periods, illustrating their variation with respect to solar activity for each element. We list our main findings  and provide explanations in the following.
    \begin{itemize}
        \item[1.] Elemental dominance in dose and dose equivalent: 
        Consistent with the findings in Section 3.2 of \cite{liu2024comprehensive}, H and He dominate the absorbed dose during both solar maximum and minimum. However, they do not dominate the dose equivalent. 
        Instead, several heavier ions, such as carbon (C), oxygen (O), magnesium (Mg), silicon (Si), calcium (Ca), titanium (Ti), chromium (Cr) and Fe, make relatively small contributions to the absorbed dose but large contributions to the dose equivalent. This contrast arises from their higher LET values, as shown in Figure \ref{fig02:allspectra201910}(c). 
        
        \item[2.] Atomic number dependence of $\langle Q\rangle_Z$: 
        The elemental radiation quality factor generally increases with atomic number for both periods, reflecting the higher LET of heavier nuclei due to their larger charge and mass. The $\langle Q\rangle_Z$ value rises from $\sim$1 to $\sim$25 for species from H to scandium (Sc), and then gradually decreases for heavier species ($Z>21$). This decline for heavier nuclei occurs because their LET falls beyond $\sim$100 keV/\textmu m, where the quality factor $Q(L)$ begins to fall, as shown in Figure \ref{fig02:allspectra201910}(c)(d). 
        
        \item[3.] Solar activity dependence of $\langle Q\rangle_Z$ across elements: 
        In addition to the atomic number dependence, we find a distinct trend in how $\langle Q\rangle_Z$ changes from solar maximum to minimum. For lighter elements, $\langle Q\rangle_Z$ increases with reduced solar activity, whereas it decreases for heavier elements starting from Ca, i.e., $Z \geqslant 20$. 
        This contrasting behavior actually reflects differences in the radiation impact caused by GCR modulation and the energy dependence of LET among light and heavy ions, which we further discuss in the next section. 
    \end{itemize}
    
    In summary, the modulation of $\langle Q\rangle_Z$ with solar activity depends strongly on elemental mass and charge. In order to elucidate the underlying physics of the contrasting behavior mentioned above, we analyze how $\langle Q \rangle_Z$ evolves with solar modulation for light and heavy species in the following.

\subsection{\texorpdfstring{$\langle Q\rangle$}{\it Q} of Lighter and Heavier GCR Elements During Solar Minimum and Maximum} \label{sec:discuss_04_qminmax_02}
    
\begin{figure*}[ht!]
    \centering {\includegraphics[width=0.99\hsize]{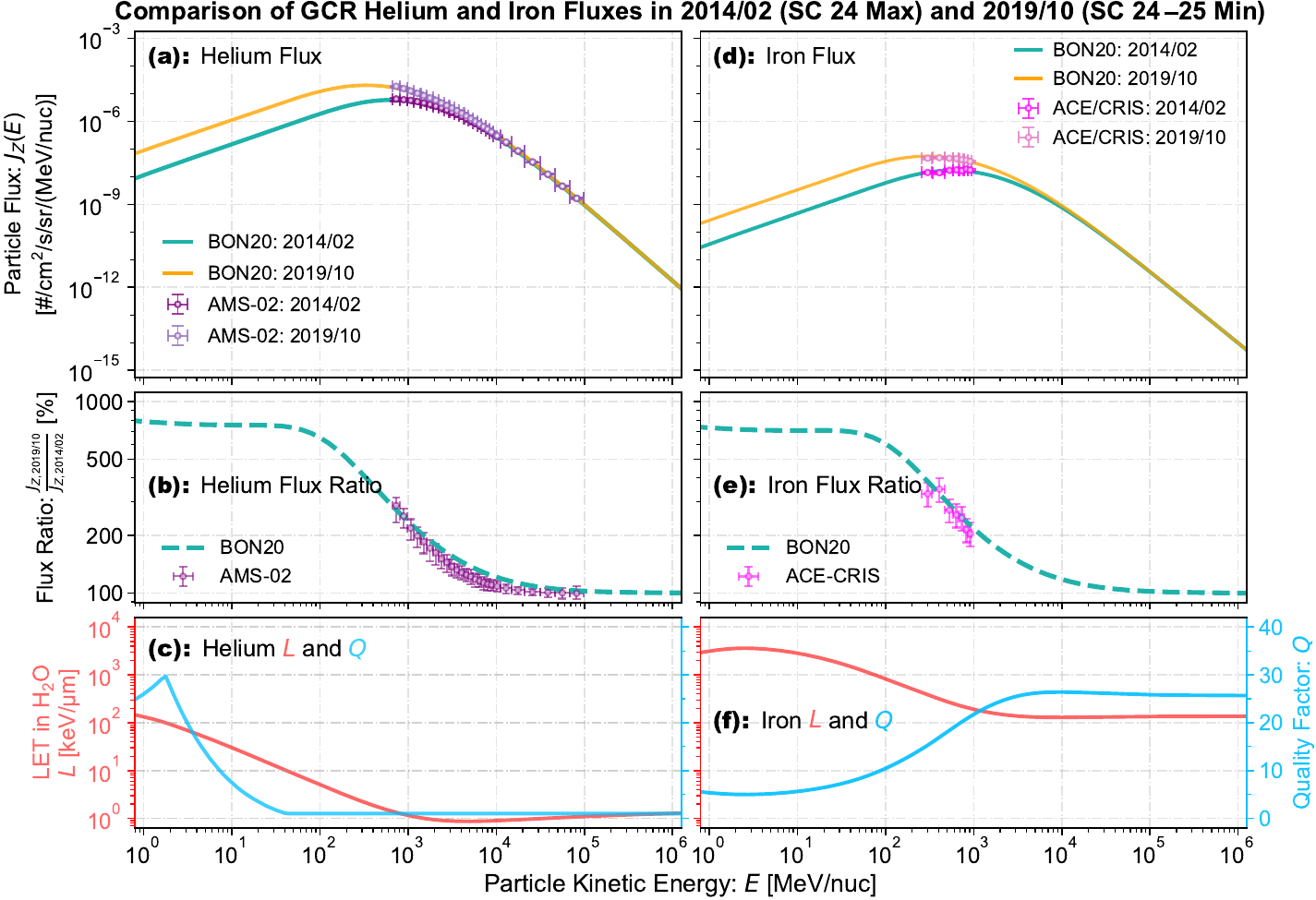}}
    \caption{Comparison of particle fluxes in February 2014 (SC 24 maximum) and October 2019 (SC 24–25 minimum) for deep-space GCR helium and iron nuclei. 
        (a) Helium fluxes. The green and orange curves represent the modeled helium particle fluxes in February 2014 and October 2019, respectively, with the light- and dark-purple scatter points indicating AMS-02 measurements during the corresponding periods. Horizontal and vertical bars on the AMS-02 points indicate the energy bin width and systematic uncertainties of the measurements, respectively. 
        (b) Ratio of helium fluxes in October 2019 to those in February 2014. The green line shows the BON20 results, and the purple scatter points show AMS-02 measurements, with horizontal and vertical bars for the energy bin width and systematic uncertainties, respectively. 
        (c) The helium LET function (red, left $y$-axis) and the radiation quality factor $Q$ (blue, right $y$-axis) as functions of the particle kinetic energy, converted based on Figure \ref{fig02:allspectra201910}(a)(b)(d). 
        (d)–(f) Iron fluxes, the flux ratio, LET function and $Q$ plotted as functions of the particle kinetic energy, following the same pattern as panels (a)–(c), except using ACE/CRIS measurements shown in light-pink and/or magenta.} \label{fig08:hefeflux}
    \end{figure*}

    To start with, we examine the He and Fe particle fluxes, representative of light and heavy species, respectively. 
    Figure \ref{fig08:hefeflux} shows the modeled He and Fe fluxes in deep space during February 2014 (SC 24 maximum) and October 2019 (SC 24–25 minimum) with corresponding measurements from the Alpha Magnetic Spectrometer \citep[AMS-02,][]{aguilar2022properties} and ACE/CRIS, respectively. 
    Panels (a) and (d) show the absolute particle fluxes ($J_Z$), and panels (b) and (e) show the flux ratios between solar minimum and maximum. The measurements are overlaid in these four panels, demonstrating good agreement with the BON20 model predictions in GCR particle fluxes. Taking the energy-dependent LET functions and the corresponding $Q(L)$ in Figures \ref{fig02:allspectra201910}(b)(d), panels (c) and (f) relate the particle kinetic energy, $E$, to LET, $L(E)$, and further the radiation quality factor $Q(L(E))$. 
    % This allows us to directly connect the particle flux modulation of light and heavy elements to their contributions to $\langle Q\rangle_Z$ and to understand why $\langle Q\rangle_Z$ responds differently for lighter and heavier ions under varying solar modulation. 

    By comparing the GCR fluxes for He and Fe during solar maximum and minimum, we can see how the GCR spectra and the energy-dependent LET function jointly determine the elemental contribution to $\langle Q\rangle_Z$. During solar maximum, low-energy GCR particles are strongly modulated, leading to a substantial reduction in their fluxes. In contrast, during solar minimum, the relative importance of low-energy ions is enhanced, leading to higher $\langle Q\rangle_Z$ for lighter elements, whose LET rises sharply at lower energies. 
    % However, for heavier elements such as Fe, the increased relative importance of lower-energy particles during solar minimum shifts the weighting toward LET values beyond the region of maximum $Q(L)$. As shown in the low-energy end in Figure \ref{fig08:hefeflux}(f), this region corresponds to a relatively smaller $Q$ value, resulting in a decrease in $\langle Q\rangle_Z$. 
    However, for heavier elements such as Fe, the relative importance of higher-energy component within the spectra, which corresponds to a higher $Q$, is reduced during solar minimum. Thus, $\langle Q\rangle_Z$ is smaller during solar minimum than solar maximum. 
    This above analysis of particle fluxes and their energy-dependent LET explains the contrasting response of $\langle Q\rangle_Z$ for light and heavy elements to solar modulation, as shown in Figure \ref{fig06:Qelement}. 
        
    \begin{figure*}[htp!]
    \centering {
        \includegraphics[width=0.8\hsize]{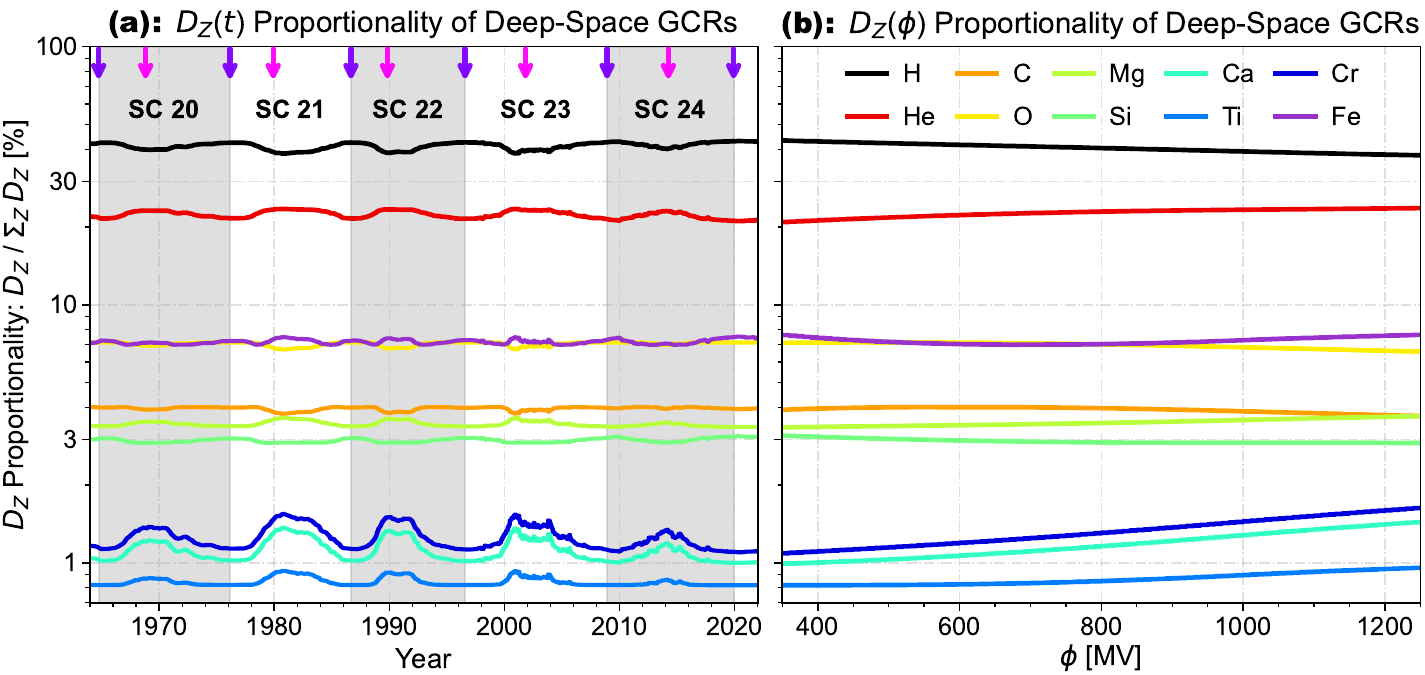} \label{fig07:Ditphi}
        \includegraphics[width=0.8\hsize]{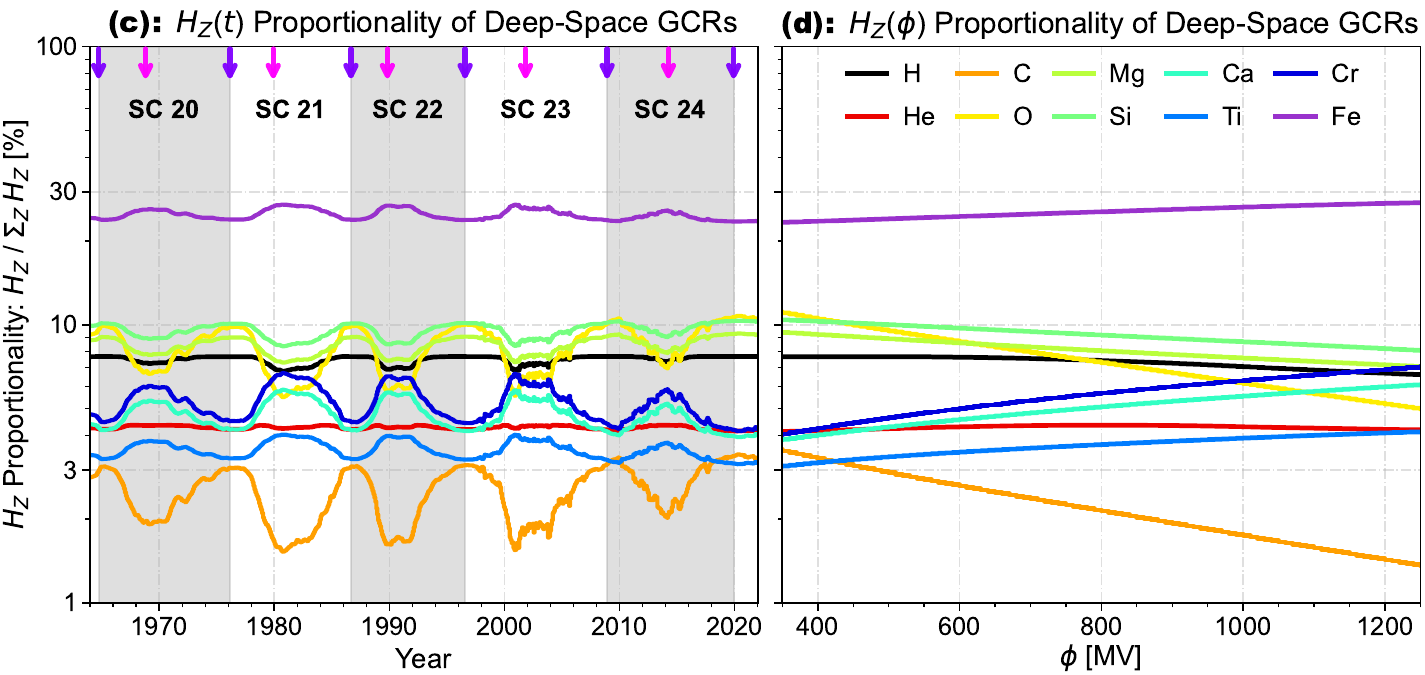} \label{fig07:Hitphi}
        \includegraphics[width=0.8\hsize]{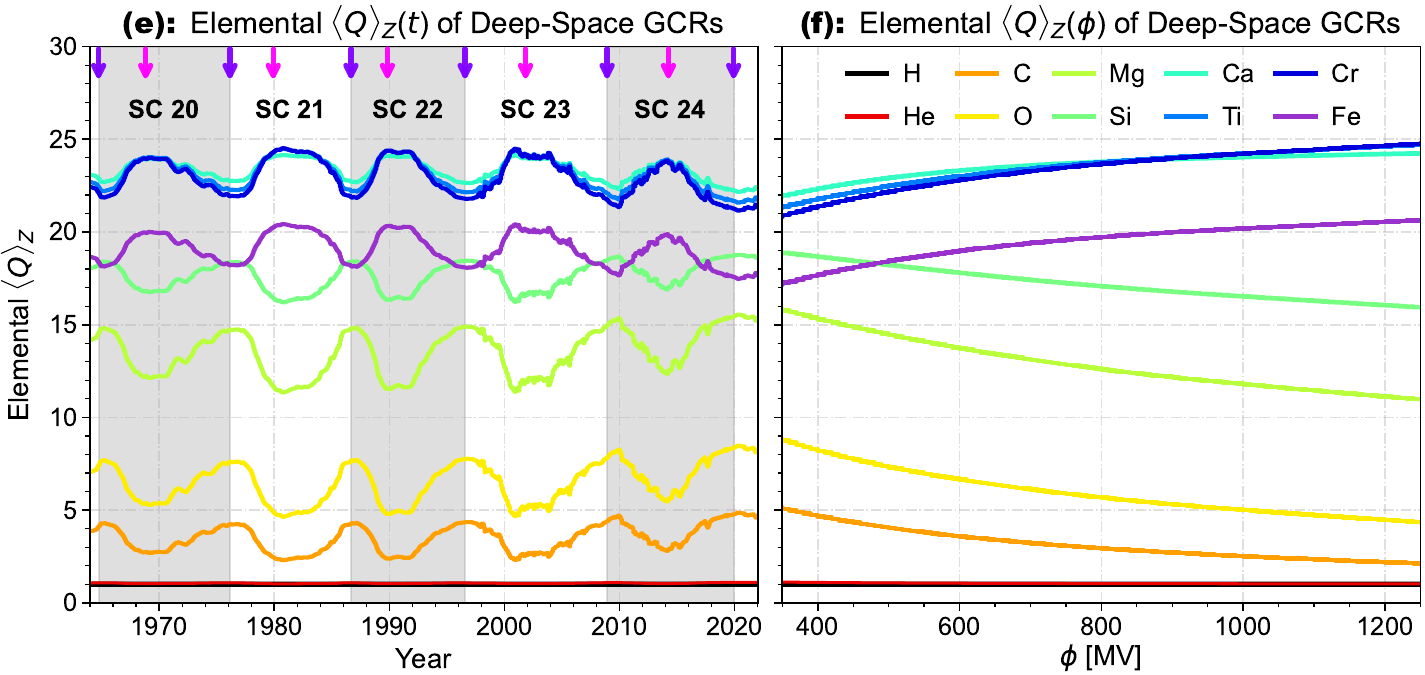} \label{fig07:Qitphi}}
    \caption{The absorbed dose rate proportionality, dose equivalent rate proportionality, and radiation quality factor for 10 selective elements of deep-space GCRs in BON20. 
        (a) Absorbed dose rate proportionalities (on a logarithmic scale for better visualization) as functions of time between 1964 and 2022 for each selected element. Alternating gray and white backgrounds indicate different solar cycles, with purple and magenta arrows at the top marking solar minima and maxima, respectively. 
        (b) Absorbed dose rate proportionalities as functions of $\phi$ for the same elements as shown in panel (a). Panels (a) and (b) share the same $y$-axis and legend, with the latter listed at the top of panel (b). 
        (c, d) Dose equivalent rate proportionalities (on a logarithmic scale, for better visualization) for each selected element, presented in the same layout as in panels (a)(b). 
        (e, f) Elemental radiation quality factor shown following the same layout as in panels (a)(b).} \label{fig07:DHQitphi}
    \end{figure*}

    Furthermore, we investigate the temporal and $\phi$-dependent variations of $D_Z$, $H_Z$ and $\langle Q \rangle_Z$ for selected GCR elements (H, He, C, O, Mg, Si, Ca, Ti, Cr and Fe) in deep space. These elements are chosen to represent a broad range of atomic numbers and to highlight the differing contributions of light and heavy ions to $D_Z$, $H_Z$ and $\langle Q \rangle_Z$ illustrated in Figure \ref{fig06:Qelement}. 
    
    Figure \ref{fig07:DHQitphi} shows these quantities in terms of their proportional contributions in different rows, from 1964 to 2022 (left column), and for $\phi$ values ranging from 350 to 1250 MV (right column). 
    Alternating gray and white backgrounds in the left-column panels indicate different solar cycles, with purple and pink arrows marking solar minima and maxima, respectively, indicating how each element responds to long-term solar modulation. 
    In panels (a) and (b), the proportionality of $D_Z$ exhibits little variation over time or $\phi$ for each selected element, reflecting similar modulation effects on the absorbed dose across GCR species.
    On the contrary, the proportionality of $H_Z$ shows clear element-dependent trends. As shown in panels (c) and (d), H and He remain nearly constant, whereas Ca, Ti, Cr and Fe increase with $\phi$, each varying between 3–10\%, while C, O, Mg and Si decrease, each varying in the range of 1–12\%. 
    Consequently, $\langle Q\rangle_Z$ increases with $\phi$ for Ca, Ti, Cr, and Fe, decreases for C, O, Mg, and Si, and remains nearly constant for H and He, which is mainly governed by the energy dependence of their LET and the corresponding $Q(L)$. 
    Taken together, the overall radiation quality factor $\langle Q\rangle$ results from a balance between these opposing trends among different GCR species, making it relatively insensitive to solar modulation. 

    It is important to note that the above results are based on GCR fluxes in deep space, whereas shielding would modify the GCR spectra with heavy ions fragmented easily into lighter ones. With shielding, the contribution of heavy ions to both absorbed dose and dose equivalent would decrease significantly. In addition, secondary particles such as neutrons generated within the shielding would make additional contributions.

\section{Summary and Conclusions} \label{sec:sumcon_05}

% Major points:\\
% 1. Reproducing the observational results by the state-of-the-art modeling tools \\
% 2. Statements of the Deep Space $\langle Q \rangle$ Dependence on Solar Activity, the reasoning, and the importance of this study for radiobiology \\
% 3. Energy range of measurements may need to extend to broader energy ranges, as one of the hints from this work, from the LET and $\langle Q \rangle$ perspective

Understanding the long-term modulation of GCRs and their resulting radiobiological effects is essential for evaluating space radiation hazards in deep space and planetary environments. In this study, we systematically model the temporal and modulation potential variations of the LET spectra, absorbed dose rate, dose equivalent rate and radiation quality factor of GCRs in both deep space and shielding environments. 

Using the SSN as input, the BON20 model is adopted to provide primary GCR spectra under different solar modulation potentials. These spectra are then used as primary particles for simulations of GCR interactions with shielding layers of different thicknesses to derive LET spectra, $D$, $H$ and $\langle Q \rangle$. The results provide a comprehensive dataset spanning a broad range of solar modulations and shielding conditions, making it convenient to quickly estimate the $D$, $H$ and $\langle Q \rangle$ of GCRs given $\phi$ and shielding thickness. In the modeled results, we find that $\langle Q \rangle$ depends strongly on shielding thickness but only weakly on solar activity. 
We then validate the model predictions against long-term measurements from the CRaTER D1 and D2 pair facing deep space at the Moon's orbit and the TGO/Liulin-MO instrument at Mars's orbit. In this comparison, we account for factors including the phantom medium where radiation dose is deposited (water/silicon), orbital altitude of the spacecraft, albedo particle contributions, ACR contributions and the radial gradients of both GCRs and ACRs. %The corresponding scaling factors of 6.3\% at 1 AU and 10.99\% at 1.5 AU are applied to the calculated absorbed dose rates. | -- a bit too detailed
With these factors taken into account, our modeled absorbed dose and $\langle Q \rangle$ show consistent temporal variations with the CRaTER and TGO measurements under thin and thick shielding conditions, respectively. 

To interpret the insensitivity of $\langle Q \rangle$ to solar activity, we further analyze the elemental contributions of $D_Z$, $H_Z$, and $\langle Q \rangle_Z$ across GCR species. Lighter elements, such as H and He, dominate the absorbed dose but contribute less to dose equivalent, whereas heavier ions, including C, O, Mg, Si, Ca, Ti, Cr and Fe, have greater influence on dose equivalent and thus $\langle Q \rangle$ because of their higher LET. 
Solar modulation primarily affects low-energy particles, thereby altering the relative weighting of particle fluxes across energies. This leads to opposite responses of $\langle Q \rangle$ for light and heavy nuclei due to their differing energy-dependent quality factors at low energies. Consequently, the overall $\langle Q \rangle$ represents a balance between these contrasting tendencies of $H_Z$ and $\langle Q \rangle_Z$ among different GCR species. Here, our analysis provides a physical basis for element-specific radiobiological variation in response to solar modulation and explains the weak dependence of the overall radiation quality factor $\langle Q \rangle$ on solar activity. 

% Overall, this study delivers a detailed and systematic characterization of LET spectra and radiation quality factors across solar and shielding conditions, offering important insights for radiation dose assessment and the design of crewed missions beyond Earth. 

\section*{Acknowledgments} \label{sec:acknowledge}
% The authors thank the anonymous reviewer for his/her time and effort in helping us improve this paper. 
The authors acknowledge the support by the National Natural Science Foundation of China (Grant Nos. 42521007, 42188101, 42474221, and W2433101). 
The authors thank Cary Zeitlin for his helpful suggestions on the CRaTER dataset and also express their gratitude to the groups developing the BON20 GCR model and the GEANT4 toolkit, the World Data Center SILSO, Royal Observatory of Belgium, Brussels, and the teams of ACE/CRIS, AMS-02, LRO/CRaTER and TGO/Liulin-MO instruments for providing valuable data involved in this study. 

%% The Appendices part is started with the command \appendix;
%% appendix sections are then done as normal sections
\appendix
% \section{Derivations of LET Spectra and \texorpdfstring{$\langle Q\rangle$}{\it Q} based on GCR Flux Spectra and Response Functions} \label{sec:calcQmodel_append01}
\section{Calculations of ACR Contributions, and GCR and ACR Radial Gradients} \label{sec:appendCal}

To account for ACR contributions and the radial gradients of GCRs and ACRs, we first consider the long-term average absorbed dose rate at 1 AU over the period 2009–2022. As stated in Section \ref{sec:method_02_GCR01}, correction factors of $+2.3\%$ for CR\`EME and $-5.9\%$ for BON20 \citep{liu2024comprehensive} are applied to remove systematic biases in the modeled absorbed dose rates compared with CRaTER measurements at 1 AU:
\begin{align}
    & D_{\rm GCR}(r=1\;{\rm AU}, t) + D_{\rm ACR}(r=1\;{\rm AU}, t) \label{eqn:Dorig} \\
    =\, & D_{\text{CR\`EME}}^*(t) \equiv \frac{{D}_{\text{CR\`EME}}(t)}{1+0.023} = \sum\limits_{Z=1}^{26} {D}_{Z, \,\text{CR\`EME}}(t) \\
    =\, & {D}_{\text{BON20}}^*(t) \equiv \frac{{D}_{\rm BON20}(t)}{1-0.059} = \sum\limits_{Z=1}^{26} {D}_{Z, \,\text{BON20}}(t)\>\! , \label{eqn:D20corr}
\end{align}
where $D_{\rm GCR}(r=1\;{\rm AU}, t)$ and $D_{\rm ACR}(r=1\;{\rm AU},t)$ denote absorbed dose rates caused by GCRs and ACRs at 1 AU, respectively, given a specific time $t$. $D_{\text{CR\`EME}}^*(t)$ and ${D}_{\text{BON20}}^*(t)$ are the bias-corrected modeled dose rates summed over all species $Z$. 

Note that BON20 only incorporates GCRs, whereas ${D}_{\text{BON20}}^*(t)$ in Equation (\ref{eqn:D20corr}) includes both ACR and GCR contributed doses: 
\begin{align}
    & {D}_{\text{BON20}}(t) = {D}_{\rm GCR}(r=1\;{\rm AU}, t), \label{eqn:D20GCR} \\
    \implies & \left\langle \frac{D_{\text{ACR}}(r=1\;{\rm AU}, t)}{D_{\text{GCR}}(r=1\;{\rm AU}, t)} \right\rangle_t = \left\langle \frac{{D}_{\text{BON20}}^*}{{D}_{\text{BON20}}} \right\rangle_t - 1 = 6.3\%, \label{eqn:DACRall}
\end{align}
in which $\langle \ldots \rangle_t$ denotes the long-term average of the given functions. Hence, Equation (\ref{eqn:D20corr}) is interpreted as including the ACR contribution at 1 AU, which amounts on average to 6.3\% of the calculated GCR dose. 

Since BON20 fluxes are modulated to 1 AU, the radial gradients of both GCRs and ACRs should be considered for an accurate estimate of radiation doses at other heliocentric distances. For GCRs, observations indicate a radial gradient ($G_{r,\,{\rm GCR}}$) of 2–4\%/AU in the inner heliosphere \citep[e.g.,][]{gieseler2016spatial, vos2016global, honig2019multi, Roussos2020}. Therefore, we adopt a representative value of 3.0\%/AU in our calculations. 
The absorbed dose for GCRs of species $Z$ at $r$ is expressed as (omitting $t$ shortly for notational simplicity):
\begin{align}
    & D_{Z,\, {\rm GCR}}(r) \simeq D_{Z,\,{\rm GCR}}(r=1\;{\rm AU}) + D'_{Z,\,{\rm GCR}}(r) \cdot \Delta r \label{eqn:DzGCR1} \\
    =\, & D_{Z,\,{\rm GCR}}(r=1\;{\rm AU}) \cdot \left[ 1 + \frac{D'_{Z,\,{\rm GCR}}(r)}{D_{Z,\,{\rm GCR}}(r=1\;{\rm AU})} \Delta r \right] \label{eqn:DzGCR2} \\
    \simeq\, & D_{Z,\,{\rm GCR}}(r=1\;{\rm AU}) \cdot \left[ 1 + \frac{J'_{Z,\,{\rm GCR}}(r)}{J_{Z,\,{\rm GCR}}(r=1\;{\rm AU})} \Delta r \right] \label{eqn:DzGCR3} \\
    \simeq\, & D_{Z,\,{\rm GCR}}(r=1\;{\rm AU}) \cdot \left( 1 + G_{r,\, Z,\, {\rm GCR}} \cdot \Delta r \right), \label{eqn:DzGCR4} 
\end{align}
in which $\Delta r = r - 1$ AU. 
% The approximation from dose to flux gradient in Equations (\ref{eqn:DzGCR2})–(\ref{eqn:DzGCR3}) assumes the same radial gradient throughout the energy range, 
% \begin{align}
%     & D_{{\rm GCR}}(r,t) = D_{{\rm GCR}}(r=1\;{\rm AU},t) \cdot \left( 1 + G_{r,\, {\rm GCR}} \cdot \Delta r \right).
% \end{align}

While spectra of H, He, C, nitrogen (N), O, neon (Ne), and argon (Ar) have been identified with ACR components at low energies during solar minimum \citep[e.g.,][]{cummings1996composition, marquardt2018energy, giacalone2022anomalous}, we primarily focus on He, N and O in this work, because H and C exhibit small model-to-model variations \citep{liu2024comprehensive}, and Ne and Ar contribute negligibly to the dose and lack reliable gradient constraints. 
% This is because, as reported in \cite{liu2024comprehensive}, H and C do not show notable differences among GCR (and ACR) models, while Ne and Ar contribute minimally to the dose and lack well-constrained radial gradients in the literature. 
% Observations by \cite{rankin2021first} indicate a radial gradient ($G_{r, \, Z,\, {\rm ACR}}$) of He in ACRs of $\sim$25\% within 0.94 AU, and \cite{rankin2022anomalous} reported a radial gradient for O of (49.4 $\pm$ 8.0)\%/AU over 0.1–0.94 AU during the SC 24–25 minimum. 
For ACRs, observations show a radial gradient ($G_{r,\,Z,\,{\rm ACR}}$) of $\sim$25\%/AU for He within 0.94 AU \citep{rankin2021first} and $(49.4 \pm 8.0)\%$/AU for O over 0.1–0.94 AU during the SC 24–25 minimum \citep{rankin2022anomalous}. 
Although our study extends slightly beyond this radial range ($r = 1.5$ AU), these values are adopted as they are representative of the period when TGO has been at Mars's orbit (since 2018), and the corresponding gradients remain relatively similar over both distance and time \citep[see][and references therein]{rankin2021first, rankin2022anomalous}. 
% Although our study extends slightly beyond this range ($r=1.5$ AU), we use these values in our calculations because they correspond to the same period when TGO has been at Mars's orbit (since 2018), and similar gradients are found during other periods \citep[see][and references therein]{rankin2021first, rankin2022anomalous}. 
For N, we adopt the same gradient as O due to their comparable mass and charge. 
% Although $r=1.5$ AU lies slightly beyond this range, we continue to refer to these values in our dose calculations because they represent the measurements (more accurate) for the period since 2018 when TGO has been at Mars's orbit. Besides, similar gradient values are expected during other periods \citep[see][and references therein]{rankin2021first, rankin2022anomalous}. For N in ACRs, we use a same radial gradient as applied to O given their similar mass and charge. 
Similar to Equations (\ref{eqn:DzGCR1})–(\ref{eqn:DzGCR3}), the absorbed dose for ACRs of species $Z$ at $(r, t)$ is:
\begin{align}
    D_{Z,\, {\rm ACR}}(r, t) & \simeq D_{Z,\, {\rm ACR}}(r=1\;{\rm AU}, t) \cdot \left( 1 + G_{r,\, Z,\, {\rm ACR}} \cdot \Delta r \right), \label{eqn:DACRr}
\end{align}
where $D_{Z,\, {\rm ACR}}(r=1\;{\rm AU}, t)$, with $Z=2$, 7 and 8, can be estimated by combining Equations (\ref{eqn:Dorig})–(\ref{eqn:D20GCR}):
\begin{align}
    & \begin{aligned}
    & D_{Z,\, {\rm ACR}}(r=1\;{\rm AU}, t) \\
    % = D_{Z,\, {\rm GCR}}(r=1\;{\rm AU}, t) \cdot \frac{D_{Z,\, {\rm ACR}}(r=1\;{\rm AU}, t)}{D_{Z,\, {\rm GCR}}(r=1\;{\rm AU}, t)} \label{eqn:DACRz1} \\
    & \simeq\, D_{Z,\, {\rm GCR}}(r=1\;{\rm AU}, t) \cdot \left\langle\frac{D_{Z,\, {\rm ACR}}(r=1\;{\rm AU}, t)}{D_{Z,\, {\rm GCR}}(r=1\;{\rm AU}, t)}\right\rangle_t \\
    \end{aligned} \label{eqn:DACRz1} \\
    & = D_{Z, \, {\rm BON20}}(t) \cdot \left\langle\frac{D_{Z,\, \text{CR\`EME}}^*(t)}{D_{Z,\, \text{BON20}}(t)}\right\rangle_t\, . \label{eqn:DACRz3}
\end{align}
% Note that the correction factor of 6.3\% in Equation (\ref{eqn:DACRall}) is applied to overall $D$, while here, we only account for specific elements $Z=2$, 7 and 8 in Equations (\ref{eqn:DACRz1})–(\ref{eqn:DACRz3}). Thus, we calculate the elemental dose by CR\`EME with corrections and BON20, and derive the long-term ratio between these two, as illustrated in Equation (\ref{eqn:DACRz3}). 

% Therefore, with Equations (\ref{eqn:DzGCR1})–(\ref{eqn:DzGCR4}) for GCRs and Equations (\ref{eqn:DACRz1})–(\ref{eqn:DACRz3}) for ACRs, together we correct the absorbed dose:
Table \ref{tab:radialCR} summarizes the parameters used for dose corrections. Applying Equations (\ref{eqn:DzGCR4}), (\ref{eqn:DACRr}) and (\ref{eqn:DACRz3}) yields the total absorbed dose by GCRs and ACRs:
{\small
\begin{align}
    & \begin{aligned}
    & D_{\rm GCR}(r, t) + D_{\rm ACR}(r, t) \\
    & = D_\mathrm{BON20}(t) \cdot \left( 1 + G_{r,\, {\rm GCR}} \cdot \Delta r \right) \, +  \\
    & \ \sum\limits_{Z=2,7,8} D_{Z, \, {\rm BON20}}(t) \cdot \left\langle\frac{D_{Z,\, \text{CR\`EME}}^*(t)}{D_{Z,\, \text{BON20}}(t)}\right\rangle_t \cdot \left( 1 + G_{r,\, Z,\, {\rm ACR}} \cdot \Delta r \right), \label{eqn:Drsum}
    \end{aligned} 
\end{align}}
and the dose equivalent $H$ is corrected in the same manner. 
% For the long-term absorbed dose and dose equivalent corrections in deep space or under different shielding conditions, we calculate the average factors for GCR and ACR contributions in (\ref{eqn:Drsum}) over time and adopt these factors for convenient use. 
For convenience, we compute the time-averaged correction factors for the GCR and ACR contributions in Equation (\ref{eqn:Drsum}) and apply these averaged factors to absorbed dose and dose equivalent estimates directly derived from BON20 in deep space or under different shielding conditions: 
\begin{align}
    \begin{aligned}
    & C_{r,\, {\rm GCR+ACR}} = C_{r,\, {\rm GCR}} + C_{r,\, {\rm ACR}} \\
    & = \left\langle\frac{D_{\rm GCR}(r, t)}{D_{\rm BON20}(t)}\right\rangle_t + \left\langle\frac{D_{\rm ACR}(r, t)}{D_{\rm BON20}(t)}\right\rangle_t . \label{eqn:DrsumAG}
    \end{aligned}
\end{align}
At $r=1.5$ AU ($\Delta r = 0.5$ AU), the correction yields a 1.5\% increase in both $D$ and $H$ due to GCR gradients, denoted as $C_{r,\, {\rm GCR}}$, and ($9.5 \pm 0.06$)\% and ($8.2 \pm 1.3$)\% increases in $D$ and $H$, respectively, due to ACR and its gradients, denoted as $C_{r,\, {\rm ACR}}$. 
Interestingly, at 1 AU, Equation (\ref{eqn:DrsumAG}) gives ($5.2 \pm 0.06$)\% and ($5.7 \pm 0.7$)\% increases in $D$ and $H$, respectively, consistent with the 6.3\% estimate from Equation (\ref{eqn:DACRall}), confirming that including only He, N, and O provides a reasonable approximation of total ACR effects. 
Therefore, we adopt correction factors of 11.0\% for $D$ and 9.7\% for $H$ for long-term deep-space and shielded dose rates at Mars, although some deviations may occur for different shielding thicknesses. 

% with $r=1.5$ AU, $\Delta r=0.5$ AU and values needed listed in Table \ref{tab:radialCR}. The dose equivalent at 1.5 AU is corrected in the same manner. Note that this correction above is applied to the deep-space radiation dose. With the correction accounted for, we calculate the ratio between the corrected dose at 1.5 AU and the original BON20-derived dose: Compared with the original BON20 dose rates for GCRs only at 1 AU, we have a 1.75\% increase in both $D$ and $H$ for GCR gradients, and a ($11.77\pm0.08$)\% increase in $D$ and a ($9.41\pm1.42$)\% increase in $H$ for ACR contributions at $r=1.5$ AU. 
% In addition, when we set $r=1$ AU, we have a ($5.67\pm0.01$)\% increase in $D$ and a ($5.25\pm0.83$)\% increase in $H$ for ACR contributions at $r=1$ AU. Compared with the value of 6.3\% in Equation (\ref{eqn:DACRall}), this again confirms that our considerations of only He, N and O can provides a reasonable approximation to the total ACR effect. 
% obtaining an increase of 11.0\% in $D$ and 9.7\% in $H$. These values represent the combined effects of ACR contributions and the radial gradients of both GCRs and ACRs. 
% In our calculation, we use these factors of 11.0\% for $D$ and 9.7\% for $H$ to correct the long-term dose rates in deep space and under shielding conditions at Mars, although this approach may introduce minor deviations with different shielding thicknesses. 

    \begin{table}[ht!]
    \begin{center}
    \caption{Parameters used to calculate the absorbed dose and dose equivalent rates in the inner heliosphere of Equation (\ref{eqn:Drsum}), accounting for the radial gradient of GCRs and ACRs in this work.} \label{tab:radialCR}
    \setlength\tabcolsep{6pt}{
    % \adjustbox{width=\textwidth}{
    \begin{tabular}{ccccc}
        \hline\hline
        \multirow{2}{*}{Particles} & $G_r$ & 
        \multirow{2}{*}{$\left\langle \frac{{D}_{Z,\, \text{CR\`EME}}^* (t)}{{D}_{Z,\, \text{BON20}}^*(t)} \right\rangle_t$} & 
        \multirow{2}{*}{$\left\langle \frac{{H}_{Z,\, \text{CR\`EME}}^*(t)}{{H}_{Z,\, \text{BON20}}^*(t)} \right\rangle_t$} \\
        & [\%/AU] & & \\
        \hline
        GCRs (All) & 3.0 & -- & -- \\
        \hline
        ACR/He & 25.0 & 1.061 & 1.288 \\
        ACR/O & 49.4 & 1.344 & 2.245 \\
        ACR/N & 49.4 & 1.552 & 2.419 \\
        \hline
    \end{tabular}}
    \end{center}
    \end{table}

%% References
%%
%% Following citation commands can be used in the body text:
%% Usage of \cite is as follows:
%%   \cite{key}         ==>>  [#]
%%   \cite[chap. 2]{key} ==>> [#, chap. 2]
%%

%% References with BibTeX database:
\bibliographystyle{elsarticle-harv}
\bibliography{paper_GCR}

\begin{thebibliography}{104}
\expandafter\ifx\csname natexlab\endcsname\relax\def\natexlab#1{#1}\fi
\providecommand{\url}[1]{\texttt{#1}}
\providecommand{\href}[2]{#2}
\providecommand{\path}[1]{#1}
\providecommand{\DOIprefix}{doi:}
\providecommand{\ArXivprefix}{arXiv:}
\providecommand{\URLprefix}{URL: }
\providecommand{\Pubmedprefix}{pmid:}
\providecommand{\doi}[1]{\href{http://dx.doi.org/#1}{\path{#1}}}
\providecommand{\Pubmed}[1]{\href{pmid:#1}{\path{#1}}}
\providecommand{\bibinfo}[2]{#2}
\ifx\xfnm\relax \def\xfnm[#1]{\unskip,\space#1}\fi
%Type = Article
\bibitem[{Adams et~al.(2012)Adams, Barghouty, Mendenhall, Reed, Sierawski,
  Warren, Watts and Weller}]{adams2012creme}
\bibinfo{author}{Adams, J.}, \bibinfo{author}{Barghouty, A.F.},
  \bibinfo{author}{Mendenhall, M.}, \bibinfo{author}{Reed, R.},
  \bibinfo{author}{Sierawski, B.}, \bibinfo{author}{Warren, K.},
  \bibinfo{author}{Watts, J.}, \bibinfo{author}{Weller, R.},
  \bibinfo{year}{2012}.
\newblock \bibinfo{title}{{CR{\`E}ME: The 2011 revision of the cosmic ray
  effects on micro-electronics code}}.
\newblock \bibinfo{journal}{IEEE Transactions on Nuclear Science}
  \bibinfo{volume}{59}, \bibinfo{pages}{3141--3147}.
\newblock \DOIprefix\doi{10.1109/TNS.2012.2218831}.
%Type = Article
\bibitem[{Agostinelli et~al.(2003)Agostinelli, Allison, Amako, Apostolakis,
  Araujo, Arce, Asai, Axen, Banerjee, Barrand et~al.}]{agostinelli2003}
\bibinfo{author}{Agostinelli, S.}, \bibinfo{author}{Allison, J.},
  \bibinfo{author}{Amako, K.}, \bibinfo{author}{Apostolakis, J.},
  \bibinfo{author}{Araujo, H.}, \bibinfo{author}{Arce, P.},
  \bibinfo{author}{Asai, M.}, \bibinfo{author}{Axen, D.},
  \bibinfo{author}{Banerjee, S.}, \bibinfo{author}{Barrand, G.}, et~al.,
  \bibinfo{year}{2003}.
\newblock \bibinfo{title}{{GEANT4}: a simulation toolkit}.
\newblock \bibinfo{journal}{Nuclear Instruments and Methods in Physics Research
  Section A} \bibinfo{volume}{506}, \bibinfo{pages}{250--303}.
\newblock \DOIprefix\doi{10.1016/S0168-9002(03)01368-8}.
%Type = Article
\bibitem[{Aguilar et~al.(2022)Aguilar, Cavasonza, Ambrosi, Arruda, Attig,
  Barao, Barrin, Bartoloni, Ba{\c{s}}e{\u{g}}mez-du Pree, Battiston
  et~al.}]{aguilar2022properties}
\bibinfo{author}{Aguilar, M.}, \bibinfo{author}{Cavasonza, L.A.},
  \bibinfo{author}{Ambrosi, G.}, \bibinfo{author}{Arruda, L.},
  \bibinfo{author}{Attig, N.}, \bibinfo{author}{Barao, F.},
  \bibinfo{author}{Barrin, L.}, \bibinfo{author}{Bartoloni, A.},
  \bibinfo{author}{Ba{\c{s}}e{\u{g}}mez-du Pree, S.},
  \bibinfo{author}{Battiston, R.}, et~al., \bibinfo{year}{2022}.
\newblock \bibinfo{title}{Properties of daily helium fluxes}.
\newblock \bibinfo{journal}{Physical review letters} \bibinfo{volume}{128},
  \bibinfo{pages}{231102}.
\newblock \DOIprefix\doi{10.1103/PhysRevLett.128.231102}.
%Type = Article
\bibitem[{Andersen et~al.(2004)Andersen, Ballarini, Battistoni, Campanella,
  Carboni, Cerutti, Empl, Fasso, Ferrari, Gadioli et~al.}]{andersen2004fluka}
\bibinfo{author}{Andersen, V.}, \bibinfo{author}{Ballarini, F.},
  \bibinfo{author}{Battistoni, G.}, \bibinfo{author}{Campanella, M.},
  \bibinfo{author}{Carboni, M.}, \bibinfo{author}{Cerutti, F.},
  \bibinfo{author}{Empl, A.}, \bibinfo{author}{Fasso, A.},
  \bibinfo{author}{Ferrari, A.}, \bibinfo{author}{Gadioli, E.}, et~al.,
  \bibinfo{year}{2004}.
\newblock \bibinfo{title}{{The FLUKA code for space applications: recent
  developments}}.
\newblock \bibinfo{journal}{Advances in Space Research} \bibinfo{volume}{34},
  \bibinfo{pages}{1302--1310}.
\newblock \DOIprefix\doi{10.1016/j.asr.2003.03.045}.
%Type = Article
\bibitem[{Banjac et~al.(2018)Banjac, Herbst and Heber}]{banjac2018}
\bibinfo{author}{Banjac, S.}, \bibinfo{author}{Herbst, K.},
  \bibinfo{author}{Heber, B.}, \bibinfo{year}{2018}.
\newblock \bibinfo{title}{The atmospheric radiation interaction simulator
  ({AtRIS}) - description and validation}.
\newblock \bibinfo{journal}{Journal of Geophysical Research: Space Physics}
  \bibinfo{volume}{123}.
\newblock \DOIprefix\doi{10.1029/2018JA026042}.
%Type = Article
\bibitem[{Battistoni et~al.(2015)Battistoni, Boehlen, Cerutti, Chin, Esposito,
  Fasso, Ferrari, Lechner, Empl, Mairani et~al.}]{battistoni2015overview}
\bibinfo{author}{Battistoni, G.}, \bibinfo{author}{Boehlen, T.},
  \bibinfo{author}{Cerutti, F.}, \bibinfo{author}{Chin, P.W.},
  \bibinfo{author}{Esposito, L.S.}, \bibinfo{author}{Fasso, A.},
  \bibinfo{author}{Ferrari, A.}, \bibinfo{author}{Lechner, A.},
  \bibinfo{author}{Empl, A.}, \bibinfo{author}{Mairani, A.}, et~al.,
  \bibinfo{year}{2015}.
\newblock \bibinfo{title}{{Overview of the FLUKA code}}.
\newblock \bibinfo{journal}{Annals of Nuclear Energy} \bibinfo{volume}{82},
  \bibinfo{pages}{10--18}.
\newblock \DOIprefix\doi{10.1016/j.anucene.2014.11.007}.
%Type = Article
\bibitem[{Bragg and Kleeman(1905)}]{bragg1905xxxix}
\bibinfo{author}{Bragg, W.H.}, \bibinfo{author}{Kleeman, R.},
  \bibinfo{year}{1905}.
\newblock \bibinfo{title}{Xxxix. on the $\alpha$ particles of radium, and their
  loss of range in passing through various atoms and molecules}.
\newblock \bibinfo{journal}{The London, Edinburgh, and Dublin Philosophical
  Magazine and Journal of Science} \bibinfo{volume}{10},
  \bibinfo{pages}{318--340}.
\newblock \URLprefix \url{10.1080/14786440509463378}.
%Type = Article
\bibitem[{Charpentier et~al.(2024)Charpentier, Ruffenach, Benacquista, Ecoffet,
  Cappe, Dossat, Varotsou, Cintas, Paillet, Boyer
  et~al.}]{charpentier2024aramis}
\bibinfo{author}{Charpentier, G.}, \bibinfo{author}{Ruffenach, M.},
  \bibinfo{author}{Benacquista, R.}, \bibinfo{author}{Ecoffet, R.},
  \bibinfo{author}{Cappe, A.}, \bibinfo{author}{Dossat, C.},
  \bibinfo{author}{Varotsou, A.}, \bibinfo{author}{Cintas, H.},
  \bibinfo{author}{Paillet, A.}, \bibinfo{author}{Boyer, L.}, et~al.,
  \bibinfo{year}{2024}.
\newblock \bibinfo{title}{{ARAMIS: a Martian radiative environment model built
  from GEANT4 simulations}}.
\newblock \bibinfo{journal}{Journal of Space Weather and Space Climate}
  \bibinfo{volume}{14}.
\newblock \DOIprefix\doi{10.1051/swsc/2024032}.
%Type = Article
\bibitem[{Chin et~al.(2007)Chin, Brylow, Foote, Garvin, Kasper, Keller, Litvak,
  Mitrofanov, Paige, Raney et~al.}]{chin2007lunar}
\bibinfo{author}{Chin, G.}, \bibinfo{author}{Brylow, S.},
  \bibinfo{author}{Foote, M.}, \bibinfo{author}{Garvin, J.},
  \bibinfo{author}{Kasper, J.}, \bibinfo{author}{Keller, J.},
  \bibinfo{author}{Litvak, M.}, \bibinfo{author}{Mitrofanov, I.},
  \bibinfo{author}{Paige, D.}, \bibinfo{author}{Raney, K.}, et~al.,
  \bibinfo{year}{2007}.
\newblock \bibinfo{title}{Lunar reconnaissance orbiter overview:
  The{\'a}instrument suite and mission}.
\newblock \bibinfo{journal}{Space Science Reviews} \bibinfo{volume}{129},
  \bibinfo{pages}{391--419}.
\newblock \DOIprefix\doi{10.1007/s11214-007-9153-y}.
%Type = Article
\bibitem[{Chowdhury et~al.(2016)Chowdhury, Kudela and
  Moon}]{chowdhury2016study}
\bibinfo{author}{Chowdhury, P.}, \bibinfo{author}{Kudela, K.},
  \bibinfo{author}{Moon, Y.J.}, \bibinfo{year}{2016}.
\newblock \bibinfo{title}{A study of heliospheric modulation and periodicities
  of galactic cosmic rays during cycle 24}.
\newblock \bibinfo{journal}{Solar Physics} \bibinfo{volume}{291},
  \bibinfo{pages}{581--602}.
\newblock \DOIprefix\doi{10.1007/s11207-016-0945-7}.
%Type = Article
\bibitem[{Cliver et~al.(2013)Cliver, Richardson and Ling}]{cliver2013solar}
\bibinfo{author}{Cliver, E.}, \bibinfo{author}{Richardson, I.},
  \bibinfo{author}{Ling, A.}, \bibinfo{year}{2013}.
\newblock \bibinfo{title}{Solar drivers of 11-yr and long-term cosmic ray
  modulation}.
\newblock \bibinfo{journal}{Space Science Reviews} \bibinfo{volume}{176},
  \bibinfo{pages}{3--19}.
\newblock \DOIprefix\doi{10.1007/s11214-011-9746-3}.
%Type = Article
\bibitem[{Cucinotta(2024)}]{cucinotta2024non}
\bibinfo{author}{Cucinotta, F.A.}, \bibinfo{year}{2024}.
\newblock \bibinfo{title}{Non-targeted effects and space radiation risks for
  astronauts on multiple international space station and lunar missions}.
\newblock \bibinfo{journal}{Life Sciences in Space Research}
  \bibinfo{volume}{40}, \bibinfo{pages}{166--175}.
\newblock \DOIprefix\doi{10.1016/j.lssr.2023.08.003}.
%Type = Article
\bibitem[{Cucinotta et~al.(2010)Cucinotta, Hu, Schwadron, Kozarev, Townsend and
  Kim}]{cucinotta2010space}
\bibinfo{author}{Cucinotta, F.A.}, \bibinfo{author}{Hu, S.},
  \bibinfo{author}{Schwadron, N.A.}, \bibinfo{author}{Kozarev, K.},
  \bibinfo{author}{Townsend, L.W.}, \bibinfo{author}{Kim, M.H.Y.},
  \bibinfo{year}{2010}.
\newblock \bibinfo{title}{Space radiation risk limits and earth-moon-mars
  environmental models}.
\newblock \bibinfo{journal}{Space Weather} \bibinfo{volume}{8}.
\newblock \DOIprefix\doi{10.1029/2010SW000572}.
%Type = Article
\bibitem[{Cucinotta et~al.(2013)Cucinotta, Kim, Chappell and
  Huff}]{cucinotta2013safe}
\bibinfo{author}{Cucinotta, F.A.}, \bibinfo{author}{Kim, M.H.Y.},
  \bibinfo{author}{Chappell, L.J.}, \bibinfo{author}{Huff, J.L.},
  \bibinfo{year}{2013}.
\newblock \bibinfo{title}{How safe is safe enough? radiation risk for a human
  mission to {Mars}}.
\newblock \bibinfo{journal}{PLoS One} \bibinfo{volume}{8},
  \bibinfo{pages}{e74988}.
\newblock \DOIprefix\doi{10.1371/journal.pone.0074988}.
%Type = Article
\bibitem[{Cummings and Stone(1996)}]{cummings1996composition}
\bibinfo{author}{Cummings, A.}, \bibinfo{author}{Stone, E.},
  \bibinfo{year}{1996}.
\newblock \bibinfo{title}{Composition of anomalous cosmic rays and implications
  for the heliosphere}.
\newblock \bibinfo{journal}{Space Science Reviews} \bibinfo{volume}{78},
  \bibinfo{pages}{117--128}.
\newblock \DOIprefix\doi{10.1007/BF00170798}.
%Type = Article
\bibitem[{De~Angelis et~al.(2006)De~Angelis, Wilson, Clowdsley, Qualls and
  Singleterry}]{deangelis2006modeling}
\bibinfo{author}{De~Angelis, G.}, \bibinfo{author}{Wilson, J.},
  \bibinfo{author}{Clowdsley, M.}, \bibinfo{author}{Qualls, G.},
  \bibinfo{author}{Singleterry, R.}, \bibinfo{year}{2006}.
\newblock \bibinfo{title}{Modeling of the martian environment for radiation
  analysis}.
\newblock \bibinfo{journal}{Radiation measurements} \bibinfo{volume}{41},
  \bibinfo{pages}{1097--1102}.
\newblock \DOIprefix\doi{10.1016/j.nuclphysbps.2006.12.035}.
%Type = Article
\bibitem[{Desai and Giacalone(2016)}]{desai2016large}
\bibinfo{author}{Desai, M.}, \bibinfo{author}{Giacalone, J.},
  \bibinfo{year}{2016}.
\newblock \bibinfo{title}{Large gradual solar energetic particle events}.
\newblock \bibinfo{journal}{Living Reviews in Solar Physics}
  \bibinfo{volume}{13}, \bibinfo{pages}{3}.
\newblock \DOIprefix\doi{10.1007/s41116-016-0002-5}.
%Type = Article
\bibitem[{Dobynde et~al.(2019)Dobynde, Effenberger, Kartashov, Shprits and
  Shurshakov}]{dobynde2019ray}
\bibinfo{author}{Dobynde, M.}, \bibinfo{author}{Effenberger, F.},
  \bibinfo{author}{Kartashov, D.}, \bibinfo{author}{Shprits, Y.Y.},
  \bibinfo{author}{Shurshakov, V.}, \bibinfo{year}{2019}.
\newblock \bibinfo{title}{Ray-tracing simulation of the radiation dose
  distribution on the surface of the spherical phantom of the matroshka-r
  experiment onboard the iss}.
\newblock \bibinfo{journal}{Life sciences in space research}
  \bibinfo{volume}{21}, \bibinfo{pages}{65--72}.
\newblock \DOIprefix\doi{10.1016/j.lssr.2019.04.001}.
%Type = Article
\bibitem[{Dobynde and Guo(2024)}]{dobynde2024guidelines}
\bibinfo{author}{Dobynde, M.}, \bibinfo{author}{Guo, J.}, \bibinfo{year}{2024}.
\newblock \bibinfo{title}{Guidelines for radiation-safe human activities on the
  moon}.
\newblock \bibinfo{journal}{Nature Astronomy} \bibinfo{volume}{8},
  \bibinfo{pages}{991--999}.
\newblock \DOIprefix\doi{10.1038/s41550-024-02287-8}.
%Type = Article
\bibitem[{Dobynde et~al.(2021)Dobynde, Shprits, Drozdov, Hoffman and
  Li}]{dobynde2021beating}
\bibinfo{author}{Dobynde, M.}, \bibinfo{author}{Shprits, Y.},
  \bibinfo{author}{Drozdov, A.Y.}, \bibinfo{author}{Hoffman, J.},
  \bibinfo{author}{Li, J.}, \bibinfo{year}{2021}.
\newblock \bibinfo{title}{Beating 1 sievert: Optimal radiation shielding of
  astronauts on a mission to mars}.
\newblock \bibinfo{journal}{Space Weather} \bibinfo{volume}{19},
  \bibinfo{pages}{e2021SW002749}.
\newblock \DOIprefix\doi{10.1029/2021SW002749}.
%Type = Article
\bibitem[{Dobynde and Guo(2021)}]{dobynde2021radiation}
\bibinfo{author}{Dobynde, M.I.}, \bibinfo{author}{Guo, J.},
  \bibinfo{year}{2021}.
\newblock \bibinfo{title}{Radiation environment at the surface and subsurface
  of the moon: Model development and validation}.
\newblock \bibinfo{journal}{Journal of Geophysical Research: Planets}
  \bibinfo{volume}{126}, \bibinfo{pages}{e2021JE006930}.
\newblock \DOIprefix\doi{10.1029/2021JE006930}.
%Type = Article
\bibitem[{Ehresmann et~al.(2014)Ehresmann, Zeitlin, Hassler,
  Wimmer-Schweingruber, B{\"o}hm, B{\"o}ttcher, Brinza, Burmeister, Guo,
  K{\"o}hler et~al.}]{ehresmann2014}
\bibinfo{author}{Ehresmann, B.}, \bibinfo{author}{Zeitlin, C.},
  \bibinfo{author}{Hassler, D.M.}, \bibinfo{author}{Wimmer-Schweingruber,
  R.F.}, \bibinfo{author}{B{\"o}hm, E.}, \bibinfo{author}{B{\"o}ttcher, S.},
  \bibinfo{author}{Brinza, D.E.}, \bibinfo{author}{Burmeister, S.},
  \bibinfo{author}{Guo, J.}, \bibinfo{author}{K{\"o}hler, J.}, et~al.,
  \bibinfo{year}{2014}.
\newblock \bibinfo{title}{Charged particle spectra obtained with the {Mars
  Science Laboratory} radiation assessment detector ({MSL}/{RAD}) on the
  surface of {Mars}}.
\newblock \bibinfo{journal}{Journal of Geophysical Research: Planets}
  \bibinfo{volume}{119}, \bibinfo{pages}{468--479}.
\newblock \DOIprefix\doi{10.1002/2013JE004547}.
%Type = Article
\bibitem[{{Fisk}(1971)}]{F1971JGR....76..221}
\bibinfo{author}{{Fisk}, L.A.}, \bibinfo{year}{1971}.
\newblock \bibinfo{title}{{Solar modulation of galactic cosmic rays, 2}}.
\newblock \bibinfo{journal}{Journal of Geophysical Research}
  \bibinfo{volume}{76}, \bibinfo{pages}{221}.
\newblock \DOIprefix\doi{10.1029/JA076i001p00221}.
%Type = Article
\bibitem[{Garcia-Munoz et~al.(1973)Garcia-Munoz, Mason and
  Simpson}]{garcia1973new}
\bibinfo{author}{Garcia-Munoz, M.}, \bibinfo{author}{Mason, G.},
  \bibinfo{author}{Simpson, J.}, \bibinfo{year}{1973}.
\newblock \bibinfo{title}{A new test for solar modulation theory: The 1972
  may-july low-energy galactic cosmic-ray proton and helium spectra}.
\newblock \bibinfo{journal}{Astrophysical Journal, vol. 182, p. L81}
  \bibinfo{volume}{182}, \bibinfo{pages}{L81}.
\newblock \DOIprefix\doi{10.1086/181224}.
%Type = Article
\bibitem[{George et~al.(2024)George, Gaza, Matthi{\"a}, Laramore, Lehti,
  Campbell-Ricketts, Kroupa, Stoffle, Marsalek, Przybyla
  et~al.}]{george2024space}
\bibinfo{author}{George, S.P.}, \bibinfo{author}{Gaza, R.},
  \bibinfo{author}{Matthi{\"a}, D.}, \bibinfo{author}{Laramore, D.},
  \bibinfo{author}{Lehti, J.}, \bibinfo{author}{Campbell-Ricketts, T.},
  \bibinfo{author}{Kroupa, M.}, \bibinfo{author}{Stoffle, N.},
  \bibinfo{author}{Marsalek, K.}, \bibinfo{author}{Przybyla, B.}, et~al.,
  \bibinfo{year}{2024}.
\newblock \bibinfo{title}{{Space radiation measurements during the Artemis I
  lunar mission}}.
\newblock \bibinfo{journal}{Nature} \bibinfo{volume}{634},
  \bibinfo{pages}{48--52}.
\newblock \DOIprefix\doi{10.1038/s41586-024-07927-7}.
%Type = Article
\bibitem[{Giacalone et~al.(2022)Giacalone, Fahr, Fichtner, Florinski, Heber,
  Hill, K{\'o}ta, Leske, Potgieter and Rankin}]{giacalone2022anomalous}
\bibinfo{author}{Giacalone, J.}, \bibinfo{author}{Fahr, H.},
  \bibinfo{author}{Fichtner, H.}, \bibinfo{author}{Florinski, V.},
  \bibinfo{author}{Heber, B.}, \bibinfo{author}{Hill, M.E.},
  \bibinfo{author}{K{\'o}ta, J.}, \bibinfo{author}{Leske, R.A.},
  \bibinfo{author}{Potgieter, M.S.}, \bibinfo{author}{Rankin, J.S.},
  \bibinfo{year}{2022}.
\newblock \bibinfo{title}{Anomalous cosmic rays and heliospheric energetic
  particles}.
\newblock \bibinfo{journal}{Space Science Reviews} \bibinfo{volume}{218},
  \bibinfo{pages}{22}.
\newblock \DOIprefix\doi{10.1007/s11214-022-00890-7}.
%Type = Article
\bibitem[{Gieseler and Heber(2016)}]{gieseler2016spatial}
\bibinfo{author}{Gieseler, J.}, \bibinfo{author}{Heber, B.},
  \bibinfo{year}{2016}.
\newblock \bibinfo{title}{{Spatial gradients of GCR protons in the inner
  heliosphere derived from Ulysses COSPIN/KET and PAMELA measurements}}.
\newblock \bibinfo{journal}{Astronomy \& Astrophysics} \bibinfo{volume}{589},
  \bibinfo{pages}{A32}.
\newblock \DOIprefix\doi{10.1051/0004-6361/201527972}.
%Type = Article
\bibitem[{Gleeson and Axford(1967)}]{gleeson1967cosmic}
\bibinfo{author}{Gleeson, L.}, \bibinfo{author}{Axford, W.},
  \bibinfo{year}{1967}.
\newblock \bibinfo{title}{Cosmic rays in the interplanetary medium}.
\newblock \bibinfo{journal}{The Astrophysical Journal} \bibinfo{volume}{149},
  \bibinfo{pages}{L115}.
\newblock \DOIprefix\doi{10.1086/180070}.
%Type = Article
\bibitem[{Gleeson and Axford(1968)}]{gleeson1968solar}
\bibinfo{author}{Gleeson, L.}, \bibinfo{author}{Axford, W.},
  \bibinfo{year}{1968}.
\newblock \bibinfo{title}{Solar modulation of galactic cosmic rays}.
\newblock \bibinfo{journal}{The Astrophysical Journal} \bibinfo{volume}{154},
  \bibinfo{pages}{1011}.
\newblock \DOIprefix\doi{10.1086/149822}.
%Type = Article
\bibitem[{Goorley et~al.(2012)Goorley, James, Booth, Brown, Bull, Cox, Durkee,
  Elson, Fensin, Forster et~al.}]{goorley2012initial}
\bibinfo{author}{Goorley, T.}, \bibinfo{author}{James, M.},
  \bibinfo{author}{Booth, T.}, \bibinfo{author}{Brown, F.},
  \bibinfo{author}{Bull, J.}, \bibinfo{author}{Cox, L.},
  \bibinfo{author}{Durkee, J.}, \bibinfo{author}{Elson, J.},
  \bibinfo{author}{Fensin, M.}, \bibinfo{author}{Forster, R.}, et~al.,
  \bibinfo{year}{2012}.
\newblock \bibinfo{title}{Initial mcnp6 release overview}.
\newblock \bibinfo{journal}{Nuclear technology} \bibinfo{volume}{180},
  \bibinfo{pages}{298--315}.
\newblock \DOIprefix\doi{10.13182/NT11-135}.
%Type = Article
\bibitem[{Grotzinger et~al.(2012)Grotzinger, Crisp, Vasavada, Anderson, Baker,
  Barry, Blake, Conrad, Edgett, Ferdowski, Gellert, Gilbert, Golombek,
  G{\'o}mez-Elvira, Hassler, Jandura, Litvak, Mahaffy, Maki, Meyer, Malin,
  Mitrofanov, Simmonds, Vaniman, Welch and Wiens}]{grotzinger2012mars}
\bibinfo{author}{Grotzinger, J.P.}, \bibinfo{author}{Crisp, J.},
  \bibinfo{author}{Vasavada, A.R.}, \bibinfo{author}{Anderson, R.C.},
  \bibinfo{author}{Baker, C.J.}, \bibinfo{author}{Barry, R.},
  \bibinfo{author}{Blake, D.F.}, \bibinfo{author}{Conrad, P.},
  \bibinfo{author}{Edgett, K.S.}, \bibinfo{author}{Ferdowski, B.},
  \bibinfo{author}{Gellert, R.}, \bibinfo{author}{Gilbert, J.B.},
  \bibinfo{author}{Golombek, M.}, \bibinfo{author}{G{\'o}mez-Elvira, J.},
  \bibinfo{author}{Hassler, D.M.}, \bibinfo{author}{Jandura, L.},
  \bibinfo{author}{Litvak, M.}, \bibinfo{author}{Mahaffy, P.},
  \bibinfo{author}{Maki, J.}, \bibinfo{author}{Meyer, M.},
  \bibinfo{author}{Malin, M.C.}, \bibinfo{author}{Mitrofanov, I.},
  \bibinfo{author}{Simmonds, J.J.}, \bibinfo{author}{Vaniman, D.},
  \bibinfo{author}{Welch, R.V.}, \bibinfo{author}{Wiens, R.C.},
  \bibinfo{year}{2012}.
\newblock \bibinfo{title}{{Mars Science Laboratory} mission and science
  investigation}.
\newblock \bibinfo{journal}{Space Science Reviews} \bibinfo{volume}{170},
  \bibinfo{pages}{5--56}.
\newblock \DOIprefix\doi{10.1007/s11214-012-9892-2}.
%Type = Article
\bibitem[{Guo et~al.(2019)Guo, Banjac, R\"ostel, Terasa, Herbst, Heber and
  Wimmer-Schweingruber}]{guo2019atris}
\bibinfo{author}{Guo, J.}, \bibinfo{author}{Banjac, S.},
  \bibinfo{author}{R\"ostel, L.}, \bibinfo{author}{Terasa, J.C.},
  \bibinfo{author}{Herbst, K.}, \bibinfo{author}{Heber, B.},
  \bibinfo{author}{Wimmer-Schweingruber, R.F.}, \bibinfo{year}{2019}.
\newblock \bibinfo{title}{Implementation and validation of the {GEANT4}/{AtRIS}
  code to model the radiation environment at {Mars}}.
\newblock \bibinfo{journal}{Journal of Space Weather and Space Climate}
  \bibinfo{volume}{9}.
\newblock \DOIprefix\doi{10.1051/swsc/2018051}.
%Type = Article
\bibitem[{Guo et~al.(2024)Guo, Wang, Whitman, Plainaki, Zhao, Bain, Cohen,
  Dalla, Dumbovic, Janvier et~al.}]{guo2024particle}
\bibinfo{author}{Guo, J.}, \bibinfo{author}{Wang, B.},
  \bibinfo{author}{Whitman, K.}, \bibinfo{author}{Plainaki, C.},
  \bibinfo{author}{Zhao, L.}, \bibinfo{author}{Bain, H.M.},
  \bibinfo{author}{Cohen, C.}, \bibinfo{author}{Dalla, S.},
  \bibinfo{author}{Dumbovic, M.}, \bibinfo{author}{Janvier, M.}, et~al.,
  \bibinfo{year}{2024}.
\newblock \bibinfo{title}{Particle radiation environment in the heliosphere:
  Status, limitations, and recommendations}.
\newblock \bibinfo{journal}{Advances in Space Research}
  \DOIprefix\doi{10.1016/j.asr.2024.03.070}.
%Type = Article
\bibitem[{Guo et~al.(2021)Guo, Zeitlin, Wimmer-Schweingruber, Hassler,
  Ehresmann, Rafkin, Freiherr~von Forstner, Khaksarighiri, Liu and
  Wang}]{guo2021review}
\bibinfo{author}{Guo, J.}, \bibinfo{author}{Zeitlin, C.},
  \bibinfo{author}{Wimmer-Schweingruber, R.F.}, \bibinfo{author}{Hassler,
  D.M.}, \bibinfo{author}{Ehresmann, B.}, \bibinfo{author}{Rafkin, S.},
  \bibinfo{author}{Freiherr~von Forstner, J.L.},
  \bibinfo{author}{Khaksarighiri, S.}, \bibinfo{author}{Liu, W.},
  \bibinfo{author}{Wang, Y.}, \bibinfo{year}{2021}.
\newblock \bibinfo{title}{Radiation environment for future human exploration on
  the surface of mars: the current understanding based on msl/rad dose
  measurements}.
\newblock \bibinfo{journal}{The Astronomy and Astrophysics Review}
  \bibinfo{volume}{29}, \bibinfo{pages}{1--81}.
\newblock \DOIprefix\doi{10.1007/s00159-021-00136-5}.
%Type = Article
\bibitem[{Hassler et~al.(2012)Hassler, Zeitlin, Wimmer-Schweingruber,
  B{\"o}ttcher, Martin, Andrews, B{\"o}hm, Brinza, Bullock, Burmeister
  et~al.}]{hassler2012}
\bibinfo{author}{Hassler, D.M.}, \bibinfo{author}{Zeitlin, C.},
  \bibinfo{author}{Wimmer-Schweingruber, R.F.}, \bibinfo{author}{B{\"o}ttcher,
  S.I.}, \bibinfo{author}{Martin, C.}, \bibinfo{author}{Andrews, J.},
  \bibinfo{author}{B{\"o}hm, E.}, \bibinfo{author}{Brinza, D.},
  \bibinfo{author}{Bullock, M.}, \bibinfo{author}{Burmeister, S.}, et~al.,
  \bibinfo{year}{2012}.
\newblock \bibinfo{title}{The {Radiation Assessment Detector} ({RAD})
  investigation}.
\newblock \bibinfo{journal}{Space Science Reviews} \bibinfo{volume}{170},
  \bibinfo{pages}{503--558}.
\newblock \DOIprefix\doi{10.1007/s11214-012-9913-1}.
%Type = Article
\bibitem[{Hassler et~al.(2014)Hassler, Zeitlin, Wimmer-Schweingruber,
  Ehresmann, Rafkin, Eigenbrode, Brinza, Weigle, B{\"o}ttcher, B{\"o}hm
  et~al.}]{hassler2014}
\bibinfo{author}{Hassler, D.M.}, \bibinfo{author}{Zeitlin, C.},
  \bibinfo{author}{Wimmer-Schweingruber, R.F.}, \bibinfo{author}{Ehresmann,
  B.}, \bibinfo{author}{Rafkin, S.}, \bibinfo{author}{Eigenbrode, J.L.},
  \bibinfo{author}{Brinza, D.E.}, \bibinfo{author}{Weigle, G.},
  \bibinfo{author}{B{\"o}ttcher, S.I.}, \bibinfo{author}{B{\"o}hm, E.}, et~al.,
  \bibinfo{year}{2014}.
\newblock \bibinfo{title}{{Mars}'s surface radiation environment measured with
  the {Mars Science Laboratory}'s curiosity rover}.
\newblock \bibinfo{journal}{Science} \bibinfo{volume}{343},
  \bibinfo{pages}{1244797}.
\newblock \DOIprefix\doi{10.1126/science.1244797}.
%Type = Article
\bibitem[{Honig et~al.(2019)Honig, Witasse, Evans, Nieminen, Kuulkers, Taylor,
  Heber, Guo and S{\'a}nchez-Cano}]{honig2019multi}
\bibinfo{author}{Honig, T.}, \bibinfo{author}{Witasse, O.G.},
  \bibinfo{author}{Evans, H.}, \bibinfo{author}{Nieminen, P.},
  \bibinfo{author}{Kuulkers, E.}, \bibinfo{author}{Taylor, M.G.},
  \bibinfo{author}{Heber, B.}, \bibinfo{author}{Guo, J.},
  \bibinfo{author}{S{\'a}nchez-Cano, B.}, \bibinfo{year}{2019}.
\newblock \bibinfo{title}{Multi-point galactic cosmic ray measurements between
  1 and 4.5 au over a full solar cycle}.
\newblock \bibinfo{journal}{Annales Geophysicae} \bibinfo{volume}{37}.
\newblock \DOIprefix\doi{10.5194/angeo-37-903-2019}.
%Type = Article
\bibitem[{ICRP(1992)}]{icrp60}
\bibinfo{author}{ICRP}, \bibinfo{year}{1992}.
\newblock \bibinfo{title}{{Recommendations of the International Commission on
  Radiological Protection (ICRP) 1990}}.
\newblock \bibinfo{journal}{European Journal of Nuclear Medicine}
  \bibinfo{volume}{19}, \bibinfo{pages}{77--79}.
\newblock \DOIprefix\doi{10.1007/BF00184120}.
%Type = Article
\bibitem[{Jun et~al.(2024)Jun, Garrett, Kim, Zheng, Fung, Corti, Ganushkina and
  Guo}]{jun2024review}
\bibinfo{author}{Jun, I.}, \bibinfo{author}{Garrett, H.}, \bibinfo{author}{Kim,
  W.}, \bibinfo{author}{Zheng, Y.}, \bibinfo{author}{Fung, S.F.},
  \bibinfo{author}{Corti, C.}, \bibinfo{author}{Ganushkina, N.},
  \bibinfo{author}{Guo, J.}, \bibinfo{year}{2024}.
\newblock \bibinfo{title}{A review on radiation environment pathways to
  impacts: Radiation effects, relevant empirical environment models, and future
  needs}.
\newblock \bibinfo{journal}{Advances in Space Research}
  \DOIprefix\doi{10.1016/j.asr.2024.03.079}.
%Type = Article
\bibitem[{Khaksarighiri et~al.(2020)Khaksarighiri, Guo, Wimmer-Schweingruber,
  Narici and Lohf}]{Khaksarighiri2020}
\bibinfo{author}{Khaksarighiri, S.}, \bibinfo{author}{Guo, J.},
  \bibinfo{author}{Wimmer-Schweingruber, R.}, \bibinfo{author}{Narici, L.},
  \bibinfo{author}{Lohf, H.}, \bibinfo{year}{2020}.
\newblock \bibinfo{title}{{Calculation of dose distribution in a realistic
  brain structure and the indication of space radiation influence on human
  brains}}.
\newblock \bibinfo{journal}{Life Sciences in Space Research}
  \bibinfo{volume}{27}, \bibinfo{pages}{33--48}.
\newblock \DOIprefix\doi{10.1016/j.lssr.2020.07.003}.
%Type = Article
\bibitem[{Kim et~al.(2014)Kim, Cucinotta, Nounu, Zeitlin, Hassler, Rafkin,
  Wimmer-Schweingruber, Ehresmann, Brinza, Böttcher, Böhm, Burmeister, Guo,
  Köhler, Martin, Reitz, Posner, Gómez-Elvira, Harri and the {MSL}
  Science~Team}]{kim2014HZETRN}
\bibinfo{author}{Kim, M.H.Y.}, \bibinfo{author}{Cucinotta, F.A.},
  \bibinfo{author}{Nounu, H.N.}, \bibinfo{author}{Zeitlin, C.},
  \bibinfo{author}{Hassler, D.M.}, \bibinfo{author}{Rafkin, S.C.R.},
  \bibinfo{author}{Wimmer-Schweingruber, R.F.}, \bibinfo{author}{Ehresmann,
  B.}, \bibinfo{author}{Brinza, D.}, \bibinfo{author}{Böttcher, S.},
  \bibinfo{author}{Böhm, E.}, \bibinfo{author}{Burmeister, S.},
  \bibinfo{author}{Guo, J.}, \bibinfo{author}{Köhler, J.},
  \bibinfo{author}{Martin, C.}, \bibinfo{author}{Reitz, G.},
  \bibinfo{author}{Posner, A.}, \bibinfo{author}{Gómez-Elvira, J.},
  \bibinfo{author}{Harri, A.M.}, \bibinfo{author}{the {MSL} Science~Team},
  \bibinfo{year}{2014}.
\newblock \bibinfo{title}{Comparison of martian surface ionizing radiation
  measurements from {MSL}-{RAD} with badhwar-{O'Neill} 2011/{HZETRN} model
  calculations}.
\newblock \bibinfo{journal}{Journal of Geophysical Research: Planets}
  \bibinfo{volume}{119}, \bibinfo{pages}{1311--1321}.
\newblock \DOIprefix\doi{10.1002/2013JE004549}.
%Type = Article
\bibitem[{Kudela(2009)}]{kudela2009energetic}
\bibinfo{author}{Kudela, K.}, \bibinfo{year}{2009}.
\newblock \bibinfo{title}{On energetic particles in space}.
\newblock \bibinfo{journal}{Acta Physica Slovaca} \bibinfo{volume}{59},
  \bibinfo{pages}{537--652}.
\newblock \DOIprefix\doi{10.2478/v10155-010-0098-4}.
%Type = Techreport
\bibitem[{Kulesza et~al.(2022)Kulesza, Adams, Armstrong, Bolding, Brown, Bull,
  Burke, Clark, Forster~III, Giron et~al.}]{kulesza2022mcnp}
\bibinfo{author}{Kulesza, J.A.}, \bibinfo{author}{Adams, T.R.},
  \bibinfo{author}{Armstrong, J.C.}, \bibinfo{author}{Bolding, S.R.},
  \bibinfo{author}{Brown, F.B.}, \bibinfo{author}{Bull, J.S.},
  \bibinfo{author}{Burke, T.P.}, \bibinfo{author}{Clark, A.R.},
  \bibinfo{author}{Forster~III, R.A.A.}, \bibinfo{author}{Giron, J.F.}, et~al.,
  \bibinfo{year}{2022}.
\newblock \bibinfo{title}{{MCNP{\textregistered} code version 6.3.0 theory \&
  user manual}}.
\newblock \bibinfo{type}{Technical Report}. Los Alamos National Laboratory
  (LANL), Los Alamos, NM (United States).
\newblock \DOIprefix\doi{10.2172/1889957}.
%Type = Article
\bibitem[{Li et~al.(2023)Li, Guo, Khaksarighiri, Dobynde, Zhang, Liu and
  Wimmer-Schweingruber}]{li2023impact}
\bibinfo{author}{Li, Y.}, \bibinfo{author}{Guo, J.},
  \bibinfo{author}{Khaksarighiri, S.}, \bibinfo{author}{Dobynde, M.I.},
  \bibinfo{author}{Zhang, J.}, \bibinfo{author}{Liu, B.},
  \bibinfo{author}{Wimmer-Schweingruber, R.F.}, \bibinfo{year}{2023}.
\newblock \bibinfo{title}{The impact of space radiation on brains of future
  martian and lunar explorers}.
\newblock \bibinfo{journal}{Space Weather} \bibinfo{volume}{21},
  \bibinfo{pages}{e2023SW003470}.
\newblock \DOIprefix\doi{10.1029/2023SW003470}.
%Type = Article
\bibitem[{Liu et~al.(2024)Liu, Guo, Wang and Slaba}]{liu2024comprehensive}
\bibinfo{author}{Liu, W.}, \bibinfo{author}{Guo, J.}, \bibinfo{author}{Wang,
  Y.}, \bibinfo{author}{Slaba, T.C.}, \bibinfo{year}{2024}.
\newblock \bibinfo{title}{{A Comprehensive Comparison of Various Galactic
  Cosmic-Ray Models to the State-of-the-art Particle and Radiation
  Measurements}}.
\newblock \bibinfo{journal}{The Astrophysical Journal Supplement Series}
  \bibinfo{volume}{271}, \bibinfo{pages}{18}.
\newblock \DOIprefix\doi{10.3847/1538-4365/ad18ad}.
%Type = Article
\bibitem[{Liu et~al.(2023)Liu, Guo, Zhang and Semkova}]{liu2023modeling}
\bibinfo{author}{Liu, W.}, \bibinfo{author}{Guo, J.}, \bibinfo{author}{Zhang,
  J.}, \bibinfo{author}{Semkova, J.}, \bibinfo{year}{2023}.
\newblock \bibinfo{title}{{Modeling the Radiation Environment of Energetic
  Particles at Mars Orbit and a First Validation against TGO Measurements}}.
\newblock \bibinfo{journal}{The Astrophysical Journal} \bibinfo{volume}{949},
  \bibinfo{pages}{77}.
\newblock \DOIprefix\doi{10.3847/1538-4357/acce3c}.
%Type = Article
\bibitem[{Looper et~al.(2020)Looper, Mazur, Blake, Spence, Schwadron, Wilson,
  Jordan, Zeitlin, Case, Kasper et~al.}]{looper2020long}
\bibinfo{author}{Looper, M.}, \bibinfo{author}{Mazur, J.},
  \bibinfo{author}{Blake, J.}, \bibinfo{author}{Spence, H.},
  \bibinfo{author}{Schwadron, N.}, \bibinfo{author}{Wilson, J.},
  \bibinfo{author}{Jordan, A.}, \bibinfo{author}{Zeitlin, C.},
  \bibinfo{author}{Case, A.}, \bibinfo{author}{Kasper, J.}, et~al.,
  \bibinfo{year}{2020}.
\newblock \bibinfo{title}{{Long-term observations of galactic cosmic ray LET
  spectra in lunar orbit by LRO/CRaTER}}.
\newblock \bibinfo{journal}{Space Weather} \bibinfo{volume}{18},
  \bibinfo{pages}{e2020SW002543}.
\newblock \DOIprefix\doi{10.1029/2020SW002543}.
%Type = Article
\bibitem[{Looper et~al.(2013)Looper, Mazur, Blake, Spence, Schwadron,
  Golightly, Case, Kasper and Townsend}]{looper2013radiation}
\bibinfo{author}{Looper, M.}, \bibinfo{author}{Mazur, J.},
  \bibinfo{author}{Blake, J.}, \bibinfo{author}{Spence, H.E.},
  \bibinfo{author}{Schwadron, N.A.}, \bibinfo{author}{Golightly, M.},
  \bibinfo{author}{Case, A.}, \bibinfo{author}{Kasper, J.},
  \bibinfo{author}{Townsend, L.}, \bibinfo{year}{2013}.
\newblock \bibinfo{title}{The radiation environment near the lunar surface:
  Crater observations and geant4 simulations}.
\newblock \bibinfo{journal}{Space Weather} \bibinfo{volume}{11},
  \bibinfo{pages}{142--152}.
\newblock \DOIprefix\doi{10.1002/swe.20034}.
%Type = Article
\bibitem[{Lyu et~al.(2024)Lyu, Qin and Shen}]{lyu2024long}
\bibinfo{author}{Lyu, D.}, \bibinfo{author}{Qin, G.}, \bibinfo{author}{Shen,
  Z.N.}, \bibinfo{year}{2024}.
\newblock \bibinfo{title}{{Long-Term Variation of the Galactic Cosmic Ray
  Radiation Dose Rates}}.
\newblock \bibinfo{journal}{Space Weather} \bibinfo{volume}{22},
  \bibinfo{pages}{e2023SW003804}.
\newblock \DOIprefix\doi{10.1029/2023SW003804}.
%Type = Article
\bibitem[{Marquardt et~al.(2018)Marquardt, Heber, Potgieter and
  Strauss}]{marquardt2018energy}
\bibinfo{author}{Marquardt, J.}, \bibinfo{author}{Heber, B.},
  \bibinfo{author}{Potgieter, M.}, \bibinfo{author}{Strauss, R.},
  \bibinfo{year}{2018}.
\newblock \bibinfo{title}{Energy spectra of carbon and oxygen with helios
  e6-radial gradients of anomalous cosmic ray oxygen within 1 au}.
\newblock \bibinfo{journal}{Astronomy \& Astrophysics} \bibinfo{volume}{610},
  \bibinfo{pages}{A42}.
\newblock \DOIprefix\doi{10.1051/0004-6361/201731490}.
%Type = Article
\bibitem[{Matthi{\"a} and Berger(2017)}]{matthiae2017radiation}
\bibinfo{author}{Matthi{\"a}, D.}, \bibinfo{author}{Berger, T.},
  \bibinfo{year}{2017}.
\newblock \bibinfo{title}{The radiation environment on the surface of
  {Mars}--numerical calculations of the galactic component with
  {GEANT4}/planetocosmics}.
\newblock \bibinfo{journal}{Life sciences in space research}
  \bibinfo{volume}{14}, \bibinfo{pages}{57--63}.
\newblock \DOIprefix\doi{10.1016/j.lssr.2017.03.005}.
%Type = Article
\bibitem[{Matthi{\"a} et~al.(2016)Matthi{\"a}, Ehresmann, Lohf, K{\"o}hler,
  Zeitlin, Appel, Sato, Slaba, Martin, Berger et~al.}]{matthia2016martian}
\bibinfo{author}{Matthi{\"a}, D.}, \bibinfo{author}{Ehresmann, B.},
  \bibinfo{author}{Lohf, H.}, \bibinfo{author}{K{\"o}hler, J.},
  \bibinfo{author}{Zeitlin, C.}, \bibinfo{author}{Appel, J.},
  \bibinfo{author}{Sato, T.}, \bibinfo{author}{Slaba, T.},
  \bibinfo{author}{Martin, C.}, \bibinfo{author}{Berger, T.}, et~al.,
  \bibinfo{year}{2016}.
\newblock \bibinfo{title}{The martian surface radiation environment--a
  comparison of models and {MSL}/{RAD} measurements}.
\newblock \bibinfo{journal}{Journal of Space Weather and Space Climate}
  \bibinfo{volume}{6}, \bibinfo{pages}{1--17}.
\newblock \DOIprefix\doi{10.1051/swsc/2016008}.
%Type = Article
\bibitem[{Mazur et~al.(2011)Mazur, Crain, Looper, Mabry, Blake, Case,
  Golightly, Kasper and Spence}]{mazur2011new}
\bibinfo{author}{Mazur, J.}, \bibinfo{author}{Crain, W.},
  \bibinfo{author}{Looper, M.}, \bibinfo{author}{Mabry, D.},
  \bibinfo{author}{Blake, J.}, \bibinfo{author}{Case, A.},
  \bibinfo{author}{Golightly, M.}, \bibinfo{author}{Kasper, J.},
  \bibinfo{author}{Spence, H.E.}, \bibinfo{year}{2011}.
\newblock \bibinfo{title}{New measurements of total ionizing dose in the lunar
  environment}.
\newblock \bibinfo{journal}{Space Weather} \bibinfo{volume}{9}.
\newblock \DOIprefix\doi{10.1029/2010SW000641}.
%Type = Article
\bibitem[{McDonald et~al.(1974)McDonald, Teegarden, Trainor and
  Webber}]{mcdonald1974anomalous}
\bibinfo{author}{McDonald, F.}, \bibinfo{author}{Teegarden, B.},
  \bibinfo{author}{Trainor, J.}, \bibinfo{author}{Webber, W.},
  \bibinfo{year}{1974}.
\newblock \bibinfo{title}{The anomalous abundance of cosmic-ray nitrogen and
  oxygen nuclei at low energies}.
\newblock \bibinfo{journal}{The Astrophysical Journal} \bibinfo{volume}{187},
  \bibinfo{pages}{L105}.
\newblock \DOIprefix\doi{10.1086/181407}.
%Type = Article
\bibitem[{{Mitrofanov} et~al.(2018){Mitrofanov}, {Malakhov}, {Bakhtin},
  {Golovin}, {Kozyrev}, {Litvak}, {Mokrousov}, {Sanin}, {Tretyakov},
  {Vostrukhin}, {Anikin}, {Zelenyi}, {Semkova}, {Malchev}, {Tomov},
  {Matviichuk}, {Dimitrov}, {Koleva}, {Dachev}, {Krastev}, {Shvetsov},
  {Timoshenko}, {Bobrovnitsky}, {Tomilina}, {Benghin} and
  {Shurshakov}}]{mitrofanov2018fine}
\bibinfo{author}{{Mitrofanov}, I.}, \bibinfo{author}{{Malakhov}, A.},
  \bibinfo{author}{{Bakhtin}, B.}, \bibinfo{author}{{Golovin}, D.},
  \bibinfo{author}{{Kozyrev}, A.}, \bibinfo{author}{{Litvak}, M.},
  \bibinfo{author}{{Mokrousov}, M.}, \bibinfo{author}{{Sanin}, A.},
  \bibinfo{author}{{Tretyakov}, V.}, \bibinfo{author}{{Vostrukhin}, A.},
  \bibinfo{author}{{Anikin}, A.}, \bibinfo{author}{{Zelenyi}, L.M.},
  \bibinfo{author}{{Semkova}, J.}, \bibinfo{author}{{Malchev}, S.},
  \bibinfo{author}{{Tomov}, B.}, \bibinfo{author}{{Matviichuk}, Y.},
  \bibinfo{author}{{Dimitrov}, P.}, \bibinfo{author}{{Koleva}, R.},
  \bibinfo{author}{{Dachev}, T.}, \bibinfo{author}{{Krastev}, K.},
  \bibinfo{author}{{Shvetsov}, V.}, \bibinfo{author}{{Timoshenko}, G.},
  \bibinfo{author}{{Bobrovnitsky}, Y.}, \bibinfo{author}{{Tomilina}, T.},
  \bibinfo{author}{{Benghin}, V.}, \bibinfo{author}{{Shurshakov}, V.},
  \bibinfo{year}{2018}.
\newblock \bibinfo{title}{{Fine Resolution Epithermal Neutron Detector (FREND)
  Onboard the ExoMars Trace Gas Orbiter}}.
\newblock \bibinfo{journal}{Space Science Reviews} \bibinfo{volume}{214},
  \bibinfo{pages}{86}.
\newblock \DOIprefix\doi{10.1007/s11214-018-0522-5}.
%Type = Article
\bibitem[{Naito and Kodaira(2022)}]{naito2022considerations}
\bibinfo{author}{Naito, M.}, \bibinfo{author}{Kodaira, S.},
  \bibinfo{year}{2022}.
\newblock \bibinfo{title}{Considerations for practical dose equivalent
  assessment of space radiation and exposure risk reduction in deep space}.
\newblock \bibinfo{journal}{Scientific Reports} \bibinfo{volume}{12},
  \bibinfo{pages}{13617}.
\newblock \DOIprefix\doi{10.1038/s41598-022-17079-1}.
%Type = Article
\bibitem[{Naito and Kodaira(2025)}]{naito2025charge}
\bibinfo{author}{Naito, M.}, \bibinfo{author}{Kodaira, S.},
  \bibinfo{year}{2025}.
\newblock \bibinfo{title}{Charge-weighted quality factors based on let for
  practical dose assessment of space radiation}.
\newblock \bibinfo{journal}{Life Sciences in Space Research}
  \DOIprefix\doi{10.1016/j.lssr.2025.10.002}.
%Type = Article
\bibitem[{Niita et~al.(2006)Niita, Sato, Iwase, Nose, Nakashima and
  Sihver}]{niita2006phits}
\bibinfo{author}{Niita, K.}, \bibinfo{author}{Sato, T.},
  \bibinfo{author}{Iwase, H.}, \bibinfo{author}{Nose, H.},
  \bibinfo{author}{Nakashima, H.}, \bibinfo{author}{Sihver, L.},
  \bibinfo{year}{2006}.
\newblock \bibinfo{title}{Phits—a particle and heavy ion transport code
  system}.
\newblock \bibinfo{journal}{Radiation measurements} \bibinfo{volume}{41},
  \bibinfo{pages}{1080--1090}.
\newblock \DOIprefix\doi{10.1016/j.radmeas.2006.07.013}.
%Type = Article
\bibitem[{Norbury et~al.(2019)Norbury, Slaba, Aghara, Badavi, Blattnig,
  Clowdsley, Heilbronn, Lee, Maung, Mertens et~al.}]{norbury2019advances}
\bibinfo{author}{Norbury, J.W.}, \bibinfo{author}{Slaba, T.C.},
  \bibinfo{author}{Aghara, S.}, \bibinfo{author}{Badavi, F.F.},
  \bibinfo{author}{Blattnig, S.R.}, \bibinfo{author}{Clowdsley, M.S.},
  \bibinfo{author}{Heilbronn, L.H.}, \bibinfo{author}{Lee, K.},
  \bibinfo{author}{Maung, K.M.}, \bibinfo{author}{Mertens, C.J.}, et~al.,
  \bibinfo{year}{2019}.
\newblock \bibinfo{title}{Advances in space radiation physics and transport at
  nasa}.
\newblock \bibinfo{journal}{Life Sciences in Space Research}
  \bibinfo{volume}{22}, \bibinfo{pages}{98--124}.
\newblock \DOIprefix\doi{10.1016/j.lssr.2019.07.003}.
%Type = Article
\bibitem[{{Parker}(1965)}]{parker1965}
\bibinfo{author}{{Parker}, E.N.}, \bibinfo{year}{1965}.
\newblock \bibinfo{title}{{The passage of energetic charged particles through
  interplanetary space}}.
\newblock \bibinfo{journal}{Planetary and Space Science} \bibinfo{volume}{13},
  \bibinfo{pages}{9--49}.
\newblock \DOIprefix\doi{10.1016/0032-0633(65)90131-5}.
%Type = Article
\bibitem[{Potgieter(2013)}]{potgieter2013cosmic}
\bibinfo{author}{Potgieter, M.}, \bibinfo{year}{2013}.
\newblock \bibinfo{title}{Cosmic rays in the inner heliosphere: Insights from
  observations, theory and models}.
\newblock \bibinfo{journal}{Space Science Reviews} \bibinfo{volume}{176},
  \bibinfo{pages}{165--176}.
\newblock \DOIprefix\doi{10.1007/s11214-011-9750-7}.
%Type = Article
\bibitem[{Rahmanian et~al.(2023)Rahmanian, Slaba, Braby, Santa~Maria,
  Bhattacharya and Straume}]{rahmanian2023galactic}
\bibinfo{author}{Rahmanian, S.}, \bibinfo{author}{Slaba, T.},
  \bibinfo{author}{Braby, L.}, \bibinfo{author}{Santa~Maria, S.},
  \bibinfo{author}{Bhattacharya, S.}, \bibinfo{author}{Straume, T.},
  \bibinfo{year}{2023}.
\newblock \bibinfo{title}{Galactic cosmic ray environment predictions for the
  nasa biosentinel mission}.
\newblock \bibinfo{journal}{Life Sciences in Space Research}
  \bibinfo{volume}{38}, \bibinfo{pages}{19--28}.
\newblock \DOIprefix\doi{10.1016/j.lssr.2023.05.001}.
%Type = Article
\bibitem[{Rahmanian et~al.(2024)Rahmanian, Slaba, George, Braby, Bhattacharya,
  Straume and Santa~Maria}]{rahmanian2024galactic}
\bibinfo{author}{Rahmanian, S.}, \bibinfo{author}{Slaba, T.C.},
  \bibinfo{author}{George, S.}, \bibinfo{author}{Braby, L.A.},
  \bibinfo{author}{Bhattacharya, S.}, \bibinfo{author}{Straume, T.},
  \bibinfo{author}{Santa~Maria, S.R.}, \bibinfo{year}{2024}.
\newblock \bibinfo{title}{Galactic cosmic ray environment predictions for the
  nasa biosentinel mission, part 2: Post-mission validation}.
\newblock \bibinfo{journal}{Life Sciences in Space Research}
  \DOIprefix\doi{10.1016/j.lssr.2024.10.006}.
%Type = Article
\bibitem[{Rankin et~al.(2021)Rankin, McComas, Leske, Christian, Cohen,
  Cummings, Joyce, Labrador, Mewaldt, Posner et~al.}]{rankin2021first}
\bibinfo{author}{Rankin, J.}, \bibinfo{author}{McComas, D.},
  \bibinfo{author}{Leske, R.}, \bibinfo{author}{Christian, E.},
  \bibinfo{author}{Cohen, C.}, \bibinfo{author}{Cummings, A.},
  \bibinfo{author}{Joyce, C.}, \bibinfo{author}{Labrador, A.},
  \bibinfo{author}{Mewaldt, R.}, \bibinfo{author}{Posner, A.}, et~al.,
  \bibinfo{year}{2021}.
\newblock \bibinfo{title}{First observations of anomalous cosmic rays in to 36
  solar radii}.
\newblock \bibinfo{journal}{The Astrophysical Journal} \bibinfo{volume}{912},
  \bibinfo{pages}{139}.
\newblock \DOIprefix\doi{10.3847/1538-4357/abec7e}.
%Type = Article
\bibitem[{Rankin et~al.(2022)Rankin, McComas, Leske, Christian, Cohen,
  Cummings, Joyce, Labrador, Mewaldt, Schwadron et~al.}]{rankin2022anomalous}
\bibinfo{author}{Rankin, J.S.}, \bibinfo{author}{McComas, D.J.},
  \bibinfo{author}{Leske, R.A.}, \bibinfo{author}{Christian, E.R.},
  \bibinfo{author}{Cohen, C.}, \bibinfo{author}{Cummings, A.C.},
  \bibinfo{author}{Joyce, C.J.}, \bibinfo{author}{Labrador, A.W.},
  \bibinfo{author}{Mewaldt, R.A.}, \bibinfo{author}{Schwadron, N.A.}, et~al.,
  \bibinfo{year}{2022}.
\newblock \bibinfo{title}{Anomalous cosmic-ray oxygen observations into 0.1
  au}.
\newblock \bibinfo{journal}{The Astrophysical Journal} \bibinfo{volume}{925},
  \bibinfo{pages}{9}.
\newblock \DOIprefix\doi{10.3847/1538-4357/ac348f}.
%Type = Article
\bibitem[{Reames(2013)}]{Reames-2013}
\bibinfo{author}{Reames, D.V.}, \bibinfo{year}{2013}.
\newblock \bibinfo{title}{{The Two Sources of Solar Energetic Particles}}.
\newblock \bibinfo{journal}{Space Science Reviews} \bibinfo{volume}{175},
  \bibinfo{pages}{53--92}.
\newblock \DOIprefix\doi{10.1007/s11214-013-9958-9}.
%Type = Article
\bibitem[{Ricco et~al.(2020)Ricco, Santa~Maria, Hanel and
  Bhattacharya}]{ricco2020biosentinel}
\bibinfo{author}{Ricco, A.J.}, \bibinfo{author}{Santa~Maria, S.R.},
  \bibinfo{author}{Hanel, R.P.}, \bibinfo{author}{Bhattacharya, S.},
  \bibinfo{year}{2020}.
\newblock \bibinfo{title}{{BioSentinel: A 6U nanosatellite for deep-space
  biological science}}.
\newblock \bibinfo{journal}{IEEE Aerospace and Electronic Systems Magazine}
  \bibinfo{volume}{35}, \bibinfo{pages}{6--18}.
\newblock \DOIprefix\doi{10.1109/MAES.2019.2953760}.
%Type = Article
\bibitem[{R\"ostel et~al.(2020)R\"ostel, Guo, Banjac, Wimmer-Schweingruber and
  Heber}]{rostel2020}
\bibinfo{author}{R\"ostel, L.}, \bibinfo{author}{Guo, J.},
  \bibinfo{author}{Banjac, S.}, \bibinfo{author}{Wimmer-Schweingruber, R.F.},
  \bibinfo{author}{Heber, B.}, \bibinfo{year}{2020}.
\newblock \bibinfo{title}{Subsurface radiation environment of {Mars} and its
  implication for shielding protection of future habitats}.
\newblock \bibinfo{journal}{Journal of Geophysical Research: Planets}
  \bibinfo{volume}{125}, \bibinfo{pages}{e2019JE006246}.
\newblock \DOIprefix\doi{10.1029/2019JE006246}.
%Type = Article
\bibitem[{Roussos et~al.(2020)Roussos, Dialynas, Krupp, Kollmann, Paranicas,
  Roelof, Yuan, Mitchell and Krimigis}]{Roussos2020}
\bibinfo{author}{Roussos, E.}, \bibinfo{author}{Dialynas, K.},
  \bibinfo{author}{Krupp, N.}, \bibinfo{author}{Kollmann, P.},
  \bibinfo{author}{Paranicas, C.}, \bibinfo{author}{Roelof, E.C.},
  \bibinfo{author}{Yuan, C.}, \bibinfo{author}{Mitchell, D.G.},
  \bibinfo{author}{Krimigis, S.M.}, \bibinfo{year}{2020}.
\newblock \bibinfo{title}{Long- and short-term variability of galactic
  cosmic-ray radial intensity gradients between 1 and 9.5 au: Observations by
  cassini, {BESS}, {BESS}-polar, {PAMELA}, and {AMS}-02}.
\newblock \bibinfo{journal}{The Astrophysical Journal} \bibinfo{volume}{904},
  \bibinfo{pages}{165}.
\newblock \DOIprefix\doi{10.3847/1538-4357/abc346}.
%Type = Article
\bibitem[{Sato et~al.(2018)Sato, Iwamoto, Hashimoto, Ogawa, Furuta, Abe, Kai,
  Tsai, Matsuda, Iwase et~al.}]{sato2018features}
\bibinfo{author}{Sato, T.}, \bibinfo{author}{Iwamoto, Y.},
  \bibinfo{author}{Hashimoto, S.}, \bibinfo{author}{Ogawa, T.},
  \bibinfo{author}{Furuta, T.}, \bibinfo{author}{Abe, S.i.},
  \bibinfo{author}{Kai, T.}, \bibinfo{author}{Tsai, P.E.},
  \bibinfo{author}{Matsuda, N.}, \bibinfo{author}{Iwase, H.}, et~al.,
  \bibinfo{year}{2018}.
\newblock \bibinfo{title}{Features of particle and heavy ion transport code
  system (phits) version 3.02}.
\newblock \bibinfo{journal}{Journal of Nuclear Science and Technology}
  \bibinfo{volume}{55}, \bibinfo{pages}{684--690}.
\newblock \DOIprefix\doi{10.1080/00223131.2017.1419890}.
%Type = Article
\bibitem[{Schwadron et~al.(2012)Schwadron, Baker, Blake, Case, Cooper,
  Golightly, Jordan, Joyce, Kasper, Kozarev, Mislinski, Mazur, Posner, Rother,
  Smith, Spence, Townsend, Wilson and Zeitlin}]{schwadron_lunar_2012}
\bibinfo{author}{Schwadron, N.A.}, \bibinfo{author}{Baker, T.},
  \bibinfo{author}{Blake, B.}, \bibinfo{author}{Case, A.W.},
  \bibinfo{author}{Cooper, J.F.}, \bibinfo{author}{Golightly, M.},
  \bibinfo{author}{Jordan, A.}, \bibinfo{author}{Joyce, C.},
  \bibinfo{author}{Kasper, J.}, \bibinfo{author}{Kozarev, K.},
  \bibinfo{author}{Mislinski, J.}, \bibinfo{author}{Mazur, J.},
  \bibinfo{author}{Posner, A.}, \bibinfo{author}{Rother, O.},
  \bibinfo{author}{Smith, S.}, \bibinfo{author}{Spence, H.E.},
  \bibinfo{author}{Townsend, L.W.}, \bibinfo{author}{Wilson, J.},
  \bibinfo{author}{Zeitlin, C.}, \bibinfo{year}{2012}.
\newblock \bibinfo{title}{Lunar radiation environment and space weathering from
  the {Cosmic} {Ray} {Telescope} for the {Effects} of {Radiation} ({CRaTER})}.
\newblock \bibinfo{journal}{Journal of Geophysical Research: Planets}
  \bibinfo{volume}{117}, \bibinfo{pages}{E00H13}.
\newblock \DOIprefix\doi{10.1029/2011JE003978}.
%Type = Article
\bibitem[{Schwadron et~al.(2014)Schwadron, Blake, Case, Joyce, Kasper, Mazur,
  Petro, Quinn, Porter, Smith, Smith, Spence, Townsend, Turner, Wilson and
  Zeitlin}]{schwadron_does_2014}
\bibinfo{author}{Schwadron, N.A.}, \bibinfo{author}{Blake, J.B.},
  \bibinfo{author}{Case, A.W.}, \bibinfo{author}{Joyce, C.J.},
  \bibinfo{author}{Kasper, J.}, \bibinfo{author}{Mazur, J.},
  \bibinfo{author}{Petro, N.}, \bibinfo{author}{Quinn, M.},
  \bibinfo{author}{Porter, J.A.}, \bibinfo{author}{Smith, C.W.},
  \bibinfo{author}{Smith, S.}, \bibinfo{author}{Spence, H.E.},
  \bibinfo{author}{Townsend, L.W.}, \bibinfo{author}{Turner, R.},
  \bibinfo{author}{Wilson, J.K.}, \bibinfo{author}{Zeitlin, C.},
  \bibinfo{year}{2014}.
\newblock \bibinfo{title}{Does the worsening galactic cosmic radiation
  environment observed by {CRaTER} preclude future manned deep space
  exploration?}
\newblock \bibinfo{journal}{Space Weather} \bibinfo{volume}{12},
  \bibinfo{pages}{622--632}.
\newblock \DOIprefix\doi{10.1002/2014SW001084}.
%Type = Article
\bibitem[{Semkova et~al.(2023a)Semkova, Benghin, Guo, Zhang, Da~Pieve, Krastev,
  Matviichuk, Tomov, Shurshakov, Drobyshev et~al.}]{semkova2023comparison}
\bibinfo{author}{Semkova, J.}, \bibinfo{author}{Benghin, V.},
  \bibinfo{author}{Guo, J.}, \bibinfo{author}{Zhang, J.},
  \bibinfo{author}{Da~Pieve, F.}, \bibinfo{author}{Krastev, K.},
  \bibinfo{author}{Matviichuk, Y.}, \bibinfo{author}{Tomov, B.},
  \bibinfo{author}{Shurshakov, V.}, \bibinfo{author}{Drobyshev, S.}, et~al.,
  \bibinfo{year}{2023}a.
\newblock \bibinfo{title}{Comparison of the particle flux measured by liulin-mo
  dosimeter in exomars tgo science orbit with those calculated by models}.
\newblock \bibinfo{journal}{Life Sciences in Space Research}
  \bibinfo{volume}{39}, \bibinfo{pages}{119--130}.
\newblock \DOIprefix\doi{10.1016/j.lssr.2022.08.007}.
%Type = Article
\bibitem[{{Semkova} et~al.(2015){Semkova}, Dachev, Maltchev, Tomov, Matviichuk,
  Dimitrov, Koleva, Mitrofanov, Malakhov, Mokrousov
  et~al.}]{semkova2015radiation}
\bibinfo{author}{{Semkova}, J.}, \bibinfo{author}{Dachev, T.},
  \bibinfo{author}{Maltchev, S.}, \bibinfo{author}{Tomov, B.},
  \bibinfo{author}{Matviichuk, Y.}, \bibinfo{author}{Dimitrov, P.},
  \bibinfo{author}{Koleva, R.}, \bibinfo{author}{Mitrofanov, I.},
  \bibinfo{author}{Malakhov, A.}, \bibinfo{author}{Mokrousov, M.}, et~al.,
  \bibinfo{year}{2015}.
\newblock \bibinfo{title}{Radiation environment investigations during exomars
  missions to mars--objectives, experiments and instrumentation}.
\newblock \bibinfo{journal}{Comptes rendus de l’Acad{\'e}mie bulgare des
  Sciences} \bibinfo{volume}{68}.
%Type = Article
\bibitem[{Semkova et~al.(2021)Semkova, Koleva, Benghin, Dachev, Matviichuk,
  Tomov, Krastev, Maltchev, Dimitrov, Bankov et~al.}]{semkova2021results}
\bibinfo{author}{Semkova, J.}, \bibinfo{author}{Koleva, R.},
  \bibinfo{author}{Benghin, V.}, \bibinfo{author}{Dachev, T.},
  \bibinfo{author}{Matviichuk, Y.}, \bibinfo{author}{Tomov, B.},
  \bibinfo{author}{Krastev, K.}, \bibinfo{author}{Maltchev, S.},
  \bibinfo{author}{Dimitrov, P.}, \bibinfo{author}{Bankov, N.}, et~al.,
  \bibinfo{year}{2021}.
\newblock \bibinfo{title}{{Results from radiation environment measurements
  aboard ExoMars Trace Gas Orbiter in Mars science orbit in May 2018--December
  2019}}.
\newblock \bibinfo{journal}{Icarus} \bibinfo{volume}{361},
  \bibinfo{pages}{114264}.
\newblock \DOIprefix\doi{10.1016/j.icarus.2020.114264}.
%Type = Article
\bibitem[{{Semkova} et~al.(2018){Semkova}, {Koleva}, {Benghin}, {Dachev},
  {Matviichuk}, {Tomov}, {Krastev}, {Maltchev}, {Dimitrov}, {Mitrofanov},
  {Malahov}, {Golovin}, {Mokrousov}, {Sanin}, {Litvak}, {Kozyrev}, {Tretyakov},
  {Nikiforov}, {Vostrukhin}, {Fedosov}, {Grebennikova}, {Zelenyi}, {Shurshakov}
  and {Drobishev}}]{Semkova2018}
\bibinfo{author}{{Semkova}, J.}, \bibinfo{author}{{Koleva}, R.},
  \bibinfo{author}{{Benghin}, V.}, \bibinfo{author}{{Dachev}, T.},
  \bibinfo{author}{{Matviichuk}, Y.}, \bibinfo{author}{{Tomov}, B.},
  \bibinfo{author}{{Krastev}, K.}, \bibinfo{author}{{Maltchev}, S.},
  \bibinfo{author}{{Dimitrov}, P.}, \bibinfo{author}{{Mitrofanov}, I.},
  \bibinfo{author}{{Malahov}, A.}, \bibinfo{author}{{Golovin}, D.},
  \bibinfo{author}{{Mokrousov}, M.}, \bibinfo{author}{{Sanin}, A.},
  \bibinfo{author}{{Litvak}, M.}, \bibinfo{author}{{Kozyrev}, A.},
  \bibinfo{author}{{Tretyakov}, V.}, \bibinfo{author}{{Nikiforov}, S.},
  \bibinfo{author}{{Vostrukhin}, A.}, \bibinfo{author}{{Fedosov}, F.},
  \bibinfo{author}{{Grebennikova}, N.}, \bibinfo{author}{{Zelenyi}, L.},
  \bibinfo{author}{{Shurshakov}, V.}, \bibinfo{author}{{Drobishev}, S.},
  \bibinfo{year}{2018}.
\newblock \bibinfo{title}{{Charged particles radiation measurements with
  Liulin-MO dosimeter of FREND instrument aboard ExoMars Trace Gas Orbiter
  during the transit and in high elliptic {Mars} orbit}}.
\newblock \bibinfo{journal}{Icarus} \bibinfo{volume}{303},
  \bibinfo{pages}{53--66}.
\newblock \DOIprefix\doi{10.1016/j.icarus.2017.12.034}.
%Type = Article
\bibitem[{Semkova et~al.(2023b)Semkova, Koleva, Benghin, Krastev, Matviichuk,
  Tomov, Maltchev, Dachev, Bankov, Mitrofanov et~al.}]{semkova2023observation}
\bibinfo{author}{Semkova, J.}, \bibinfo{author}{Koleva, R.},
  \bibinfo{author}{Benghin, V.}, \bibinfo{author}{Krastev, K.},
  \bibinfo{author}{Matviichuk, Y.}, \bibinfo{author}{Tomov, B.},
  \bibinfo{author}{Maltchev, S.}, \bibinfo{author}{Dachev, T.},
  \bibinfo{author}{Bankov, N.}, \bibinfo{author}{Mitrofanov, I.}, et~al.,
  \bibinfo{year}{2023}b.
\newblock \bibinfo{title}{Observation of the radiation environment and solar
  energetic particle events in mars orbit in may 2018-june 2022}.
\newblock \bibinfo{journal}{Life Sciences in Space Research}
  \DOIprefix\doi{10.1016/j.lssr.2023.03.006}.
%Type = Article
\bibitem[{{SILSO World Data Center}(2006–2022)}]{sidc}
\bibinfo{author}{{SILSO World Data Center}}, \bibinfo{year}{2006–2022}.
\newblock \bibinfo{title}{{The International Sunspot Number}}.
\newblock \bibinfo{journal}{International Sunspot Number Monthly Bulletin and
  online catalogue} \URLprefix \url{http://www.sidc.be/silso/,}.
%Type = Article
\bibitem[{Simonsen et~al.(2020)Simonsen, Slaba, Guida and
  Rusek}]{simonsen2020nasa}
\bibinfo{author}{Simonsen, L.C.}, \bibinfo{author}{Slaba, T.C.},
  \bibinfo{author}{Guida, P.}, \bibinfo{author}{Rusek, A.},
  \bibinfo{year}{2020}.
\newblock \bibinfo{title}{Nasa’s first ground-based galactic cosmic ray
  simulator: Enabling a new era in space radiobiology research}.
\newblock \bibinfo{journal}{PLoS biology} \bibinfo{volume}{18},
  \bibinfo{pages}{e3000669}.
\newblock \DOIprefix\doi{10.1371/journal.pbio.3000669}.
%Type = Article
\bibitem[{Simpson(1983)}]{simpson1983}
\bibinfo{author}{Simpson, J.}, \bibinfo{year}{1983}.
\newblock \bibinfo{title}{Elemental and isotopic composition of the galactic
  cosmic rays}.
\newblock \bibinfo{journal}{Annual Review of Nuclear and Particle Science}
  \bibinfo{volume}{33}, \bibinfo{pages}{323--382}.
\newblock \DOIprefix\doi{10.1146/annurev.ns.33.120183.001543}.
%Type = Article
\bibitem[{Slaba et~al.(2017)Slaba, Bahadori, Reddell, Singleterry, Clowdsley
  and Blattnig}]{slaba2017optimal}
\bibinfo{author}{Slaba, T.C.}, \bibinfo{author}{Bahadori, A.A.},
  \bibinfo{author}{Reddell, B.D.}, \bibinfo{author}{Singleterry, R.C.},
  \bibinfo{author}{Clowdsley, M.S.}, \bibinfo{author}{Blattnig, S.R.},
  \bibinfo{year}{2017}.
\newblock \bibinfo{title}{Optimal shielding thickness for galactic cosmic ray
  environments}.
\newblock \bibinfo{journal}{Life Sciences in space research}
  \bibinfo{volume}{12}, \bibinfo{pages}{1--15}.
\newblock \DOIprefix\doi{10.1016/j.lssr.2016.12.003}.
%Type = Article
\bibitem[{Slaba et~al.(2010)Slaba, Blattnig and Badavi}]{slaba2010faster}
\bibinfo{author}{Slaba, T.C.}, \bibinfo{author}{Blattnig, S.R.},
  \bibinfo{author}{Badavi, F.F.}, \bibinfo{year}{2010}.
\newblock \bibinfo{title}{{Faster and more accurate transport procedures for
  HZETRN}}.
\newblock \bibinfo{journal}{Journal of Computational Physics}
  \bibinfo{volume}{229}, \bibinfo{pages}{9397--9417}.
\newblock \DOIprefix\doi{10.1016/j.jcp.2010.09.010}.
%Type = Article
\bibitem[{Slaba and Whitman(2020)}]{slaba2020badhwar}
\bibinfo{author}{Slaba, T.C.}, \bibinfo{author}{Whitman, K.},
  \bibinfo{year}{2020}.
\newblock \bibinfo{title}{{The Badhwar-O'Neill 2020 GCR Model}}.
\newblock \bibinfo{journal}{Space Weather} \bibinfo{volume}{18},
  \bibinfo{pages}{e2020SW002456}.
\newblock \DOIprefix\doi{10.1029/2020SW002456}.
%Type = Inproceedings
\bibitem[{Smith et~al.(2020)Smith, Craig, Herrmann, Mahoney, Krezel, McIntyre
  and Goodliff}]{smith2020artemis}
\bibinfo{author}{Smith, M.}, \bibinfo{author}{Craig, D.},
  \bibinfo{author}{Herrmann, N.}, \bibinfo{author}{Mahoney, E.},
  \bibinfo{author}{Krezel, J.}, \bibinfo{author}{McIntyre, N.},
  \bibinfo{author}{Goodliff, K.}, \bibinfo{year}{2020}.
\newblock \bibinfo{title}{The artemis program: an overview of nasa's activities
  to return humans to the moon}, in: \bibinfo{booktitle}{2020 IEEE aerospace
  conference}, \bibinfo{organization}{IEEE}. pp. \bibinfo{pages}{1--10}.
\newblock \DOIprefix\doi{10.1109/AERO47225.2020.9172323}.
%Type = Article
\bibitem[{Spence et~al.(2010)Spence, Case, Golightly, Heine, Larsen, Blake,
  Caranza, Crain, George, Lalic et~al.}]{spence2010}
\bibinfo{author}{Spence, H.E.}, \bibinfo{author}{Case, A.},
  \bibinfo{author}{Golightly, M.}, \bibinfo{author}{Heine, T.},
  \bibinfo{author}{Larsen, B.}, \bibinfo{author}{Blake, J.},
  \bibinfo{author}{Caranza, P.}, \bibinfo{author}{Crain, W.},
  \bibinfo{author}{George, J.}, \bibinfo{author}{Lalic, M.}, et~al.,
  \bibinfo{year}{2010}.
\newblock \bibinfo{title}{Crater: The cosmic ray telescope for the effects of
  radiation experiment on the lunar reconnaissance orbiter mission}.
\newblock \bibinfo{journal}{Space science reviews} \bibinfo{volume}{150},
  \bibinfo{pages}{243--284}.
\newblock \DOIprefix\doi{10.1007/s11214-009-9584-8}.
%Type = Article
\bibitem[{Stone et~al.(1998a)Stone, Frandsen, Mewaldt, Christian, Margolies,
  Ormes and Snow}]{stone1998advanced}
\bibinfo{author}{Stone, E.}, \bibinfo{author}{Frandsen, A.},
  \bibinfo{author}{Mewaldt, R.}, \bibinfo{author}{Christian, E.},
  \bibinfo{author}{Margolies, D.}, \bibinfo{author}{Ormes, J.},
  \bibinfo{author}{Snow, F.}, \bibinfo{year}{1998}a.
\newblock \bibinfo{title}{The advanced composition explorer}.
\newblock \bibinfo{journal}{Space Science Reviews} \bibinfo{volume}{86},
  \bibinfo{pages}{1--22}.
\newblock \DOIprefix\doi{10.1023/A:1005082526237}.
%Type = Book
\bibitem[{Stone et~al.(1998b)Stone, Cohen, Cook, Cummings, Gauld, Kecman,
  Leske, Mewaldt, Thayer, Dougherty, Grumm, Milliken, Radocinski, Wiedenbeck,
  Christian, Shuman, Trexel, Von~Rosenvinge, Binns, Crary, Dowkontt, Epstein,
  Hink, Klarmann, Lijowski and Olevitch}]{Stone1998}
\bibinfo{author}{Stone, E.C.}, \bibinfo{author}{Cohen, C.M.S.},
  \bibinfo{author}{Cook, W.R.}, \bibinfo{author}{Cummings, A.C.},
  \bibinfo{author}{Gauld, B.}, \bibinfo{author}{Kecman, B.},
  \bibinfo{author}{Leske, R.A.}, \bibinfo{author}{Mewaldt, R.A.},
  \bibinfo{author}{Thayer, M.R.}, \bibinfo{author}{Dougherty, B.L.},
  \bibinfo{author}{Grumm, R.L.}, \bibinfo{author}{Milliken, B.D.},
  \bibinfo{author}{Radocinski, R.G.}, \bibinfo{author}{Wiedenbeck, M.E.},
  \bibinfo{author}{Christian, E.R.}, \bibinfo{author}{Shuman, S.},
  \bibinfo{author}{Trexel, H.}, \bibinfo{author}{Von~Rosenvinge, T.T.},
  \bibinfo{author}{Binns, W.R.}, \bibinfo{author}{Crary, D.J.},
  \bibinfo{author}{Dowkontt, P.}, \bibinfo{author}{Epstein, J.},
  \bibinfo{author}{Hink, P.L.}, \bibinfo{author}{Klarmann, J.},
  \bibinfo{author}{Lijowski, M.}, \bibinfo{author}{Olevitch, M.A.},
  \bibinfo{year}{1998}b.
\newblock \bibinfo{title}{The Cosmic-Ray Isotope Spectrometer for the Advanced
  Composition Explorer}.
\newblock \bibinfo{publisher}{Springer}, \bibinfo{address}{Dordrecht}.
\newblock \DOIprefix\doi{10.1007/978-94-011-4762-0_14}.
%Type = Article
\bibitem[{Tylka et~al.(1997)Tylka, Adams, Boberg, Brownstein, Dietrich,
  Flueckiger, Petersen, Shea, Smart and Smith}]{tylka1997creme96}
\bibinfo{author}{Tylka, A.J.}, \bibinfo{author}{Adams, J.H.},
  \bibinfo{author}{Boberg, P.R.}, \bibinfo{author}{Brownstein, B.},
  \bibinfo{author}{Dietrich, W.F.}, \bibinfo{author}{Flueckiger, E.O.},
  \bibinfo{author}{Petersen, E.L.}, \bibinfo{author}{Shea, M.A.},
  \bibinfo{author}{Smart, D.F.}, \bibinfo{author}{Smith, E.C.},
  \bibinfo{year}{1997}.
\newblock \bibinfo{title}{Creme96: A revision of the cosmic ray effects on
  micro-electronics code}.
\newblock \bibinfo{journal}{IEEE Transactions on Nuclear Science}
  \bibinfo{volume}{44}, \bibinfo{pages}{2150--2160}.
\newblock \DOIprefix\doi{10.1109/23.659030}.
%Type = Article
\bibitem[{Vago et~al.(2015)Vago, Witasse, Svedhem, Baglioni, Haldemann,
  Gianfiglio, Blancquaert, McCoy and De~Groot}]{vago2015esa}
\bibinfo{author}{Vago, J.}, \bibinfo{author}{Witasse, O.},
  \bibinfo{author}{Svedhem, H.}, \bibinfo{author}{Baglioni, P.},
  \bibinfo{author}{Haldemann, A.}, \bibinfo{author}{Gianfiglio, G.},
  \bibinfo{author}{Blancquaert, T.}, \bibinfo{author}{McCoy, D.},
  \bibinfo{author}{De~Groot, R.}, \bibinfo{year}{2015}.
\newblock \bibinfo{title}{Esa exomars program: the next step in exploring
  mars}.
\newblock \bibinfo{journal}{Solar System Research} \bibinfo{volume}{49},
  \bibinfo{pages}{518--528}.
\newblock \DOIprefix\doi{10.1134/S0038094615070199}.
%Type = Article
\bibitem[{Vondrak et~al.(2010)Vondrak, Keller, Chin and
  Garvin}]{vondrak2010lunar}
\bibinfo{author}{Vondrak, R.}, \bibinfo{author}{Keller, J.},
  \bibinfo{author}{Chin, G.}, \bibinfo{author}{Garvin, J.},
  \bibinfo{year}{2010}.
\newblock \bibinfo{title}{{Lunar Reconnaissance Orbiter (LRO): Observations for
  lunar exploration and science}}.
\newblock \bibinfo{journal}{Space science reviews} \bibinfo{volume}{150},
  \bibinfo{pages}{7--22}.
\newblock \DOIprefix\doi{10.1007/s11214-010-9631-5}.
%Type = Article
\bibitem[{Vos and Potgieter(2016)}]{vos2016global}
\bibinfo{author}{Vos, E.}, \bibinfo{author}{Potgieter, M.},
  \bibinfo{year}{2016}.
\newblock \bibinfo{title}{Global gradients for cosmic-ray protons in the
  heliosphere during the solar minimum of cycle 23/24}.
\newblock \bibinfo{journal}{Solar Physics} \bibinfo{volume}{291},
  \bibinfo{pages}{2181--2195}.
\newblock \DOIprefix\doi{10.1007/s11207-016-0945-7}.
%Type = Article
\bibitem[{Wilson et~al.(2014)Wilson, Slaba, Badavi, Reddell and
  Bahadori}]{wilson2014advances}
\bibinfo{author}{Wilson, J.W.}, \bibinfo{author}{Slaba, T.C.},
  \bibinfo{author}{Badavi, F.F.}, \bibinfo{author}{Reddell, B.D.},
  \bibinfo{author}{Bahadori, A.A.}, \bibinfo{year}{2014}.
\newblock \bibinfo{title}{Advances in nasa radiation transport research:
  3d{HZETRN}}.
\newblock \bibinfo{journal}{Life Sciences in Space Research}
  \bibinfo{volume}{2}, \bibinfo{pages}{6--22}.
\newblock \DOIprefix\doi{10.1016/j.lssr.2014.05.003}.
%Type = Article
\bibitem[{Wilson et~al.(1991)Wilson, Townsend, Schimmerling, Khandelwal, Khan,
  Nealy, Cucinotta, Simonsen, Shinn and Norbury}]{wilson1991}
\bibinfo{author}{Wilson, J.W.}, \bibinfo{author}{Townsend, L.W.},
  \bibinfo{author}{Schimmerling, W.S.}, \bibinfo{author}{Khandelwal, G.S.},
  \bibinfo{author}{Khan, F.S.}, \bibinfo{author}{Nealy, J.E.},
  \bibinfo{author}{Cucinotta, F.A.}, \bibinfo{author}{Simonsen, L.C.},
  \bibinfo{author}{Shinn, J.L.}, \bibinfo{author}{Norbury, J.W.},
  \bibinfo{year}{1991}.
\newblock \bibinfo{title}{Transport methods and interactions for space
  radiations}.
\newblock \bibinfo{journal}{NASA RP} \bibinfo{volume}{1257}.
\newblock \URLprefix
  \url{https://spaceradiation.larc.nasa.gov/nasapapers/RP1257.pdf}.
%Type = Article
\bibitem[{Wimmer-Schweingruber et~al.(2020)Wimmer-Schweingruber, Yu,
  B{\"o}ttcher, Zhang, Burmeister, Lohf, Guo, Xu, Schuster, Seimetz
  et~al.}]{wimmer2020lunar}
\bibinfo{author}{Wimmer-Schweingruber, R.F.}, \bibinfo{author}{Yu, J.},
  \bibinfo{author}{B{\"o}ttcher, S.I.}, \bibinfo{author}{Zhang, S.},
  \bibinfo{author}{Burmeister, S.}, \bibinfo{author}{Lohf, H.},
  \bibinfo{author}{Guo, J.}, \bibinfo{author}{Xu, Z.},
  \bibinfo{author}{Schuster, B.}, \bibinfo{author}{Seimetz, L.}, et~al.,
  \bibinfo{year}{2020}.
\newblock \bibinfo{title}{\protect{The Lunar Lander Neutron and Dosimetry (LND)
  Experiment on Chang'E 4}}.
\newblock \bibinfo{journal}{Space Science Reviews} \bibinfo{volume}{216},
  \bibinfo{pages}{1--40}.
\newblock \DOIprefix\doi{10.1007/s11214-020-00725-3}.
%Type = Article
\bibitem[{Xu et~al.(2022)Xu, Guo, Wimmer-Schweingruber, Dobynde, Kühl,
  Khaksarighiri and Zhang}]{Xu_2022}
\bibinfo{author}{Xu, Z.}, \bibinfo{author}{Guo, J.},
  \bibinfo{author}{Wimmer-Schweingruber, R.F.}, \bibinfo{author}{Dobynde,
  M.I.}, \bibinfo{author}{Kühl, P.}, \bibinfo{author}{Khaksarighiri, S.},
  \bibinfo{author}{Zhang, S.}, \bibinfo{year}{2022}.
\newblock \bibinfo{title}{Primary and albedo protons detected by the lunar
  lander neutron and dosimetry experiment on the lunar farside}.
\newblock \bibinfo{journal}{Frontiers in Astronomy and Space Sciences}
  \bibinfo{volume}{9}.
\newblock \DOIprefix\doi{10.3389/fspas.2022.974946}.
%Type = Article
\bibitem[{Zaman et~al.(2022)Zaman, Townsend, de~Wet, Looper, Brittingham,
  Burahmah, Spence, Schwadron and Smith}]{zaman2022modeling}
\bibinfo{author}{Zaman, F.}, \bibinfo{author}{Townsend, L.},
  \bibinfo{author}{de~Wet, W.}, \bibinfo{author}{Looper, M.},
  \bibinfo{author}{Brittingham, J.}, \bibinfo{author}{Burahmah, N.},
  \bibinfo{author}{Spence, H.}, \bibinfo{author}{Schwadron, N.},
  \bibinfo{author}{Smith, S.}, \bibinfo{year}{2022}.
\newblock \bibinfo{title}{{Modeling the lunar radiation environment: A
  comparison among FLUKA, Geant4, HETC-HEDS, MCNP6, and PHITS}}.
\newblock \bibinfo{journal}{Space Weather} \bibinfo{volume}{20},
  \bibinfo{pages}{e2021SW002895}.
\newblock \DOIprefix\doi{10.1029/2021SW002895}.
%Type = Article
\bibitem[{Zaman et~al.(2020)Zaman, Townsend, de~Wet, Schwadron, Spence, Wilson,
  Jordan, Smith and Looper}]{zaman2020absorbed}
\bibinfo{author}{Zaman, F.}, \bibinfo{author}{Townsend, L.W.},
  \bibinfo{author}{de~Wet, W.C.}, \bibinfo{author}{Schwadron, N.A.},
  \bibinfo{author}{Spence, H.E.}, \bibinfo{author}{Wilson, J.K.},
  \bibinfo{author}{Jordan, A.P.}, \bibinfo{author}{Smith, S.S.},
  \bibinfo{author}{Looper, M.D.}, \bibinfo{year}{2020}.
\newblock \bibinfo{title}{Absorbed doses from gcr and albedo particles emitted
  by the lunar surface}.
\newblock \bibinfo{journal}{Acta Astronautica} \bibinfo{volume}{175},
  \bibinfo{pages}{185--189}.
\newblock \DOIprefix\doi{10.1016/j.actaastro.2020.05.040}.
%Type = Article
\bibitem[{Zeitlin et~al.(2013a)Zeitlin, Case, Spence, Schwadron, Golightly,
  Wilson, Kasper, Blake, Looper, Mazur et~al.}]{zeitlin2013measurements}
\bibinfo{author}{Zeitlin, C.}, \bibinfo{author}{Case, A.},
  \bibinfo{author}{Spence, H.E.}, \bibinfo{author}{Schwadron, N.A.},
  \bibinfo{author}{Golightly, M.}, \bibinfo{author}{Wilson, J.K.},
  \bibinfo{author}{Kasper, J.}, \bibinfo{author}{Blake, J.},
  \bibinfo{author}{Looper, M.}, \bibinfo{author}{Mazur, J.}, et~al.,
  \bibinfo{year}{2013}a.
\newblock \bibinfo{title}{Measurements of galactic cosmic ray shielding with
  the crater instrument}.
\newblock \bibinfo{journal}{Space Weather} \bibinfo{volume}{11},
  \bibinfo{pages}{284--296}.
\newblock \DOIprefix\doi{10.1002/swe.20043}.
%Type = Article
\bibitem[{{Zeitlin} et~al.(2016){Zeitlin}, {Case}, {Schwadron}, {Spence},
  {Mazur}, {Joyce}, {Looper}, {Jordan}, {Rios}, {Townsend}, {Kasper}, {Blake},
  {Smith}, {Wilson} and {Iwata}}]{zeitlin2016CRaTER}
\bibinfo{author}{{Zeitlin}, C.}, \bibinfo{author}{{Case}, A.W.},
  \bibinfo{author}{{Schwadron}, N.A.}, \bibinfo{author}{{Spence}, H.E.},
  \bibinfo{author}{{Mazur}, J.E.}, \bibinfo{author}{{Joyce}, C.J.},
  \bibinfo{author}{{Looper}, M.D.}, \bibinfo{author}{{Jordan}, A.},
  \bibinfo{author}{{Rios}, R.R.}, \bibinfo{author}{{Townsend}, L.W.},
  \bibinfo{author}{{Kasper}, J.C.}, \bibinfo{author}{{Blake}, J.B.},
  \bibinfo{author}{{Smith}, S.}, \bibinfo{author}{{Wilson}, J.},
  \bibinfo{author}{{Iwata}, Y.}, \bibinfo{year}{2016}.
\newblock \bibinfo{title}{{Solar modulation of the deep space galactic cosmic
  ray lineal energy spectrum measured by CRaTER, 2009-2014}}.
\newblock \bibinfo{journal}{Space Weather} \bibinfo{volume}{14},
  \bibinfo{pages}{247--258}.
\newblock \DOIprefix\doi{10.1002/2015SW001314}.
%Type = Article
\bibitem[{Zeitlin et~al.(2019)Zeitlin, Hassler, Ehresmann, Rafkin, Guo,
  Wimmer-Schweingruber, Berger and Matthi{\"a}}]{zeitlin2019quality}
\bibinfo{author}{Zeitlin, C.}, \bibinfo{author}{Hassler, D.},
  \bibinfo{author}{Ehresmann, B.}, \bibinfo{author}{Rafkin, S.},
  \bibinfo{author}{Guo, J.}, \bibinfo{author}{Wimmer-Schweingruber, R.F.},
  \bibinfo{author}{Berger, T.}, \bibinfo{author}{Matthi{\"a}, D.},
  \bibinfo{year}{2019}.
\newblock \bibinfo{title}{Measurements of radiation quality factor on {Mars}
  with the {Mars Science Laboratory} {Radiation Assessment Detector}}.
\newblock \bibinfo{journal}{Life Sciences in Space Research}
  \bibinfo{volume}{22}, \bibinfo{pages}{89--97}.
\newblock \DOIprefix\doi{10.1016/j.lssr.2019.07.010}.
%Type = Article
\bibitem[{Zeitlin et~al.(2013b)Zeitlin, Hassler, Cucinotta, Ehresmann,
  Wimmer-Schweingruber, Brinza, Kang, Weigle, B{\"o}ttcher, B{\"o}hm,
  Burmeister, Guo, K{\"o}hler, Martin, Posner, Rafkin and Reitz}]{zeitlin2013}
\bibinfo{author}{Zeitlin, C.}, \bibinfo{author}{Hassler, D.M.},
  \bibinfo{author}{Cucinotta, F.A.}, \bibinfo{author}{Ehresmann, B.},
  \bibinfo{author}{Wimmer-Schweingruber, R.F.}, \bibinfo{author}{Brinza, D.E.},
  \bibinfo{author}{Kang, S.}, \bibinfo{author}{Weigle, G.},
  \bibinfo{author}{B{\"o}ttcher, S.}, \bibinfo{author}{B{\"o}hm, E.},
  \bibinfo{author}{Burmeister, S.}, \bibinfo{author}{Guo, J.},
  \bibinfo{author}{K{\"o}hler, J.}, \bibinfo{author}{Martin, C.},
  \bibinfo{author}{Posner, A.}, \bibinfo{author}{Rafkin, S.},
  \bibinfo{author}{Reitz, G.}, \bibinfo{year}{2013}b.
\newblock \bibinfo{title}{{Measurements of Energetic Particle Radiation in
  Transit to {Mars} on the {Mars Science Laboratory}}}.
\newblock \bibinfo{journal}{Science} \bibinfo{volume}{340},
  \bibinfo{pages}{1080--1084}.
\newblock \DOIprefix\doi{10.1126/science.1235989}.
%Type = Article
\bibitem[{Zhang et~al.(2022)Zhang, Guo, Dobynde, Wang and
  Wimmer-Schweingruber}]{ZhangJ2022JGR}
\bibinfo{author}{Zhang, J.}, \bibinfo{author}{Guo, J.},
  \bibinfo{author}{Dobynde, M.I.}, \bibinfo{author}{Wang, Y.},
  \bibinfo{author}{Wimmer-Schweingruber, R.F.}, \bibinfo{year}{2022}.
\newblock \bibinfo{title}{{From the Top of Martian Olympus to Deep Craters and
  Beneath: Mars Radiation Environment Under Different Atmospheric and Regolith
  Depths}}.
\newblock \bibinfo{journal}{Journal of Geophysical Research: Planets}
  \bibinfo{volume}{127}, \bibinfo{pages}{e2021JE007157}.
\newblock \DOIprefix\doi{10.1029/2021JE007157}.
%Type = Article
\bibitem[{Zhang et~al.(2020)Zhang, Wimmer-Schweingruber, Yu, Wang, Fu, Zou,
  Sun, Wang, Hou, B{\"o}ttcher et~al.}]{zhang2020first}
\bibinfo{author}{Zhang, S.}, \bibinfo{author}{Wimmer-Schweingruber, R.F.},
  \bibinfo{author}{Yu, J.}, \bibinfo{author}{Wang, C.}, \bibinfo{author}{Fu,
  Q.}, \bibinfo{author}{Zou, Y.}, \bibinfo{author}{Sun, Y.},
  \bibinfo{author}{Wang, C.}, \bibinfo{author}{Hou, D.},
  \bibinfo{author}{B{\"o}ttcher, S.I.}, et~al., \bibinfo{year}{2020}.
\newblock \bibinfo{title}{First measurements of the radiation dose on the lunar
  surface}.
\newblock \bibinfo{journal}{Science Advances} \bibinfo{volume}{6},
  \bibinfo{pages}{eaaz1334}.
\newblock \DOIprefix\doi{10.1126/sciadv.aaz1334}.
%Type = Article
\bibitem[{Zheng et~al.(2019)Zheng, Ganushkina, Jiggens, Jun, Meier, Minow,
  O'Brien, Pitchford, Shprits, Tobiska et~al.}]{zheng2019space}
\bibinfo{author}{Zheng, Y.}, \bibinfo{author}{Ganushkina, N.Y.},
  \bibinfo{author}{Jiggens, P.}, \bibinfo{author}{Jun, I.},
  \bibinfo{author}{Meier, M.}, \bibinfo{author}{Minow, J.I.},
  \bibinfo{author}{O'Brien, T.P.}, \bibinfo{author}{Pitchford, D.},
  \bibinfo{author}{Shprits, Y.}, \bibinfo{author}{Tobiska, W.K.}, et~al.,
  \bibinfo{year}{2019}.
\newblock \bibinfo{title}{Space radiation and plasma effects on satellites and
  aviation: Quantities and metrics for tracking performance of space weather
  environment models}.
\newblock \bibinfo{journal}{Space Weather} \bibinfo{volume}{17},
  \bibinfo{pages}{1384--1403}.
\newblock \DOIprefix\doi{10.1029/2018SW002042}.

\end{thebibliography}

%% Authors are advised to use a BibTeX database file for their reference list.
%% The provided style file elsarticle-num.bst formats references in the required Procedia style

%% For references without a BibTeX database:
% \begin{thebibliography}{00}
%% \bibitem must have the following form:
%%   \bibitem{key}...
%%
% \bibitem{}
% \end{thebibliography}

\end{document}